\documentclass[]{extarticle}
\usepackage{amssymb, bbm, amsmath, amsfonts, amsthm, amsbsy, mathtools, mathrsfs, tensor,enumerate,upgreek,dsfont}
\usepackage{graphicx,subfigure,float,subcaption,import}
\graphicspath{ {Images/} }
\usepackage{color}
\usepackage{hyperref}
\usepackage[marginal, multiple]{footmisc}
\usepackage{cleveref}
\usepackage{upgreek}
\usepackage{euscript}
\def \mathes {\EuScript}
\usepackage{epsfig}
\usepackage{cancel}
\usepackage{cite}
\usepackage{simpler-wick}
\usepackage{empheq}
\newcommand*\widefbox[1]{\fbox{\hspace{0.5em}#1\hspace{0.5em}}}

\newlength{\abstractwidth}
\newcommand {\pd} [2] {\frac {\partial #1} {\partial #2}}

\newcommand {\tdrm} [2] {\frac {\mathrm{d} #1} {\mathrm{d} #2}}
\newcommand {\tdrmord} [3] {\frac {\mathrm{d}^{#1} #2} {\mathrm{d} #3^{#1}}}

\newcommand {\intf} [2] {\int_{#1}^{#2}}

\newcommand{\inp} [1] {\left( #1 \right)}

\newcommand{\insb} [1] {\left[ #1 \right]}
\newcommand {\bs}[1] {\textcolor{blue}{#1}} 

\newcommand {\ket} [1] {\left|\, #1 \,\right>}
\newcommand {\bra} [1] {\left <\, #1 \,\right|}

\newcommand{\be}{\begin{equation}}
\newcommand{\bea}{\begin{eqnarray}}
\newcommand{\eea}{\end{eqnarray}}
\newcommand{\beq}{\begin{equation}}
\newcommand{\ee}{\end{equation}}

\def\ellds{\ell_{\mathrm{dS}}}
\def\V{\Omega_{(D-2)}}
\def \ve {\varepsilon\,}

\def \real {\text{Re}\,}

\def\C{\mathbb{C}}

\def\R{\mathbb{R}}
\def\Z{\mathbb{Z}}

\numberwithin{equation}{section}

\theoremstyle{definition}

\theoremstyle{remark}

\usepackage{authblk}
\title{New Modes for Vector Bosons in the Static Patch}
\author[1]{Adel Rahman}
\author[1,2]{Leonard Susskind}
\affil[1]{Stanford Institute for Theoretical Physics and Department of Physics\\ Stanford University, Stanford, CA 94305-4060, USA \vspace{1em}}
\affil[2]{Google, Mountain View, CA}
\date{}
\begin{document}

\maketitle
\begin{abstract}
We consider a massive vector Boson in a static patch of $D$-dimensional de Sitter space (dS$_D$). We argue that this field is controlled by an effective physical (squared) mass ${\mu_{\mathrm{v}}^2 = m_{\mathrm{v}}^2 + 2\inp{D-1}\ell_{\mathrm{dS}}^{-2}}$ which differs from the na\"ive ``Lagrangian" (squared) mass $m_{\mathrm{v}}^2$ that appears in the usual form of the Proca Lagrangian/action. In particular, we conjecture that the theory remains well-defined in the na\"ively tachyonic Lagrangian mass range $-2\inp{D-1} < m_{\mathrm{v}}^2\ellds^2 < 0$. The width of this range and the discrepancy between the physical and Lagrangian masses vanishes in the flat space limit, but is nontrivial for finite cosmological constant. We identify several interesting physical features of the ``edge of stability" $m_{\mathrm{v}}^2\ellds^2 = -2\inp{D-1}$. Fixing a static patch breaks the $D$-dimensional de Sitter isometries down to a ``static patch subgroup", which explains why our theory may continue to be well-defined within the above mass range despite not fitting into a unitary irreducible representation of SO$(D,1)$. We conjecture that for situations such as ours, the usual $\mathrm{SO}(D,1)$ ``Higuchi bound" on unitarity is replaced by the concept of the edge of stability. In $D = 3$ spacetime dimensions, the $s$-wave sector of our theory remarkably simplifies, becoming equivalent to the $p$-wave sector of an ordinary massive scalar. In this case we can explicitly check that the $D = 3$ $s$-wave sector remains well-defined---both classically and quantum mechanically---in the above mass range. In the course of our analysis, we will derive the general classical solution and the quasinormal frequency spectrum for the massive vector Boson in the static patch of dS$_D$, generalizing previous work by Higuchi \cite{Higuchi:1986ww} which was done for the special case $D = 4$. While this work was being completed, we became aware of upcoming work by Grewal, Law, and Lochab \cite{Albert} which will contain a similar derivation.

\end{abstract}

\newpage 
\setcounter{tocdepth}{2}
\tableofcontents
\newpage 

\section{Introduction}
\quad \
$D$-dimensional Lorentzian de Sitter space
\begin{equation}
	\mathrm{dS}_{D} = (\mathcal{M}_{D},\mathsf{g}_{\mu\nu})
\end{equation}
is a maximally symmetric spacetime with  isometry group
\begin{equation}
	\mathrm{Iso}(\mathrm{dS}_D) = \mathrm{O}(D,1)
	\label{IsodS}
\end{equation}
In the usual analysis of field theory in de Sitter space, the isometry group \eqref{IsodS} serves to constrain possible forms of matter, which are required to furnish unitary irreducible representations of $\mathrm{Iso}(\mathrm{dS}_D)$ (or at least of its identity component $\mathrm{SO}^+(D,1)$). In this paper we will discuss a situation in which this symmetry group is explicitly broken, thereby freeing us to consider novel parameter ranges for matter fields which would otherwise be disallowed by the constraints of $\mathrm{SO}(D,1)$ representation theory. Specifically, we will consider what happens when one fixes a particular static patch of de Sitter space, thereby breaking the isometry group \eqref{IsodS} down to a ``static patch subgroup" $\mathrm{O}(1,1)\times\mathrm{O}(D-1)$. 

There are several situations in which one might wish to fix a particular static patch of de Sitter space. Classically, one may wish to study a ``compactified" theory, in which one isolates a warped $\mathbb{S}^{(D-2)}$ factor of the geometry to then supress; as we will explain below, picking any particular notion of ``warped $\mathbb{S}^{(D-2)}$ factor" implicitly singles out a particular choice of (antipodal pair of) static patch(es). Quantum mechanically, it is expected that the de Sitter isometries \eqref{IsodS} are gauge symmetries of a complete quantum mechanical description of de Sitter space (see e.g. \cite{Goheer:2002vf,Marolf:2008hg,Susskind:2021omt, Chandrasekaran:2022cip}); fixing a particular static patch can therefore be viewed as a form of (partial) gauge-fixing. Indeed, such a gauge fixing has served as the starting point for several recent studies on semiclassical and quantum aspects of de Sitter space including the conjectured duality between high-temperature double-scaled SYK (DSSYK$_{\infty}$) and dimensionally reduced 3D de Sitter space \cite{HV, Susskind:2021esx,Susskind:2022dfz,Susskind:2022bia,Susskind:2023hnj,Narovlansky:2023lfz,Rahman:2023pgt,Rahman:2024vyg,Verlinde:2024znh,Rahman:2024iiu} (see for example \cite{Rahman:2022jsf}) as well as recent work on algebras of observables \cite{Chandrasekaran:2022cip}. See also \cite{Banks:2003cg,Banks:2006rx,Anninos:2011af, Gorbenko:2018oov,Coleman:2021nor,Batra:2024kjl,Batra:2024qju,Ball:2024hqe,Ball:2024xhf,Grewal:2024emf,Banks:2022irh,A:2023psv,Banks:2024cqo,Banks:2024lvl} for related situations involving a fixed static patch, such as the ``$T\bar{T} + \Lambda_2$" approach to de Sitter holography \cite{Gorbenko:2018oov,Coleman:2021nor,Batra:2024kjl} and various works over the years by Banks and collaborators \cite{Banks:2003cg,Banks:2006rx,Banks:2022irh,A:2023psv,Banks:2024cqo,Banks:2024lvl}.

In this paper we will discuss an intriguing feature of massive minimally-coupled ``spin-1" vector Bosons in de Sitter space which arises when they are studied relative to a fixed static patch frame. We find that this field appears to be controlled by an effective physical (squared) mass ${\mu_{\mathrm{v}}^2 = m_{\mathrm{v}}^2 + 2\inp{D-1}\ell_{\mathrm{dS}}^{-2}}$ which differs from the na\"ive ``Lagrangian" mass $m_{\mathrm{v}}^2$ appearing in the usual form of the ``Proca" Lagrangian/action \eqref{IA}. We will argue, via an anlysis of the quasinormal frequency spectrum, that this theory remains (at least classically) well-defined in the na\"ively tachyonic mass range $-2\inp{D-1} < m_{\mathrm{v}}^2\ellds^2 < 0$. The width of this range and the discrepancy between the physical and Lagrangian masses vanishes in the flat space limit, but is nontrivial for finite cosmological constant. In $D = 3$ spacetime dimensions, the $s$-wave sector of this theory remarkably simplifies, becoming equivalent to the $p$-wave sector of an ordinary massive scalar with a particular mass. In this case we can explicitly check that the $D = 3$ $s$-wave sector remains well-defined---both classically and quantum mechanically---in the above mass range.  

We will also identify several interesting physical features of the ``edge of stability" $m_{\mathrm{v}}^2\ellds^2 = -2\inp{D-1}$, namely the emergence of static solutions, zero modes, and global shift symmetries. The last of these was previously reported in \cite{Bonifacio:2018zex}, and we will provide a fresh perspective using the ``edge of stability" as a unifying concept. The edge of stability can be thought of as the concept which replaces the usual $\mathrm{SO}(D,1)$ ``Higuchi bound" \cite{Higuchi:1986py} on unitarity (see also \cite{Anninos:2020hfj,Lust:2019lmq}) once we break the symmetries of the problem down to just the symmetries of a static patch. In the solvable case of the $s$-wave mode in $D = 3$ spacetime dimensions, we also identify the emergence of ``infrared divergences"/quantization ambiguities analogous to those which appear in the massless limit of the minimally-coupled scalar field in de Sitter space (see e.g. \cite{Allen:1985ux,Allen:1987tz,Grewal:2024emf}).

While the role of the effective physical mass $\mu_{\mathrm{v}}$ first came to our attention in the context of the conjectured DSSYK$_{\infty}$/dS duality, the phenomena that we will report on here are generic features of ordinary bulk de Sitter space (of any spacetime dimension $D\geq 3$) that should be true independent of any possible holographic duality. In particular, we expect that the results which will be reported here will be of general interest to anyone working on the physics of de Sitter space and/or cosmology. In the course of our analysis, we find the general classical solution and quasinormal frequency spectrum for the massive minimally-coupled vector Boson in the static patch, generalizing previous work by Higuchi \cite{Higuchi:1986ww} (which was done for the special case $D = 4$). While this work was being completed, we became aware of upcoming work by Grewal, Law, and Lochab \cite{Albert} which will contain a similar derivation of the classical solutions.

\subsection{Some Notation and Conventions}
\label{Notation}
\subsubsection*{Notation for $s$-Wave Modes}
\quad \ In this paper, we will adapt the following notation:
\begin{quote}
	\begin{center}
		\emph{We will denote the metric, coordinates, general scalar fields, and general vector fields on dS$_D$\\ by $\mathsf{g}_{\mu\nu}$, $\mathsf{x}^{\mu}$, $\upphi$, and $\mathsf{A}_{\mu}$ respectively.}
	\end{center} 
\end{quote} 
All quantities derived from these objects (e.g. field strength tensors, Green's functions etc.) will be similarly denoted by $\mathsf{serif}$ font. All $D$-dimensional indices ($\mu,\nu$ etc.) will be raised and lowered using the $D$-dimensional metric $\mathsf{g}_{\mu\nu}(\mathsf{x})$. We will denote the covariant derivative operator associated to the Levi-Civita connection of $\mathsf{g}_{\mu\nu}$ by $\nabla_{\mu}$ and we will denote $D$-dimensional densities of weight one by boldfaced capital letters, e.g. $\mathbf{J}^{\mu}$.

\begin{quote}
	\begin{center}
		\emph{We will denote the $s$-wave parts of the metric, coordinates, and vector fields on dS$_D$\\ by $g_{ab}$, $x^a$, 
		and $A_a$ respectively.
		}
	\end{center} 
\end{quote}
All quantities derived from these objects will appear in ordinary font. All $(1 + 1)$-dimensional indices ($a, b$ etc.) will be raised and lowered using the $(1 + 1)$-dimensional metric $g_{ab}(x)$. We will denote the \emph{$s$-wave reduction of} $D$-dimensional densities of weight one by boldfaced lowercase letters, e.g. $\mathbf{j}^{a}$. We will never make use of the covariant derivative operator associated to the Levi-Civita connection of $g_{ab}$, nor will we have ocassion to single out the $s$-wave part of a scalar field.

\section{Preliminaries: Static Patches and Spherical Decomposition}
\label{BulkPrelim}
\subsection{Static Patches}
\label{StaticPatch}
\begin{figure}[H]
	\centering
	\begin{normalsize}
		\scalebox{0.85}{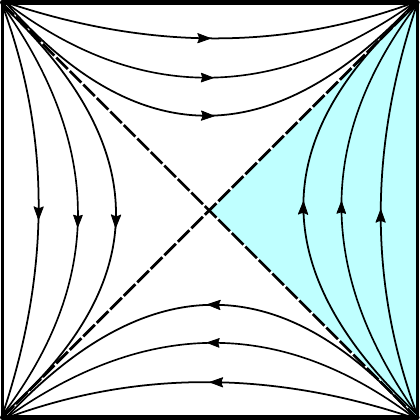}
	\end{normalsize}
	\caption{Orbits of the $\mathrm{SO}(1,1)$ symmetry of dS$_D$ associated with a particular choice of static patch (shaded in blue), which is the region where these orbits are timelike and (for definiteness) future-directed.}
	\label{fig:orbits}
\end{figure}

\quad \
A ``static patch" (SP) of dS$_D$ is the domain of dependence of a complete timelike curve (worldline), modeling the spacetime region causally accessible to a localized massive observer. Acting with the isometry group \eqref{IsodS}, one can transform different worldlines/static patches into one another; conversely, fixing a particular static patch breaks the de Sitter isometry group \eqref{IsodS} down to a ``static patch subgroup"
\begin{equation}
	\mathrm{Iso}(\mathrm{SP}) \ \simeq \ \underbrace{\mathrm{SO}(1,1)\rtimes\Z_2}_{\mathrm{O}(1,1)} \,\times\, \mathrm{O}(D-1)
	\label{IsoSP}
\end{equation}
The $\mathrm{O}(D-1)$ ``spherical symmetry" expresses the isotropy of de Sitter space relative to a given observer. Orbits of $\mathrm{O}(D-1)$ are round codimension-2 spheres which we will refer to as the ``local $(D-2)$-spheres"\footnote{The local $(D-2)$-spheres are precisely the objects which are suppressed when writing down ``the" dS$_D$ Penrose diagram. We therefore see that the process of fixing a static patch is precisely the same as the process of writing down a concrete Penrose diagram for dS$_D$ (which is equivalently nonunique).}. We should think of the notion of spherical symmetry and of local $(D-2)$-spheres as being ``observer-dependent" in the same way that the de Sitter cosmological horizon (see below) is observer-dependent. Orbits of the $\mathrm{SO}(1,1) \simeq \R$ factor---which we will refer to as the ``boost" symmetry of dS$_D$---are shown on the associated de Sitter Penrose diagram in figure \ref{fig:orbits}. The static patch under consideration is the region where these orbits are timelike and (for definiteness) future-directed. 

Within a given static patch, we may erect ``static patch coordinates" $\mathsf{x}^{\mu} = (t, r, \theta^A)$ adapted to the static patch isometry group \eqref{IsoSP}, in terms of which the metric takes the form
\begin{equation}
	\boxed{\mathsf{g}_{\mu\nu}(\mathsf{x})\,\mathrm{d}\mathsf{x}^{\mu}\mathrm{d}\mathsf{x}^{\nu}\Big|_{\text{static patch}} = -f(r)\,\mathrm{d}t^2 + \frac{\mathrm{d}r^2}{f(r)} + r^2\,\Omega_{AB}(\theta)\,\mathrm{d}\theta^A\mathrm{d}\theta^B}
	\label{dS}
\end{equation}
with the blueshift factor $f(r)$ given by
\begin{equation}
	\boxed{f(r) = 1 - \frac{r^2}{\ell_{\mathrm{dS}}^2}}
\end{equation}
The ``radial" coordinate $r \in [0,\ellds)$ parameterizes the distance from the defining worldline and also serves to label the local $(D-2)$-spheres, which are the codimension-2 surfaces of fixed $x^a = (t,r)$. The ``angular" coordinates $\theta^A$ are dimensionless coordinates on the local $(D-2)$-spheres, so that
\begin{equation}
	\mathrm{d}\Omega^2_{(D-2)} \equiv \Omega_{AB}(\theta)\,\mathrm{d}\theta^A\mathrm{d}\theta^B
\end{equation}
is the metric of the round unit $(D-2)$-sphere. Geometrically, the local $(D-2)$-sphere at fixed $x^a = (t,r)$ is a round $(D-2)$-sphere of radius $r$ centered about the point $r = 0$; similarly, surfaces of constant $r$ are spherically-symmetric round ``world-tubes" centered about the worldline $r = 0$. Following previous works by the authors, we will refer to the point $r = 0$ (at fixed $t$) as the ``pode" and we will refer to the full worldline $r = 0$ as the worldline of the pode.

``Static patch time" $t$ is a dimensionful coordinate on the orbits of the boost symmetry which runs over $\R$ and which is normalized to agree with proper time along the worldline of the pode\footnote{The $\Z_2$ factor in $\mathrm{O}(1,1) \simeq \mathrm{SO}(1,1)\rtimes\Z_2$ corresponds to time reversals $t \to - t$. More accurately, it corresponds to the ``CPT" operation which flips time and also exchanges the static patch with its ``antipodal" partner.}. The static patch is surrounded by an event horizon---the \emph{cosmological horizon}---at $r = \ell_{\mathrm{dS}}$ which is a bifurcate Killing horizon for the boost Killing field $(\partial/\partial t)$. Indeed, the coordinate $t$ breaks down along the horizon, where $t \to \pm\infty$ and the Killing field $(\partial/\partial t)$ becomes null (as its orbits transition from being timelike within the static patch to spacelike ``behind the horizon").   Surfaces of constant $t$ are round $(D-1)$-disks centered about the pode and bounded by the bifurcation surface of the cosmological horizon.

\begin{figure}[H]
	\centering
	\begin{normalsize}
		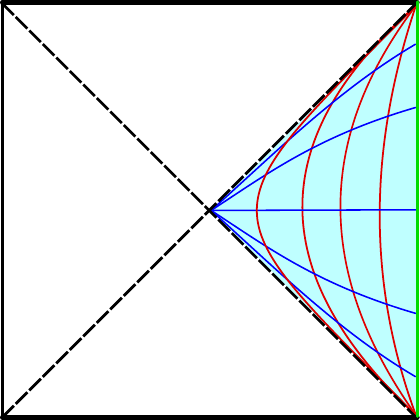
	\end{normalsize}
	\caption{Static patch coordinates cover the region shaded in light blue on the assocaited dS$_D$ Penrose diagram. The solid green line is the worldline $r = 0$ which defines the static patch; the red lines are surfaces of constant $r$; the dashed black lines denote the bifurcate cosmological horizon; and the blue lines are surfaces of constant $t$. As usual, each point on the Penrose diagram represents a suppressed local $(D-2)$-sphere.}
	\label{fig:static}
\end{figure}

\noindent Note that in the limit $\ellds \to \infty$ with all else being held fixed, the static patch approaches flat $D$-dimensional Minkowski space (in polar coordinates), with the cosmological horizon going over to the flat space asymptotic boundary\footnote{Note that this is different from the near-horizon limit of dS$_D$, in which the near-horizon geometry approaches that of $D$-dimensional Rindler space, with the cosmological horizon going over to the Rindler horizon.}. In what follows, we will refer to this limit as the ``flat space limit" of dS$_D$.

\subsection{The Spherical Decomposition}
\label{dSDdS2p}
\quad \
We can extend the notion of spherical symmetry and local $(D-2)$-spheres to global de Sitter space, though this extension must be done relative to a fixed initial static patch frame. For example, we can define a global time coordinate $\tau$ relative to the initial static patch frame \eqref{dS} via 
\begin{equation}
	\sinh\bigg(\frac{\tau}{\ellds}\bigg) = \sqrt{f(r)}\,\sinh\bigg(\frac{t}{\ellds}\bigg)
\end{equation}
We can similarly define a global distance coordinate $x$ relative to the initial static patch frame via
\begin{equation}
	\cos(x) = \frac{1}{\sqrt{1 + f(r)\sinh^2(t/\ell)}}\inp{\frac{r}{\ellds}}
\end{equation}
In terms of these coordinates, the metric in \emph{global} dS$_D$ takes the $O(D-1)$-symmetric form 
\begin{equation}
	\mathsf{g}_{\mu\nu}(\mathsf{x})\,\mathrm{d}\mathsf{x}^{\mu}\mathrm{d}\mathsf{x}^{\nu} = -\mathrm{d}\tau^2 + \cosh^2\bigg(\frac{\tau}{\ellds}\bigg)\inp{\mathrm{d}x^2 + \sin^2(x)\,\mathrm{d}\Omega^2_{(D-2)}}
	\label{global}
\end{equation}
In general, we can discuss the spherical symmetry and local $(D-2)$-spheres in a way that in covariant in the direction transversal to the local $(D-2)$-spheres. Specifically, we can \emph{globally} decompose the metric of dS$_D$ as
\begin{equation}
	\boxed{\mathsf{g}_{\mu\nu}(\mathsf{x})\,\mathrm{d}\mathsf{x}^{\mu}\mathrm{d}\mathsf{x}^{\nu} = g_{ab}(x)\,\mathrm{d}x^a\mathrm{d}x^b + r(x)^2\mathrm{d}\Omega^2_{(D-2)}}
	\label{LTDecomp}
\end{equation}
with $x^a$ a coordinate on the $(1 + 1)$-dimensional spacetime transversal to the local $(D- 2)$-spheres. This expresses the geometry of dS$_D$ as a warped product of a round $(D-2)$-sphere $\mathbb{S}^{(D-2)}$ with a $(1+1)$-dimensional spacetime which, following \cite{Rahman:2022jsf}, we will call $\mathrm{dS}_2{}'$. We will call the decomposition \eqref{LTDecomp} the \emph{spherical decomposition} of dS$_D$ and refer to the directions along the local $(D-2)$-spheres as the ``spherical" directions. The $D$-dimensional volume element similarly decomposes as
\begin{equation}
	\sqrt{|\mathsf{g}(\mathsf{x})|}\,\mathrm{d}^D\mathsf{x} = r(x)^{D-2}\,\sqrt{|g(x)|}\,\mathrm{d}^2x\,\sqrt{\Omega}\ \mathrm{d}^{D-2}\theta
\end{equation}
where we have defined, as usual,
\begin{equation}
	\mathsf{g} \equiv \det(\mathsf{g}_{\mu\nu}), \qquad \Omega \equiv \det\inp{\Omega_{AB}}, \qquad g \equiv \det(g_{ab})
\end{equation}

The radial function $r(x)$ appearing in \eqref{LTDecomp} encodes the sizes of the local $(D-2)$-spheres. In particular, the local $(D-2)$-sphere at the point $x$ has area 
\begin{equation}
	\mathrm{Area}(x) = \Omega_{(D-2)}\,r(x)^{D-2}
	\label{Adef}
\end{equation}
---and therefore radius $r(x)$---in the full dS$_D$ ``parent spacetime", where we have defined
\begin{equation}
	\Omega_{(D-2)} \equiv \int_{\mathbb{S}^{(D-2)}}\mathrm{d}^{D-2}\theta\,\sqrt{\Omega}
\end{equation}
to be the $(D-2)$-volume of the round unit $(D-2)$-sphere. Note that picking any particular notion of ``local $(D-2)$-spheres"---i.e. any particular spherical/warped product decomposition \eqref{LTDecomp} of dS$_D$---breaks the de Sitter isometries \eqref{IsodS} down to a static patch subgroup. One way to see this is to notice that fixing a particular choice of spherical decomposition \eqref{LTDecomp} in turn fixes a particular choice of (antipodal pair of) static patch(es). This is due to the fact that the spherical decompositon \eqref{LTDecomp} degenerates along the two antipodal timelike worldlines where $r(x) = 0$ (and where the size of the local spheres consequently goes to zero). These two distinguished antipodal worldlines can then be used to define an antipodal pair of static patches. We see that any given spherical decomposition \eqref{LTDecomp} determines a corresponding antipodal pair of static patches (and vice-versa).

\subsection{The Spherical Decomposition with Matter}
\quad \ 
Once we have established a particular choice of spherical decomposition \eqref{LTDecomp} (or, equivalently, a particular choice of static patch), matter fields in dS$_D$ can be correspondingly decomposed into modes of fixed angular momentum on the local $(D-2)$-spheres. For example, for a scalar field $\upphi(\mathsf{x})$ on dS$_D$, we can write
\begin{equation}
	\upphi(\mathsf{x}) = \sum_{lm}\phi_{lm}(x)\,Y_{lm}(\theta)
	\label{Scalarl}
\end{equation}
where $Y_{lm}$ are the scalar spherical harmonics on $\mathbb{S}^{(D-2)}$, obeying
\begin{equation}
	-\frac{1}{\sqrt{\Omega}}\,\partial_A\inp{\sqrt{\Omega}\,\Omega^{AB}\partial_BY_{lm}} = l\inp{l + D - 3}Y_{lm}
	\label{SH}
\end{equation}
In this paper we will choose to normalize the spherical harmonics via\footnote{The more common normalization convention omits the factor of $1/\Omega_{(D-2)}$ from the left hand side of \eqref{NC} so that the lowest spherical harmonic is given by $Y_{00}^{(\mathrm{usual})} = \Omega_{(D-2)}^{-1/2}$. For example, taking $D = 4$, we find the familiar expression $Y_{00}^{(\mathrm{usual})} = 1/\sqrt{4\pi}$. In this paper we have normalized our spherical harmonics so that $Y_{00} = 1$ independent of dimension.} 
\begin{equation}
	\frac{1}{\Omega_{(D-2)}}\int\ \mathrm{d}^{D-2}\theta\,\sqrt{\Omega}\,Y^*_{lm}(\theta)\,Y^{}_{l'm'}(\theta) = \delta_{ll'}\delta_{mm'}
	\label{NC}
\end{equation}
so that 
\begin{equation}
	\phi_{lm}(x) = \frac{1}{\Omega_{(D-2)}}\int\ \mathrm{d}^{D-2}\theta\,\sqrt{\Omega}\,Y^*_{lm}(\theta)\,\upphi(x,\theta)
\end{equation}
With this normalization convention \eqref{NC}, we have $Y_{00} = 1$ independent of dimension, so that a spherically-symmetric field configuration is automatically equal to its $s$-wave reduction (see below). Decomposing the field in this way is sometimes known as ``compactifying" the theory on the local $(D-2)$-sphere. Compactification turns a $D$-dimensional local field theory into a $(1 + 1)$-dimensional local field theory on the transversal spacetime dS$_2{}'$, albiet with an infinite number of fields (each of which is individually nonlocal on the supressed local $(D-2)$-sphere). 

As the example which will be most relevant for this paper, let us also consider a spin-1 vector field on dS$_D$, which we begin by decomposing into nonspherical and spherical parts
\begin{equation}
	\mathsf{A}_{\mu}(\mathsf{x})\,\mathrm{d}\mathsf{x}^{\mu} = \mathsf{A}_a(\mathsf{x})\,\mathrm{d}x^a + \mathsf{A}_{A}(\mathsf{x})\,\mathrm{d}\theta^A
	\label{split}
\end{equation}
The nonspherical part is a $(D-2)$-sphere scalar and can correspondingly be decomposed as 
\begin{equation}
	\mathsf{A}_a(\mathsf{x}) = \sum_{lm}A_{a}^{lm}(x)\,Y_{lm}(\theta)
	\label{Aal}
\end{equation}
The spherical part is a $(D-2)$-sphere vector, which can first be decomposed into ``sphere-longitudinal" ($D_A\mathes{A}$) and ``sphere-transverse" ($\tilde{\mathsf{A}}_A$) parts 
\begin{equation}
	\mathsf{A}_A = D_A\mathes{A} + \tilde{\mathsf{A}}_A
	\label{LTVector}
\end{equation}
Here $\mathes{A}$ is a scalar field and $\tilde{\mathsf{A}}_A$ is a vector field which is ``sphere-transverse" (i.e. sphere divergence-free) 
\begin{equation}
	\Omega^{AB}D_A\tilde{\mathsf{A}}_B = 0
	\label{TDef}
\end{equation}
Here and below $D_A$ denotes the covariant derivative operator associated to the Levi-Civita connection of $\Omega_{AB}$, i.e. of the round unit sphere. Note that the sphere-longitudinal and sphere-transverse parts of $\mathsf{A}_A$ are orthogonal with respect to the natural inner product (by virtue of \eqref{TDef})
\begin{equation}
	\int\mathrm{d}^{D-2}\theta\,\sqrt{\Omega}\,\Omega^{AB}D_A\mathes{A}\,\tilde{\mathsf{A}}_B = - \int\mathrm{d}^{D-2}\theta\,\sqrt{\Omega}\,\mathes{A}\,\Omega^{AB}D_A\tilde{\mathsf{A}}_B 
	= 0
	\label{LTO}
\end{equation}

The sphere-longitudinal part $D_A\mathes{A}$ can be decomposed as 
\begin{equation}
	D_A\mathes{A}(\mathsf{x}) = \sum_{l\geq 1,m}\mathes{A}^{lm}(x)\,D_AY_{lm}(\theta)
\end{equation}
Note that the sum starts at $l = 1$ by virtue of the fact that $Y_{00} = 1 \implies D_AY_{00} = 0$. For $D = 3$, the sphere-transverse part $\tilde{\mathsf{A}}_A$ is simply given by a multiple of the volume form 
\begin{equation}
	\tilde{\mathsf{A}}_{\theta}(\mathsf{x})\,\mathrm{d}\theta = C(x)\,\mathrm{d}\theta, \qquad \inp{D = 3}
\end{equation}
For $D > 3$, the sphere-transverse part can be decomposed in terms of the ``spin-1 transverse vector harmonics" $\tilde{\mathsf{Y}}_A^{lm}$ on $\mathbb{S}^{(D-2)}$ (see Appendix \eqref{SHApp}):
\begin{equation}
	\tilde{\mathsf{A}}_A(\mathsf{x})\,\mathrm{d}\theta^A = \sum_{l\geq 1, m}\tilde{A}_{lm}(x)\,\tilde{\mathsf{Y}}_A^{lm}(\theta)\,\mathrm{d}\theta^A
	\label{TD}
\end{equation}
which satisfy 
\begin{equation}
	-\Omega^{BC}D_BD_C\tilde{\mathsf{Y}}^{lm}_A = \insb{l\inp{l + D - 3} - 1}\tilde{\mathsf{Y}}^{lm}_A
\end{equation}
and 
\begin{equation}
	\Omega^{AB}D_A\tilde{\mathsf{Y}}_B^{lm} = 0
	\label{Ttilde}
\end{equation}
Note that these are defined only for $l \geq 1$ (see Appendix \ref{SHApp} for further details). In this paper we will choose to normalize the vector spherical harmonics via
\begin{align}
	\frac{1}{\V}\int\ \mathrm{d}^{D-2}\theta\,\sqrt{\Omega}\,\Omega^{AB}\,\tilde{\mathsf{Y}}_A^{lm}\tilde{\mathsf{Y}}_B^{l'm'} = \delta^{ll'}\delta^{mm'}
\end{align}
Note that we will also automatically have (by virtue of \eqref{SH})
\begin{align}
	\frac{1}{\V}\int\ \mathrm{d}^{D-2}\theta\,\sqrt{\Omega}\,\Omega^{AB}\,D_AY_{lm}D_BY_{l'm'} = l\inp{l + D - 3}\delta_{ll'}\delta_{mm'}
\end{align}
as well as (by virtue of \eqref{Ttilde})
\begin{align}
	\frac{1}{\V}\int\ \mathrm{d}^{D-2}\theta\,\sqrt{\Omega}\,\Omega^{AB}\,\tilde{\mathsf{Y}}_A^{lm}D_BY_{lm'} = 0
	\label{orth}
\end{align}

As explained in \S\ref{dSDdS2p} above, the spherical decomposition degenerates at the pode. This necessitates supplementing the fields $\upphi(\mathsf{x})$, $A_a^{lm}$, $\mathes{A}^{lm}$, $\tilde{A}_{lm}$ etc. with appropriate boundary conditions there. These boundary conditions should reflect the fact that these compactification modes descend from ``parent" fields $\upphi(\mathsf{x})$, $\mathsf{A}_{\mu}$ which live in global dS$_D$, for which the pode is not a distinguished point. In other words, these boundary conditions should reflect the fact that the parent fields $\upphi(\mathsf{x})$, $\mathsf{A}_{\mu}$ smoothly pass through the center of the parent static patch. We will describe these boundary conditions in more detail in \S\ref{Prelim2} below.

\subsubsection{The ``$s$-Wave" Mode}
\quad \ 
A key point of focus for this paper will be the spherically symmetric---i.e. $\mathrm{O}(D-1)$-invariant/$\mathrm{O}(D-1)$ singlet---``\emph{s-wave}" mode of $\mathsf{A}_{\mu}$.  For scalar fields and for vector fields in $D > 3$ spacetime dimensions, ``$s$-wave reduction" (i.e. projecting out all modes except the $s$-wave mode) is equivalent to simply restricting to the $l = 0$ sector. For vector fields in $D = 3$ spacetime dimensions, $s$-wave reduction additionally requires projecting out the $l = 0$ ``circularly polarized mode" proportional to $\mathrm{d}\theta$, which fails to be invariant under the $\mathrm{d}\theta \to -\mathrm{d}\theta$ antipodal/parity symmetry of the local circle\footnote{In other words, this mode is $\mathrm{SO}(2)$ invariant but not $\mathrm{O}(2)$ invariant.}. In any case, we will adapt the notation that the $s$-wave part of a field is denoted by removing the $lm$ index in \eqref{Scalarl}, e.g.
\begin{equation}
	\mathsf{A}_{\mu}(\mathsf{x})\,\mathrm{d}\mathsf{x}^{\mu}\big|_{\text{$s$-wave}} = A_{a}(x)\,\mathrm{d}x^a, \qquad A_a(x) \equiv A_{a}^{00}(x)
\end{equation}
For fields which are $(D-2)$-sphere scalars (such as $\upphi$ and $\mathsf{A}_a$), $s$-wave reduction is equivalent to homogenization on the local $(D-2)$-sphere, e.g.
\begin{equation}
	\mathsf{A}_a(\mathsf{x}) \ \longrightarrow \ A_a(x) = \frac{1}{\Omega_{(D-2)}}\int\mathrm{d}^{D-2}\theta\,\sqrt{\Omega}\,\mathsf{A}_a(x,\theta)
	\label{phi}
\end{equation}
For fields which are $(D-2)$-sphere scalar \emph{densities} (of weight one), $s$-wave reduction is equivalent to \emph{coordinate} homogenization on the local circle, e.g.
\begin{equation}
	\boldsymbol{\mathsf{J}}_{a}(\mathsf{x}) \ \longrightarrow \ \frac{1}{\Omega_{(D-2)}}\int\mathrm{d}^{(D-2)}\theta\,\boldsymbol{\mathsf{J}}_a(x,\theta)
	\label{A}
\end{equation}
Our notational conventions for $s$-wave modes are summarized in \S\ref{Notation} above.

\section{Matter Fields in dS$_D$}
\label{Prelim2}
\subsection{Warm Up: Scalar Fields in dS$_D$}
\label{Scalars}
\quad \ 
While we will ultimately be interested in the physics of the the massive vector Boson, we will find it helpful to first consider the physics of the ``spin-0" massive real scalar field $\upphi$, minimally-coupled to the metric of dS$_{D}$. This field is governed by the familiar action
\begin{equation}
	I[\upphi] = -\frac{1}{2}\int_{\mathcal{M}_D}\mathrm{d}^{D}\mathsf{x}\sqrt{|\mathsf{g}|}\inp{\mathsf{g}^{\mu\nu}\nabla_{\mu}\upphi\nabla_{\nu}\upphi + m_{\mathrm{s}}^2\upphi^2}
	\label{IPhi}
\end{equation}
where we have denoted by $m_{\mathrm{s}}$ the mass of the scalar field and we remind the reader that \emph{$\nabla_{\mu}$ is the covariant derivative operator associated to the Levi-Civita connection of the full $D$-dimensional metric $\mathsf{g}_{\mu\nu}$}. We would like to understand the compactification of this theory on the local $(D-2)$-spheres (defined relative to some static patch frame). We begin by performing the spherical decomposition \eqref{Scalarl}
\begin{equation}
	\upphi(\mathsf{x}) 
	= \sum_{lm}\phi_{l}(x)\,Y_{lm}(\theta)
\end{equation}
In terms of these modes, the action becomes
\begin{equation}
	\frac{I[\upphi]}{\Omega_{(D-2)}} = -\frac{1}{2}\sum_{lm}\int_{\mathrm{dS}_2{}'} \mathrm{d}^2x\,\sqrt{|g|}\,r(x)^{D-2}\inp{g^{ab}\,\nabla_a\phi_{lm}^{}\nabla_b\phi_{lm}^{*} + \inp{\frac{l\inp{l + D - 3}}{r(x)^2} + m_{\mathrm{s}}^2}\phi_{lm}^{}\phi_{lm}^*}
	\label{IStatic}
\end{equation}
The various modes decouple, and are classically governed by the independent 
equations 
\begin{equation}
	\inp{-\nabla_{(\mathrm{s})}^2 + \frac{l\inp{l + D - 3}}{r(x)^2} + m_{\mathrm{s}}^2}\phi_{lm} = 0
	\label{lEOM}
\end{equation}
Here we have denoted by $\nabla^2_{(\mathrm{s})}$ the Laplace-Beltrami operator (covariant Laplacian) $\mathsf{g}^{\mu\nu}\nabla_{\mu}\nabla_{\nu}$ of dS$_D$ \emph{acting on s-wave scalars/zero forms}
\begin{equation}
	\boxed{\nabla^2_{(\mathrm{s})} \equiv \frac{1}{r^{D-2}\sqrt{|g|}}\,\partial_{a}\inp{r^{D-2}\sqrt{|g|}\,g^{ab}\partial_{b}}}
\end{equation}
In order to capture the fact that the ``parent" field $\upphi(\mathsf{x})$ lives in global dS$_D$, we must impose that each of the modes $\phi_{lm}$ be smooth at the pode\footnote{It is not hard to see that this leads to a good variational problem. Varying the action with respect to $\upphi$ gives rise to a bulk term which vanishes on shell plus a boundary term:
	\begin{equation}
		\updelta I[\upphi] 
		= -\int_{\mathcal{M}_D}\mathrm{d}^{D}\mathsf{x}\sqrt{|\mathsf{g}|}\Big(\inp{-
			\nabla^{\mu}\nabla_{\mu} + m_{\mathrm{s}}^2}\upphi\Big)\updelta\upphi- \int_{\partial\mathcal{M}_D}\mathrm{d}^{D-1}\mathsf{x}\sqrt{|\upgamma|}\inp{n^{\mu}\nabla_{\mu}\upphi}\updelta\upphi
		\label{dIS}
	\end{equation}
	where $\sqrt{|\upgamma|}$ and $n^{\mu}$ are respectively the induced volume element and outward facing unit normal of the boundary $\partial\mathcal{M}_D$. Working in static patch coordinates and making use of the fact that the pode is a surface of constant $r$, we find that
	\begin{equation}
		\frac{\updelta I[\upphi]}{\Omega_{(D-2)}} 
		\ \supset \ - \sum_{lm}\int_{\mathrm{pode}}\mathrm{d}t\,\Big(r^{D-2}\,f(r)\,\partial_r\phi_{lm}\Big)\,\updelta\phi_{lm}^*
		\label{dIS}
	\end{equation}
	This term vanishes---and our variational problem is well-defined---since smoothness at the pode ensures the Dirichlet condition
	\begin{equation}
		r^{D-2}\,f(r)\,\phi^*_{lm}\big|_{\mathrm{pode}} = 0
	\end{equation}
	which in turn ensures that the variational condition
	\begin{equation}
		r^{D-2}\,f(r)\,\updelta\phi^*_{lm}\big|_{\mathrm{pode}} = \updelta\inp{r^{D-2}\,f(r)\,\phi^*_{lm}}\big|_{\mathrm{pode}} = 0
	\end{equation}
	holds everywhere on the relevant field configuration space.
}. In terms of static patch coordinates \eqref{dS}, the action \eqref{IStatic} takes the form
\begin{equation}
	\boxed{\frac{I[\upphi]}{\V} = \frac{1}{2}\sum_{lm}\int \mathrm{d}t\,\mathrm{d}r\,r^{D-2}\inp{\frac{1}{f(r)}\|\partial_t\phi_{lm}\|^2 - f(r)\|\partial_r\phi_{lm}\|^2 - \inp{\frac{l\inp{l + D-3}}{r^2} + m_{\mathrm{s}}^2}\|\phi_{lm}\|^2}}
	\label{IStaticSc}
\end{equation}
The operator $\nabla^2_{(\mathrm{s})}$ takes the form 
\begin{equation}
	\nabla^2_{(\mathrm{s})} = -\frac{1}{f(r)}\,\partial_t^2 + f(r)\,\partial_r^2 + \inp{\frac{1}{r} - \frac{Dr}{\ellds^2}}\partial_r
\end{equation}
and the mode-by-mode equations of motion take the form
\begin{equation}
	\boxed{\inp{\frac{1}{f(r)}\,\partial_t^2 -f(r)\,\partial_r^2 - \inp{\frac{1}{r} - \frac{Dr}{\ellds^2}}\partial_r + \frac{l\inp{l + D - 3}}{r^2} + m_{\mathrm{s}}^2}\phi_{lm} = 0}
	\label{KGl}
\end{equation}

\subsection{Massive Vector Bosons in dS$_D$}
\label{Prelim2}
\quad \ 
The type of field that we will be primarily concerned with in this paper is the ``spin-1" massive real vector Boson $\mathsf{A}_{\mu}$ minimally-coupled to the metric of dS$_{D}$. This field is governed by the dS-Proca action 
\begin{equation}
	I[\mathsf{A}] = -\frac{1}{2}\int\mathrm{d}^{D}\mathsf{x}\sqrt{|\mathsf{g}|}\inp{\frac{1}{2}\,\mathsf{F}^{\mu\nu}\mathsf{F}_{\mu\nu} + m_{\mathrm{v}}^2\,\mathsf{A}^{\mu}\mathsf{A}_{\mu}}
	\label{IA}
\end{equation}
where the field strength tensor $\mathsf{F}_{\mu\nu}$ is given, as usual, by 
\begin{equation}
	\mathsf{F}_{\mu\nu} 
	= \partial_{\mu}\mathsf{A}_{\nu} - \partial_{\nu}\mathsf{A}_{\mu}
\end{equation}
We will call the parameter $m_{\mathrm{v}}$ appearing in the action \eqref{IA} the ``Lagrangian mass" in order to distinguish it from a---as we will argue---more physical notion of mass to be introduced in \S\ref{Asymptotics} below. When $m_{\mathrm{v}} = 0$ the theory develops a local U(1) gauge symmetry and the theory becomes dS-Maxwell theory. For reasons to be explained in the next section, we will \emph{not} be concerned with this point in parameter space, and
\begin{center}
	 \emph{we will assume from here on out that $m_{\mathrm{v}}^2 \neq 0$.}
\end{center}
The classical equation of motion for $\mathsf{A}_{\mu}$ is given by\footnote{\label{waveApp}
	This is the exact Euler-Lagrange equation for the action \eqref{IA}.
	Conditioned on the Lorenz constraint \eqref{LC} being fulfilled, the equation of motion can be reduced to the Klein-Gordon-like equation
	\begin{equation}
		\inp{-\mathsf{g}^{\mu\nu}\nabla_{\mu}\nabla_{\nu} + \inp{m_{\mathrm{v}}^2 + \inp{D-1}\ellds^{-2}}}\mathsf{A}_{\mu} = 0
		\label{KGv}
\end{equation}} 
\begin{equation}
	\frac{1}{\sqrt{|\mathsf{g}|}}\,\partial_{\nu}\inp{\sqrt{|\mathsf{g}|}\,\mathsf{F}^{\nu\mu}} = m_{\mathrm{v}}^2\,\mathsf{A}^{\mu}
	\label{Proca}
\end{equation}
or, equivalently (due to the antisymmetry of $\mathsf{F}_{\mu\nu}$),
\begin{equation}
	\nabla^{\nu}\mathsf{F}_{\nu\mu} = m_{\mathrm{v}}^2\,\mathsf{A}_{\mu}
	\label{ProcaCov}
\end{equation}
For $m_{\mathrm{v}} \neq 0$, this equation of motion contains a constraint (the ``Lorenz constraint" or ``transversality constraint") 
\begin{equation}
	m_{\mathrm{v}}^2\,\partial_{\mu}\inp{\sqrt{|\mathsf{g}|}\,\mathsf{A}^{\mu}} = 0
	\label{LC}
\end{equation}
or, equivalently,
\begin{equation}
	m_{\mathrm{v}}^2\,\nabla^{\mu}\mathsf{A}_{\mu} = 0
	\label{LC}
\end{equation}
which can be found by taking the coordinate divergence of \eqref{Proca} or the covariant divergence of \eqref{ProcaCov}.

\subsection{Spherical Decomposition of the Vector Boson}
\quad In terms of the split
\begin{equation}
	\mathsf{A}_{\mu}(\mathsf{x})\,\mathrm{d}\mathsf{x}^{\mu} = \mathsf{A}_a(\mathsf{x})\,\mathrm{d}x^a + D_A\mathes{A}(\mathsf{x})\,\mathrm{d}\theta^A +  \tilde{\mathsf{A}}_{A}(\mathsf{x})\,\mathrm{d}\theta^A
	\label{split}
\end{equation}
of the vector Boson into nonspherical, sphere-longitudinal, and sphere-tranverse parts, the dS-Proca action becomes 
\begin{multline}
	I[\mathsf{A}] = -\frac{1}{2}\int\mathrm{d}^{D}\mathsf{x}\sqrt{|\mathsf{g}|}\,\bigg(\frac{1}{2}\,\mathsf{F}^{ab}\mathsf{F}_{ab} + \frac{1}{r(x)^2}\,g^{ab}\Omega^{AB}\partial_A\mathsf{A}_a\partial_B\mathsf{A}_b + m_{\mathrm{v}}^2\,\mathsf{A}^{a}\mathsf{A}_{a}\\ 
	+ \frac{1}{2}\,\mathsf{F}^{AB}\mathsf{F}_{AB} + \frac{1}{r(x)^2}\,g^{ab}\Omega^{AB}\partial_a\mathsf{A}_A\partial_b\mathsf{A}_B  + m_{\mathrm{v}}^2\,\mathsf{A}^{A}\mathsf{A}_{A}\\
	+ \frac{2}{r(x)^2}\,g^{ab}\mathsf{A}_a\partial_b\inp{\Omega^{AB}D_AD_B\mathes{A}}\bigg)
\end{multline}
We see that the (non $s$-wave) nonspherical and sphere-longitudinal parts of the vector Boson are coupled via the term 
\begin{align}
	-\frac{1}{2}\int\mathrm{d}^{D-2}\theta\,\sqrt{\Omega}\,F^{aA}F_{aA} 
	\ &\supset \  
	+\frac{1}{r(x)^2}\int\mathrm{d}^{D-2}\theta\,\sqrt{\Omega}\,g^{ab}\Omega^{AB}\partial_A\mathsf{A}_a\,\partial_b\mathsf{A}_B\\
	\ & = \ -\frac{1}{r(x)^2}\int\mathrm{d}^{D-2}\theta\,\sqrt{\Omega}\,g^{ab}\mathsf{A}_a\partial_b\inp{\Omega^{AB}D_A\mathsf{A}_B}\\
	\ & = \ -\frac{1}{r(x)^2}\int\mathrm{d}^{D-2}\theta\,\sqrt{\Omega}\,g^{ab}\mathsf{A}_a\partial_b\inp{\Omega^{AB}D_AD_B\mathes{A}}\\
	\ & = \ +\frac{1}{r(x)^2}\,g^{ab}\sum_{l\geq 1, m}l\inp{l + D - 3}A_a^{lm}(x)\,\partial_b\mathes{A}_{lm}(x)
\end{align}
The $s$-wave parts and the sphere-transverse part of the vector Boson each remain uncoupled to the others. We see that our theory describes three different types of field configurations, which are mutually decoupled from one another:
\begin{enumerate}
	\item The $s$-wave mode: $$\mathsf{A}_{\mu}(\mathsf{x})\,\mathrm{d}\mathsf{x}^{\mu} = A_{a}(x)\,\mathrm{d}x^a$$
	
	\item Sphere-transverse modes $$\mathsf{A}_{\mu}(\mathsf{x})\,\mathrm{d}\mathsf{x}^{\mu} = \tilde{\mathsf{A}}_A(\mathsf{x})\,\mathrm{d}\theta^A, \qquad
		\Omega^{AB}D_A\tilde{\mathsf{A}}_B = 0$$
	
	\item Non-$s$-wave nonspherical modes coupled to sphere-longitudinal modes:
	\begin{equation}
		\mathsf{A}_{\mu}(\mathsf{x})\,\mathrm{d}\mathsf{x}^{\mu} = \mathsf{A}_a(\mathsf{x})\,\mathrm{d}x^a + D_A\mathes{A}(\mathsf{x})\,\mathrm{d}\theta^A, \qquad \partial_A\mathsf{A}_a \neq 0
	\end{equation}
	
\end{enumerate}
We will now discuss each of these classes of field configuration in turn.

\section{$s$-Wave Massive Vector Bosons in dS$_D$}
\label{Longitudinal}
\quad \
We would like to begin by focusing on the $s$-wave mode 
\begin{equation}
	\mathsf{A}_{\mu}(\mathsf{x})\,\mathrm{d}\mathsf{x}^{\mu} = A_{a}(x)\,\mathrm{d}x^a
	\label{swave}
\end{equation}
which is effectively a $(1 + 1)$-dimensional vector field on the $(1 + 1)$-dimensional space transversal to the local $(D-2)$-spheres. The field strength tensor of this mode is similarly an effective $(1 + 1)$-dimensional two-form:
\begin{equation}
	\frac{1}{2}\,\mathsf{F}_{\mu\nu}(\mathsf{x})\,\mathrm{d}\mathsf{x}^{\mu}\wedge\mathrm{d}\mathsf{x}^{\nu} = \frac{1}{2}\,F_{ab}(x)\,\mathrm{d}x^{a}\wedge\mathrm{d}x^{b}
\end{equation}
where 
\begin{equation}
	F_{ab} = \partial_aA_b - \partial_bA_a
\end{equation}
We can write a ``dimensionally reduced" effective action for the $s$-wave mode $A_a$, given by
\begin{equation}
	\frac{I[A]}{\Omega_{(D-2)}} = -\frac{1}{2}\int_{\mathcal{M}_2{}'} \mathrm{d}^2x\,r(x)^{D-2}\sqrt{|g|}\inp{\frac{1}{2}\,F^{ab}F_{ab} + m_{\mathrm{v}}^2\,A^aA_a }
	\label{IADimRed}
\end{equation}
or, specializing to static patch coordinates
\begin{equation}
	\frac{I[A]}{\V} = \frac{1}{2}\int \mathrm{d}t\,\mathrm{d}r\,r^{D-2}\inp{\inp{\partial_tA_r}^2 + \inp{\partial_rA_t}^2 - 2\,\partial_tA_r\partial_rA_t + \frac{m_{\mathrm{v}}^2}{f(r)}\,A_t^2 - f(r)\,m_{\mathrm{v}}^2\,A_r^2}
	\label{IStatic}
\end{equation}
Either by reading off of \eqref{IStatic} or by plugging into \eqref{Proca}, one finds that the effective equations of motion are given by
\begin{empheq}[box=\widefbox]{align}
	\frac{f(r)}{r^{D-2}}\,\partial_r\insb{r^{D-2}\inp{\partial_rA_t - \partial_tA_r}} &= m_{\mathrm{v}}^2A_t
	\label{AtStatic}\\
	\frac{1}{f(r)}\,\partial_t\insb{\inp{\partial_rA_t-\partial_tA_r}} &= m_{\mathrm{v}}^2A_r
	\label{ArStatic}
\end{empheq}
while the Lorenz constraint \eqref{LC} reads\footnote{The constraint \eqref{LCStatic} can also be obtained from the remaining equations of motion by simply equating $\partial_t\inp{\frac{1}{f(r)}\times\eqref{AtStatic}}$ with $\partial_r\inp{\Phi f(r)\times\eqref{ArStatic}}$ (assuming that our fields are sufficiently smooth as to allow the equating of mixed partials).}
\begin{equation}
	\boxed{-\frac{1}{f(r)}\,\partial_tA_t + \frac{1}{r^{D-2}}\,\partial_r\inp{r^{D-2}f(r)A_r} = 0}
	\label{LCStatic}
\end{equation}
For reasons to be explained below, we impose the Dirichlet-like boundary condition for $A_r$ at the pode
\begin{equation}
	\boxed{r^{D-2}\,f(r)\,A_r\big|_{\mathrm{pode}} = 0}
	\label{Dirichlet}
\end{equation}
and (when applicable) the Dirichlet-like boundary condition for $A_t$ at the cosmological horizon
\begin{equation}
	\boxed{r^{D-2}\,f(r)\,A_t\big|_{\text{horizon}} = 0}
	\label{DirichletH}
\end{equation}
This latter boundary condition is the projection to the $s$-wave sector of the ``dynamical edge mode" boundary conditions of \cite{Ball:2024hqe}, which were shown to be ``shrinkable" (i.e. leading to a partition function equivalent to the one determined by the usual ``no boundary" path integral \cite{Hartle:1983ai}). It is also of course the projection of the common ``electrically conducting" boundary condition which gives the horizon the properties of an electrically conducting membrane, with nontrivial normal component of the electric field $\propto E_r \equiv F_{rt}$ (the $s$-wave mode cannot distinguish between these boundary conditions).

As we will explain presently, the $s$-wave sector of dS-Proca theory \eqref{IA} is a constrained system, with only one $(1+1)$-dimensional field's worth of independent physical degrees of freedom. It is for this reason that we only prescribe a single field's worth of boundary conditions per boundary. The physical meaning of these boundary conditions will become clear over the course of our analysis, and we will verify that they lead to a good variational problem in \S\ref{EField} below.

\subsection{Constraints and Physical Degrees of Freedom}
\quad \ 
The constraint \eqref{LCStatic} tells us that, within the $s$-wave sector, we only have one $(1 + 1)$-dimensional field's worth of independent physics degrees of freedom contained within the $s$-wave mode $A_a(x)\,\mathrm{d}x^a$. Indeed, the component $A_t$ has vanishing canonical momentum with respect to the static patch frame, and we should take the physical field to be\footnote{Indeed, by manipulating equations \eqref{AtStatic}, \eqref{ArStatic}, and \eqref{LCStatic}, one can find a ``wave-like equation" for just the mode $A_r$ by itself 
	\begin{equation}
		-\frac{1}{f(r)}\,\partial_t^2A_r + \frac{1}{f(r)}\,\partial_r\inp{\frac{f(r)}{r}\,\partial_r\inp{rf(r)A_r}}- m_{\mathrm{v}}^2\,A_r = 0
		\label{ArEq}
\end{equation}} $A_r$ or a related quantity.
The canonical momentum conjugate to $A_r$ is given by
\begin{equation}
	\boldsymbol{\pi}^r \equiv \frac{\updelta I}{\updelta(\partial_tA_r)} 
	= 
	\V\,r^{D-2}\,\underbrace{\inp{\partial_tA_r-\partial_rA_t}}_{-F^{tr}}
	\label{pir}
\end{equation}
or
\begin{equation}
	\boldsymbol{\pi}^r = -\V\,r^{D-2}\,E_r
	\label{pirE}
\end{equation}
where we have recognized the electric field measured relative to the static patch frame
\begin{equation}
	\boxed{E_r \equiv F_{rt} }
	\label{Edef}
\end{equation}
	The ``equation of motion" \eqref{AtStatic}, being first order in time derivatives, should also be thought of as a constraint; \eqref{AtStatic} and \eqref{LCStatic} together determine the ``constraint submanifold" of the $s$-wave phase space, i.e. the physical phase space on which the canonical formulation of our $s$-wave dS-Proca field theory is well-defined. $A_t$ and its time derivative are nontrivial functions on the physical phase space, which can be given in terms of the canonical coordinates $A_r$ and $\boldsymbol{\pi}^r$ as 
	\begin{align}
A_t = -\frac{1}{ m_{\mathrm{v}}^2\ellds}\frac{1}{\V}\frac{f(r)}{r^{D-2}}\,\partial_r\boldsymbol{\pi}^r, \qquad \dot{A}_t = +\frac{f(r)}{r^{D-2}}\,\partial_r\inp{r^{D-2}\,f(r)A_r}
\end{align}
We can also write them in terms of the conjugate (but not canonically so) coordinates $A_r$ and $E_r$ as
\begin{equation}
A_t = +\frac{1}{m_{\mathrm{v}}^2}\frac{f(r)}{r^{D-2}}\,\partial_r\inp{r^{D-2}E_r}, \qquad \dot{A}_t = +\frac{f(r)}{r^{D-2}}\,\partial_r\inp{r^{D-2}f(r)A_r}
\label{Constraints}
\end{equation} 
We will now explain the physical interpretation of this canonical momentum as a natural observable in the static patch.

\subsection{Current and Charge} 
\label{charge}
\quad \ 
The field $\mathsf{A}^{\mu}$ acts as a source for the field tensor $\mathsf{F}_{\mu\nu}$ with coupling given by the squared Lagrangian mass $m_{\mathrm{v}}^2$. We can therefore view the field equation \eqref{Proca} as being Maxwell's equations with a vector (self-)source 
\begin{equation}
	\mathsf{J}^{\mu} \equiv m_{\mathrm{v}}^2\,\mathsf{A}^{\mu}
	\label{JDef}
\end{equation}
This ``source" is conserved
\begin{equation}
	\nabla_{\mu}\mathsf{J}^{\mu} = 0
\end{equation}
by virtue of the Lorenz/transversality constraint \eqref{LC}. When studying the isolated physics of the $s$-wave mode $A_a(x)$, we can $s$-wave reduce both sides of \eqref{JDef}, which then reads
\begin{equation}
	J^a(x) \equiv m_{\mathrm{v}}^2\,A^a(x)
	\label{Js}
\end{equation}
We can now understand the physical meaning of the Dirichlet-like boundary condition on $A_r$ at the pode: \emph{\eqref{Dirichlet} is simply the requirement that current not spontaneously ``leak" out of (or into) the center of the static patch}, i.e. that that there be no net flux of current at the pode:
\begin{equation}
	\underbrace{-\Omega_{(D-2)}\,r^{D-2}\,J^r(x)}_{\mathrm{flux}}\Big|_{\text{pode}} = 0 \ \implies \ r^{D-2}\,f(r)\,A_r\big|_{\mathrm{pode}} = 0
\end{equation}

\begin{figure}
	\centering
	\begin{normalsize}
\begingroup%
  \makeatletter%
  \providecommand\color[2][]{%
    \errmessage{(Inkscape) Color is used for the text in Inkscape, but the package 'color.sty' is not loaded}%
    \renewcommand\color[2][]{}%
  }%
  \providecommand\transparent[1]{%
    \errmessage{(Inkscape) Transparency is used (non-zero) for the text in Inkscape, but the package 'transparent.sty' is not loaded}%
    \renewcommand\transparent[1]{}%
  }%
  \providecommand\rotatebox[2]{#2}%
  \newcommand*\fsize{\dimexpr\f@size pt\relax}%
  \newcommand*\lineheight[1]{\fontsize{\fsize}{#1\fsize}\selectfont}%
  \ifx\svgwidth\undefined%
    \setlength{\unitlength}{365.29770258bp}%
    \ifx\svgscale\undefined%
      \relax%
    \else%
      \setlength{\unitlength}{\unitlength * \real{\svgscale}}%
    \fi%
  \else%
    \setlength{\unitlength}{\svgwidth}%
  \fi%
  \global\let\svgwidth\undefined%
  \global\let\svgscale\undefined%
  \makeatother%
  \begin{picture}(1,0.61782455)%
    \lineheight{1}%
    \setlength\tabcolsep{0pt}%
    \put(0.48732793,0.59621063){\color[rgb]{0,0,0}\makebox(0,0)[lt]{\lineheight{1.25}\smash{\begin{tabular}[t]{l}$\mathcal{I}^+$\end{tabular}}}}%
    \put(0.49081397,0.00649885){\color[rgb]{0,0,0}\makebox(0,0)[lt]{\lineheight{1.25}\smash{\begin{tabular}[t]{l}$\mathcal{I}^-$\end{tabular}}}}%
    \put(0,0){\includegraphics[width=\unitlength,page=1]{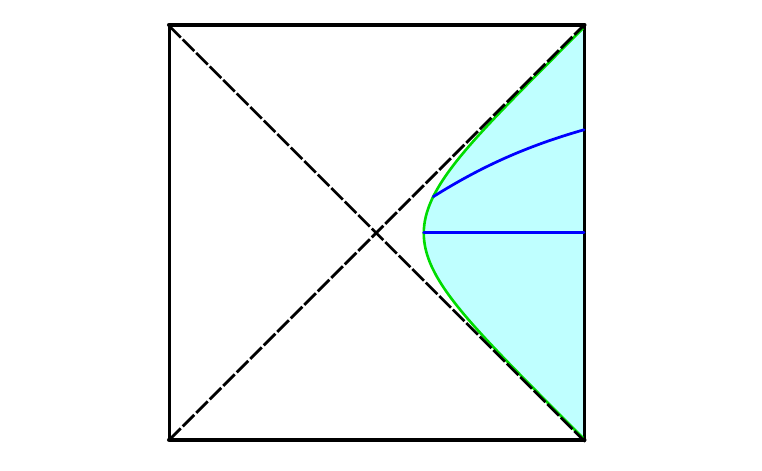}}%
    \put(0.78650899,0.30345652){\color[rgb]{0,0,0}\makebox(0,0)[lt]{\lineheight{1.25}\smash{\begin{tabular}[t]{l}$\Sigma_0$\end{tabular}}}}%
    \put(0.78665434,0.43884904){\color[rgb]{0,0,0}\makebox(0,0)[lt]{\lineheight{1.25}\smash{\begin{tabular}[t]{l}$\Sigma_{t > 0}$\end{tabular}}}}%
  \end{picture}%
\endgroup%

	\end{normalsize}
	\caption{Two examples of the types of slices $\Sigma_t$ which are used to define the charge $Q(t,r)$. The solid green line is the surface of constant $r$.}
	\label{Sigmat}
\end{figure}
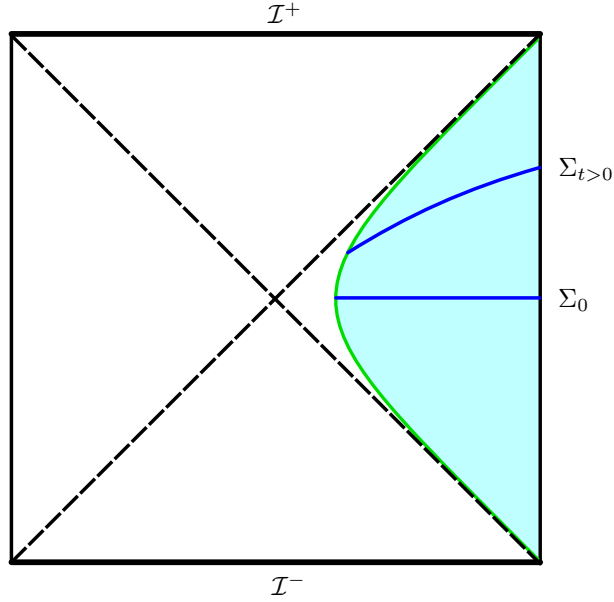

A simple observable of the \emph{full} dS-Proca theory \eqref{IA} which only depends on the $s$-wave mode is the charge $Q(t)$ contained within a static patch, i.e. the charge on a constant $t$ slice $\Sigma_t$ running from the pode to the horizon (or, more accurately, \emph{centered} at the pode and \emph{bounded} by the horizon). It is given by
\begin{equation}
	Q(t) \equiv Q(t,r)\big|_{\mathrm{horizon}}
\end{equation}
where we have defined
\begin{align}
	Q(t,r) 
	&\equiv \int_{\Sigma_t}\sqrt{\mathsf{h}}\,n_{\mu}\mathsf{J}^{\mu}
	\label{Q(t)}\\
	&= - \intf{0}{r}\mathrm{d}r'\,r'^{D-2}\int\mathrm{d}^{D-2}\theta\,\sqrt{\Omega}\,\mathsf{J}^t(t,r',\theta)\\
	&= -\V\intf{0}{r}\mathrm{d}r'\,r'^{D-2}\,J^t(t,r')
\end{align}
(see figure \ref{Sigmat}). Here we have recognized that integrating over the spherically-symmetric slice $\Sigma_t$ automatically projects us into the $s$-wave sector: 
\begin{equation}
	\int\mathrm{d}^{D-2}\theta\,\sqrt{\Omega}\,\mathsf{J}^{a}(x,\theta) = \V\,J^a(x)
\end{equation}
Plugging in the definition \eqref{Js} of the $s$-wave current, we find that 
\begin{align}
	Q(t,r) 
	&= +\V\,m_{\mathrm{v}}^2\intf{0}{r}\frac{\mathrm{d}r'\,r'^{D-2}}{f(r')}\,A_t(t,r')
	\label{QA}
\end{align}
Using the first of the constraints \eqref{Constraints}, we see that we can rewrite this as
\begin{align}
	\boxed{Q(t,r) = \V\,r^{D-2}\,E_r(t,r)}
	\label{Gauss}
\end{align}

\noindent Recognizing $\V\,r^{D-2}$ as the codimension-2 ``area" \eqref{Adef} of the bounding surface, we see that \eqref{Gauss} is simply Gauss's law, \emph{which holds as as constraint} (just as in the gauge theory case). In terms of canonical coordinates, the charge is simply (up to sign) the horizon value of the momentum conjugate to $A_r$: 
\begin{equation}
	Q(t,r) = - \boldsymbol{\pi}^r(t,r)
\end{equation}

Note that, by using the second constraint of \eqref{Constraints}, we can also rewrite \eqref{QA} as
\begin{align}
	Q(t,r) -Q(0,r)
	&= \V\,r^{D-2}\,m_{\mathrm{v}}^2\intf{0}{t}\mathrm{d}t'A^r(t',r)
	\label{QAr}\\
	&= \intf{0}{t}\mathrm{d}t'\Big(\V\,r^{D-2}\,J^r(t',r)\Big)
\end{align}
The quantity $\V\,r^{D-2}\,J^r(t',r)$ is of course just the total flux of current out of the region of interest (i.e. falling through the surface of constant $r$); integrating this over time gives the change in the charge, as expected. We can convert the integral relations \eqref{QA}, \eqref{QAr} into differential relations
\begin{align}
	A_t(t,r) &= +\frac{1}{m_{\mathrm{v}}^2}\frac{1}{\V\,r^{D-2}}\,f(r)\,\partial_rQ(t,r) \label{AtQ}\\
	A_r(t,r) &= +\frac{1}{m_{\mathrm{v}}^2}\frac{1}{\V\,r^{D-2}}\,\frac{1}{f(r)}\,\partial_tQ(t,r) \label{ArQ}
\end{align}
or, using Gauss's law \eqref{Gauss}, 
\begin{align}
	\boxed{A_t = +\frac{1}{m_{\mathrm{v}}^2}\,\frac{f(r)}{r^{D-2}}\,\partial_r\big(r^{D-2}E_r\big), \qquad A_r = -\frac{1}{m_{\mathrm{v}}^2}\,\frac{1}{f(r)}\,\partial_tE_r}
	\label{ArFromEr}
\end{align}
We recognize the first equation as the first constraint of \eqref{Constraints}, while the second equation tells us that the contraints allow us to trade the conjugate pair $(A_r, E_r)$ for the conjugate pair $(E_r, \partial_tE_r)$ with $A_r$ then determined via
\begin{equation}
	\boxed{\partial_tE_r = -f(r)\,m_{\mathrm{v}}^2\,A_r}
	\label{dtE}
\end{equation}

Using \eqref{dtE}, we see that boundary condition \eqref{Dirichlet} implies the same Dirichlet-like boundary condition for $E_r$ at the pode
\begin{equation}
	\boxed{r^{D-2}f(r)\,E_r\big|_{\mathrm{pode}} = 0}
	\label{DirichletE}
\end{equation}
Meanwhile, using \eqref{Constraints}, we see that the boundary condition \eqref{DirichletH} implies the Neumann-like boundary condition for $E_r$ at the horizon
\begin{equation}
	\boxed{f(r)\,\partial_r\big(r^{D-2}\,E_r\big)\Big|_{\text{horizon}} = 0}
	\label{NeumannE}
\end{equation}
Using Gauss's law \eqref{Gauss}, we see that \eqref{DirichletE} is simply the requirement that the charge mode $Q(t,r)$ vanish as $r \to 0$
\begin{equation}
	r^{D-2}\,E_r\big|_{\text{pode}} = 0 \quad \Longleftrightarrow \quad Q\big|_{\text{pode}} = 0
\end{equation}
In other words, it is simply the requirement that there be no point charges at the center of the static patch and that the charge density $\frac{r^{D-2}}{f(r)}\,A_t$ be smooth. Meanwhile \eqref{NeumannE} is simply the natural condition that the electric field and charge mode be less singular than $\log(1-r/\ellds)$ as we approach the horizon. To see that these boundary conditions lead to a good variational problem, note that the first order variation in the action is given by a bulk term which vanishes on shell plus a boundary term:
\begin{align}
	\updelta I[\mathsf{A}] 
	&= -\int\mathrm{d}^{D}\mathsf{x}\sqrt{|\mathsf{g}|}\inp{\frac{1}{2}\,\mathsf{F}^{\mu\nu}\updelta\mathsf{F}_{\mu\nu} + m_{\mathrm{v}}^2\,\mathsf{A}^{\mu}\updelta\mathsf{A}_{\mu}}\\
	&= \int\mathrm{d}^{D}\mathsf{x}\sqrt{|\mathsf{g}|}\underbrace{\inp{\nabla_{\nu}\mathsf{F}^{\nu\mu} - m_{\mathrm{v}}^2\,\mathsf{A}^{\mu}}}_{\mathrm{EOM}}\updelta\mathsf{A}_{\mu}
	-\int_{\mathrm{boundaries}}\mathrm{d}^{D-1}\mathsf{x}\sqrt{|\upgamma|}\,n_{\mu}\,\mathsf{F}^{\mu\nu}\updelta \mathsf{A}_{\nu}
	\label{dIA}
\end{align}
Here $\sqrt{|\upgamma|}$ and $n^{\mu}$ are the induced volume element and outward facing unit normal of the boundary $\partial\mathcal{M}_D$, respectively.  Working in static patch coordinates, making use of the fact that our boundaries are surfaces of constant $r$, and restricting to the $s$-wave mode, we find that 
\begin{equation}
	\frac{\updelta I[\upphi]}{\Omega_{(D-2)}} = \int_{\mathrm{boundaries}}\mathrm{d}t\inp{r^{D-2}\,E_r}\,\updelta A_{t}
\end{equation}
This term vanishes at the pode due to the Dirichlet-like boundary condition \eqref{DirichletE} and at the horizon due to the Dirichlet-like boundary condition \eqref{DirichletH}. We see that our boundary conditions indeed give rise to a good variational problem.

\subsection{A Simplification for $D = 3$}
\label{EField}
\quad \ 
The differential relations \eqref{ArFromEr} are a simple consequences of the constraints \eqref{Constraints} and hold everywhere on the constraint surface, i.e. everywhere on the physical phase space.  They are another reflection of the fact that there is only one fields's worth of degrees of freedom contained within the $s$-wave mode $A_a(x)$. We can take this field to be $A_r$, or we can alternatively take it to be $\boldsymbol{\pi}^r$, $Q$, or $E_r$. 

Let us now set the simplest nontrivial case of $D = 3$. We will see that in this case our theory remarkably simplifies. Let us choose to work in terms of the electric field mode $E_r$, with the charge then determined by Gauss's law \eqref{Gauss} and the $s$-wave vector Boson component $A_r$ determined by \eqref{dtE}. Using the definition \eqref{Edef} of the electric field as well as the constraints in the form \eqref{ArFromEr}, we can rewrite the effective $s$-wave action \eqref{IADimRed} purely in terms of $E_r$ and its derivatives:
\begin{equation}
	\boxed{\frac{I[E_r]}{2\pi} = -\frac{1}{2m_{\mathrm{v}}^2}\int\mathrm{d}t\,\mathrm{d}r\,r\,\bigg[\frac{1}{f(r)}\inp{\partial_tE_r}^2-f(r)\inp{\partial_rE_r}^2-\inp{\inp{m_{\mathrm{v}}^2 + \frac{1}{\ellds^2}} + \frac{1}{r^2}}E_r^2\bigg]}
	\label{IEr}
\end{equation}
From \eqref{IEr}, we can immediately obtain the equation of motion for $E_r$ as 
\begin{equation}
	\boxed{\insb{- \nabla^2_{(\mathrm{s})} + \frac{1}{r^2} + \inp{m_{\mathrm{v}}^2 + \ellds^{-2}}}E_r = 0}
	\label{EOMEr}
\end{equation} 
We stress that the above equations only hold for the special case of $D = 3$ spacetime dimensions. 
From the action in the form \eqref{IEr}, we see that the momentum canonically conjugate to $E_r$ is given by 
\begin{equation}
	\boldsymbol{\pi}_E \equiv \frac{\updelta I}{\updelta (\partial_tE_r)} = -\frac{2\pi}{m_{\mathrm{v}}^2}\frac{r}{f(r)}\,\partial_tE_r
	\label{piE}
\end{equation}
or, using \eqref{dtE}, 
\begin{equation}
	\boldsymbol{\pi}_E = 2\pi r\,A_r
\end{equation}
which we see is consistent with (and in fact required by) \eqref{pirE}. 

Comparing to \eqref{IStaticSc} and \eqref{KGl}, we recognize \eqref{IEr} and \eqref{EOMEr} as (up to an overall dimensionful scaling) the action and equation of motion for the $l = 1$ component (i.e. the ``$p$-wave mode") of a scalar field of mass 
\begin{equation}
	\boxed{m_{\mathrm{s}}^2 
	\equiv m_{\mathrm{v}}^2 + \ellds^{-2}}
	\label{ms}
\end{equation}
with the same Dirichlet-like boundary condition \eqref{DirichletE} at the pode. This observation will allow us to quantize the $s$-wave phase space of the $D = 3$ vector Boson using the well-understood quantization of the massive scalar. In forthcoming work \cite{Add}, we will show that the $s$-wave mode of the vector Boson in $D > 3$ spacetime dimensions remains related to the $l = 1$ mode of a scalar field of mass 
\begin{equation}
	\boxed{m^2_{\mathrm{s}} = m_{\mathrm{v}}^2 + \inp{D-2}\ellds^{-2}}
\end{equation}
though the relationship of this scalar to the basic fields of the theory becomes much more complicated. This was previously explained for the special case of $D = 4$ by \cite{Higuchi:1986ww}.

As an aside, note that, using \eqref{IEr} and \eqref{piE}, the Hamiltonian of our theory can be written (again in the special case of $D = 3$ spacetime dimensions) as
\begin{equation}
	\boxed{\frac{H[E_r]}{2\pi\ellds} = -\frac{1}{2m_{\mathrm{v}}^2}\int\mathrm{d}^2x\,r\,\bigg[\frac{1}{f(r)}\inp{\partial_tE_r}^2+f(r)\inp{\partial_rE_r}^2+\inp{\inp{m_{\mathrm{v}}^2 + \ellds^{-2}} + \frac{1}{r^2}}E_r^2\bigg]}
	\label{HEr}
\end{equation}
Rather surprisingly, the energy is only bounded below for $m_{\mathrm{v}}^2 < 0$. We will explain in the following section the reason why this corresponds to a natural mass range given by $-2\inp{D-1} < m_{\mathrm{v}}^2\ellds^2 < 0$. 

\subsection{A Brief Aside: Working in Terms of the Charge Mode}
\quad \ 
For completeness and also for later reference, let us also work out the action and equation of motion in terms of the charge mode (again in the simple case of $D = 3$)
\begin{equation}
	Q(t,r) \equiv 2\pi r\,E_r(t,r)
\end{equation}
In terms of this mode, the action \eqref{IEr} can be written as
\begin{equation}
	2\pi\,I[Q] = -\frac{1}{2m_{\mathrm{v}}^2}\int\mathrm{d}^2x\,\frac{1}{r}\inp{\frac{1}{f(r)}\inp{\partial_tQ}^2-f(r)\inp{\partial_rQ}^2-m_{\mathrm{v}}^2\,Q^2}
	\label{IQ}
\end{equation}
The corresponding equation of motion for $Q$ is given by
\begin{equation}
	\frac{1}{f(r)}\,\partial_t^2Q - r\,\partial_r\inp{\frac{f(r)}{r}\,\partial_rQ} + m_{\mathrm{v}}^2Q = 0
	\label{EOMQ}
\end{equation}
Looking at the action \eqref{IQ} and resulting equation of motion, we see that the charge mode feels the \emph{T-dual} geometry, with local circles of radius $\propto 1/r$ and with $(2 + 1)$-dimensional parent metric
\begin{equation}
	\mathrm{d}s^2\big|_{\text{T-dual}} = -f(r)\,\mathrm{d}t^2 + \frac{\mathrm{d}r^2}{f(r)} + \inp{\frac{\ellds}{2\pi}}^4\frac{1}{r^2}\,\mathrm{d}\theta^2
\end{equation}
Let us denote by $\Delta^2_{(\mathrm{s})}$ the Laplace-Beltrami operator (covariant Laplacian) of this \emph{T-dual geometry} acting on $s$-wave zero-forms/scalars
\begin{align}
	\Delta^2_{(\mathrm{s})} 
	&\equiv -\frac{1}{f(r)}\,\partial_t^2 + r\,\partial_r\inp{\frac{f(r)}{r}\,\partial_r}
	\label{DeltaTdual}
\end{align}
We can then write the equation of motion \eqref{EOME} as 
\begin{equation}
	\inp{-\Delta^2_{(\mathrm{s})} + m_{\mathrm{v}}^2}Q = 0
\end{equation}

\section{The ``Edge of Stability"}
\label{Asymptotics}
\quad \ 
In this section we will argue that, in dS$_D$, the actual ``physical" mass which governs the behavior of the compactified vector Boson $\mathsf{A}_{\mu}$ is not the ``Lagrangian mass" $m_{\mathrm{v}}$ (which shows up in the usual form \eqref{IA} of the dS-Proca action) but rather the quantity 
\begin{equation}
	\boxed{\mu^2_{\mathrm{v}} \equiv m_{\mathrm{v}}^2 + 2\inp{D-1}\ell_{\mathrm{dS}}^{-2}}
\end{equation}
Note that this differs from the effective scalar mass \eqref{ms} controlling the action \eqref{IEr} for the $D = 3$ electric field. We will explain the relation between the two in \S\ref{EdgeE} below.

\subsection{The Flat Slicing}
\label{flatDef}
\quad \
We will begin by studying the late time behavior of solutions to the equations of motion. For this purpose, it is helpful to define the ``flat slicing" time coordinate
\begin{equation}
	T \equiv t + \frac{\ellds}{2}\log\big(f(r)\big)
	\label{Tflat}
\end{equation}
which is well defined throughout the ``expanding patch", i.e. throughout the interior of the causal future of the pode. Surfaces of constant $T$ are spherically-symmetric infinite flat $(D-1)$-planes centered about the pode. If we coordinatize these ``flat slices" with polar coordinates adapted to the local $(D-2)$-spheres, the metric in the expanding patch takes the form
\begin{equation}
	\mathsf{g}_{\mu\nu}(\mathsf{x})\,\mathrm{d}\mathsf{x}^{\mu}\mathrm{d}\mathsf{x}^{\nu}\Big|_{\text{flat slicing}} = -\mathrm{d}T^2 + e^{2T/\ellds}\inp{\mathrm{d}R^2 + R^2\mathrm{d}\Omega^2_{(D-2)}}
	\label{flat}
\end{equation}
The two clocks $T$ and $t$ agree at the pode. 
\begin{figure}
	\centering
	\begin{normalsize}
\begingroup%
  \makeatletter%
  \providecommand\color[2][]{%
    \errmessage{(Inkscape) Color is used for the text in Inkscape, but the package 'color.sty' is not loaded}%
    \renewcommand\color[2][]{}%
  }%
  \providecommand\transparent[1]{%
    \errmessage{(Inkscape) Transparency is used (non-zero) for the text in Inkscape, but the package 'transparent.sty' is not loaded}%
    \renewcommand\transparent[1]{}%
  }%
  \providecommand\rotatebox[2]{#2}%
  \newcommand*\fsize{\dimexpr\f@size pt\relax}%
  \newcommand*\lineheight[1]{\fontsize{\fsize}{#1\fsize}\selectfont}%
  \ifx\svgwidth\undefined%
    \setlength{\unitlength}{201.42109067bp}%
    \ifx\svgscale\undefined%
      \relax%
    \else%
      \setlength{\unitlength}{\unitlength * \real{\svgscale}}%
    \fi%
  \else%
    \setlength{\unitlength}{\svgwidth}%
  \fi%
  \global\let\svgwidth\undefined%
  \global\let\svgscale\undefined%
  \makeatother%
  \begin{picture}(1,1.05448041)%
    \lineheight{1}%
    \setlength\tabcolsep{0pt}%
    \put(0.48601136,1.01528135){\color[rgb]{0,0,0}\makebox(0,0)[lt]{\lineheight{1.25}\smash{\begin{tabular}[t]{l}$\mathcal{I}^+$\end{tabular}}}}%
    \put(0,0){\includegraphics[width=\unitlength,page=1]{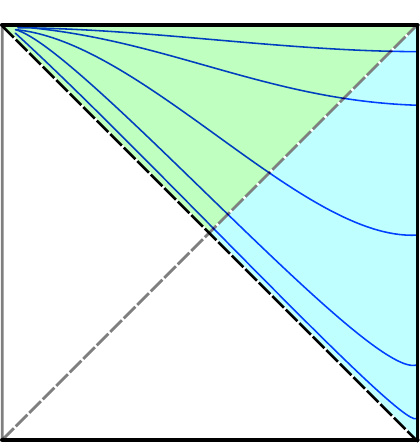}}%
  \end{picture}%
\endgroup%

	\end{normalsize}
	\caption{Flat slicing coordinates cover the interior of the causal future of the static patch (i.e. the static patch plus the region shaded in light green). The blue lines are surfaces of constant $T$---each isometric to an infinite flat $(D-1)$-plane---which extend all the way up to the global asymptotic future $\mathcal{I}^+$.}
	\label{dSFlat}
\end{figure}

In terms of the spherical decomposition, we have 
\begin{align}
	g_{ab}(x)\,\mathrm{d}x^a\mathrm{d}x^b\Big|_{\text{flat slicing}} &= -\mathrm{d}T^2 + e^{2T/\ellds}\,\mathrm{d}R^2\\[0.5em] r(x)\Big|_{\text{flat slicing}} &= e^{T/\ellds}R
\end{align}
For fixed static patch radial position $r$, the late time limit $t \gg \ell_{\mathrm{dS}}$ involves taking $T \to \infty$, $R \sim e^{-T/\ellds} \to 0$.

\subsection{Warm Up: The Scalar Field}
\quad \ 
Let's start by studying the late time asymptotics of the massive minimally-coupled real scalar field $\upphi$. In terms of flat slicing coordinates, the action \eqref{IPhi} is given by
\begin{equation}
	I[\upphi] = \frac{1}{2}\int\mathrm{d}^D\mathsf{x}\,\sqrt{\Omega}\, e^{(D-1)T/\ell_{\mathrm{dS}}}\inp{\dot{\upphi}^2 - e^{-2T/\ell_{\mathrm{dS}}}(\partial_{\vec{x}}\upphi)^2 - m_{\mathrm{s}}^2\upphi^2}
	\label{IPhiFlat}
\end{equation}
The corresponding classical equation of motion is given by
\begin{equation}
	-\ddot{\upphi} + \frac{\inp{D-1}}{\ell_{\mathrm{dS}}}\,\dot{\upphi} + e^{-2T/\ell_{\mathrm{dS}}}\,\partial_{\vec{x}}^2\upphi + m_{\mathrm{s}}^2\upphi = 0
	\label{LTEOM}
\end{equation}
At late times we can ignore the spatial gradient term so that the equation of motion reduces to the ODE 
\begin{equation}
	-\ddot{\upphi} + \frac{\inp{D-1}}{\ell_{\mathrm{dS}}}\,\dot{\upphi} + m^2\upphi = 0
\end{equation}
This is the equation of a damped harmonic oscillator, with the ``damping" term $\frac{\inp{D-1}}{\ell_{\mathrm{dS}}}\,\dot{\upphi}$ encoding the ``Hubble friction" due to a nonzero cosmological constant. This equation has leading late time solution
\begin{equation}
	\upphi(T,\vec{x}) \ \underset{T \to \infty}{\sim} \ \upphi^{(-)}\,e^{-\delta_{-} T/\ell_{\mathrm{dS}}}
\end{equation}
with $\upphi^{(-)}$ a constant and with
\begin{equation}
	\delta_{-} \equiv \frac{\inp{D-1}}{2} - \sqrt{\inp{\frac{D-1}{2}}^2 - m_{\mathrm{s}}^2\ell_{\mathrm{dS}}^2}
\end{equation}
In terms of static patch coordinates, we have 
\begin{equation}
	\upphi(t,r) \ \underset{t \to \infty}{\sim} \ \frac{\upphi^{(-)}}{f(r)^{\delta_-/2}}\,e^{-\delta_{-} t/\ell_{\mathrm{dS}}}
	\label{ScalarDecay}
\end{equation}

As $m_{\mathrm{s}} \to 0$ (at fixed $\ell_{\mathrm{dS}}$), we have that
\begin{equation}
	\delta_{-} \ \underset{m_{\mathrm{s}}\ellds \to 0}{\sim} \ \frac{1}{\inp{D-1}}\,m_{\mathrm{s}}^2\ell_{\mathrm{dS}}^2 \ \to \ 0
	\label{deltaminus}
\end{equation}
and the system develops an asymptotically static mode. For positive but ``light" masses
\begin{equation}
	0 < m_{\mathrm{s}}^2\ell_{\mathrm{dS}}^2 < \inp{\frac{D-1}{2}}^2
\end{equation}
the system is ``critically damped" (the Hubble term dominates over the mass term) and asympotically exponentially decays; for sufficiently large masses
\begin{equation}
	m_{\mathrm{s}}^2\ell_{\mathrm{dS}}^2 > \inp{\frac{D-1}{2}}^2
\end{equation}
the system is ``underdamped" (the mass term dominates over the Hubble term) and the decay is accompanied by oscillations. For negative $m_{\mathrm{s}}^2$ there is unstable tachyonic behavior, i.e., asymptotic exponential growth. We can express this by saying that $m_{\mathrm{s}}^2 = 0$ is the ``edge of stability": the system only makes sense for $m_{\mathrm{s}}^2 \geq 0$. In the flat space limit $\ell_{\mathrm{dS}} \to \infty$ (at fixed $m_{\mathrm{s}}^2$) we recover the expected oscillations controlled by $m_{\mathrm{s}}$
\begin{equation}
	\upphi^{(-)}(t,r) \ \underset{\ell_{\mathrm{dS}} \to \infty}{\sim} \ \upphi^{(-)}\,e^{\mathrm{i}m_{\mathrm{s}}t}
\end{equation}
so we see that the late time decay \eqref{ScalarDecay} is the finite $\ell_{\mathrm{dS}}$ analog of the usual flat space oscillations. In de Sitter space---where we have no notion of representations of the translation group---we should take the parameter defining the edge of stability to \emph{define} what we operationally mean by the physical mass of a field.

\subsubsection{Individual Fourier Modes of the Scalar Field}
\label{1Decay}
\quad \ 
The late time decay of the full scalar field $\upphi$ is controlled by the decay of lowest angular momentum mode---i.e. the $l = 0$ ``$s$-wave" mode---which is the mode of slowest asymptotic decay. 
To see this, let us spherically-decompose our field 
\begin{equation}
	\upphi(T,R,\theta) = \sum_{lm}\phi_{lm}(T,R)\,Y_{lm}(\theta)
\end{equation}
As we will explain in detail in \S \ref{QNM} below, the higher angular momentum modes of $\upphi$ decay via subleading quasinormal resonances; in particular, the mode with angular momentum $l$ decays with leading decay exponent
\begin{align}
	\delta_{-,l} 
	&\equiv \delta_- + l\\
	&= \frac{\inp{D-1}}{2}  + l - \sqrt{\inp{\frac{D-1}{2}}^2 - \ell_{\mathrm{dS}}^2m_{\mathrm{s}}^2}
\end{align}

If we were to study the isolated late-time dynamics of a single mode of fixed angular momentum $l$, its edge of stability would be given by the na\"ively tachyonic mass
\begin{equation}
	 \boxed{m_{\mathrm{s}}^2\ell_{\mathrm{dS}}^2\Big|_{\text{edge},l}= - l\big(l + D - 1\big)}
	 \label{edgelScalar}
\end{equation}
When we consider a superposition of different angular momentum modes, the edge of stability of the superposition will be given by that of the lowest contributing angular momentum mode (otherwise the superposition will be exponentially divergent---i.e. unstable---at late times). It is for this reason that the edge of stability of the full scalar field is given by $m_{\mathrm{s}}^2 = 0$.

\subsection{The $s$-Wave Mode of the Vector Field}
\label{AsympVec}
\quad \ 
Consider now the $s$-wave mode $A_a$ of massive minimally-coupled real vector field $\mathsf{A}_{\mu}$, whose action \eqref{IADimRed} is given in terms of flat slicing coordinates by
\begin{equation}
	\frac{I[A]}{\V}
	= \frac{1}{2}\int
	\mathrm{d}^2x\,e^{\inp{D-3}T/\ellds}R^{D-2}\,\Big(\inp{\partial_TA_R}^2 + \inp{\partial_RA_T}^2 - 2\partial_TA_R\partial_RA_T + m_{\mathrm{v}}^2e^{2T/\ell_{\mathrm{dS}}}A_T^2 - m_{\mathrm{v}}^2A_R^2\Big)
\end{equation}
We can easily read off the effective equations of motion as
\begin{align}
	\frac{1}{e^{2T/\ell_{\mathrm{dS}}}R^{D-2}}\,\partial_R\inp{R^{D-2}\inp{\partial_RA_T-\partial_TA_R}}  &= m_{\mathrm{v}}^2A_T
	\label{AT0}\\[0.5em]
	\frac{1}{e^{\inp{D-3}T/\ell_{\mathrm{dS}}}}\,\partial_T\inp{e^{\inp{D-3}T}\inp{\partial_RA_T-\partial_TA_R}} &= m_{\mathrm{v}}^2A_R
	\label{AR0}
\end{align}
Meanwhile, ($e^{\inp{D-1}T/\ellds}$ times) the Lorenz constraint \eqref{LC} reads
\begin{equation}
	0 
	= -\partial_T\inp{e^{-\inp{D-1}T/\ell_{\mathrm{dS}}}A_T} + \frac{e^{-\inp{D-3}T/\ell_{\mathrm{dS}}}}{R^{D-2}}\,\partial_R\inp{R^{D-2}A_R}
	\label{LCTR0}
\end{equation}
One can manipulate these equations to find 
\begin{equation}
	\insb{-\partial_T^2A_T + \frac{\inp{D + 1}}{\ellds}\,\partial_TA_T + \inp{m_{\mathrm{v}}^2 + \frac{2\inp{D-1}}{\ellds^2}}A_T} - \frac{e^{-\inp{D-1}T/\ellds}}{R^{D-2}}\,\partial_R\inp{R^{D-2}\,\partial_RA_T}= 0
	\label{AT1}
\end{equation}
At late times we can again ignore the spatial gradient term so that \eqref{AT1} reduces to the ODE
\begin{equation}
	-\partial_T^2A_T + \frac{\inp{D + 1}}{\ell_{\mathrm{dS}}}\,\partial_TA_T + \mu^2_{\mathrm{v}}A_T = 0
	\label{ATLate}
\end{equation}
Here we have defined the effective (squared) mass
\begin{equation}
	\boxed{\mu^2_{\mathrm{v}} \equiv m_{\mathrm{v}}^2 + \frac{2\inp{D-1}}{\ell_{\mathrm{dS}}^{2}}}
	\label{mu2}
\end{equation}
We see that, at late times, the component $A_T$ behaves like a scalar field in dimension $D + 1$ with mass $\mu_{\mathrm{v}}$. \eqref{ATLate} therefore has leading late time solution of the form
\begin{equation}
	A_T \ \underset{T\to\infty}{\sim} \ A^{(-)}\,e^{-\Delta_-T/\ell_{\mathrm{dS}}}
	\label{AT}
\end{equation}
with $A^{(-)}$ a constant and with 
\begin{equation}
		\boxed{\Delta_{-} \equiv \frac{D + 1}{2} \pm \sqrt{\inp{\frac{D+1}{2}}^2 -\mu_{\mathrm{v}}^2\ell_{\mathrm{dS}}^2}}
		\label{Deltam}
\end{equation}
In terms of the Lagrangian mass, we have 
\begin{equation}
	\Delta_{-} \equiv \frac{D + 1}{2} \pm \sqrt{\inp{\frac{D-3}{2}}^2 -m_{\mathrm{v}}^2\ell_{\mathrm{dS}}^2}
	\label{DeltaLag}
\end{equation}
At the pode in static patch coordinates \eqref{dS}, one therefore finds that\footnote{We can also solve for the leading late time behavior of $A_R$ by plugging \eqref{AT} back into \eqref{LCTR0}, to find 
\begin{equation}
	\partial_RA_R + \frac{D-2}{R}\,A_R + \frac{\inp{\Delta_{-} + D-1}}{\ellds}\,A^{(-)}\,e^{-2T/\ellds}\,e^{-\Delta_-T/\ellds}  \ \underset{T \to \infty}{\sim} \   0
\end{equation}
This equation has solution (consistent with the Dirichlet boundary condition \eqref{Dirichlet})
\begin{equation}
	A_R \ \underset{T \to \infty}{\sim} \ -\frac{\inp{\Delta_{-} + D-1}}{\inp{D-1}\ellds}\,A^{(-)}\,R\,e^{-2T/\ellds}\,e^{-\Delta_-T/\ellds}
\end{equation}}
\begin{equation}
	\boxed{A_t \ \underset{t \to \infty}{\sim} \ \frac{A^{(-)}}{f(r)^{\Delta_-/2}}\,e^{-\Delta_-t/\ell_{\mathrm{dS}}}}
\end{equation}
as well as that 
\begin{equation}
	\boxed{A_r \ \underset{t \to \infty}{\sim} \ -\frac{r}{\ellds}\frac{A^{(-)}}{f(r)^{\Delta_-/2}}\,e^{-\Delta_-t/\ell_{\mathrm{dS}}}}
\end{equation}
Here we have used that 
\begin{equation}
	A_t = A_T - \frac{R}{\ellds}\,A_R \ \underset{T \to \infty}{\sim} \ A_T
\end{equation}
and 
\begin{equation}
	f(r)A_r = \frac{e^{-t/\ellds}}{\sqrt{f(r)}}\,A_R - \frac{r}{\ellds}\,A_T \ \underset{T \to \infty}{\sim} \ -\frac{r}{\ellds}\,A_T
\end{equation}

\subsubsection{The Edge of Stability for $s$-Wave Vectors}
\quad \
Now we encounter something that we found surprising. Unlike for scalars, the edge of stability for $s$-wave vectors does not correspond to vanishing ``Lagrangian mass" $m_{\mathrm{v}}^2 = 0$, as one might have na\"ively expected. From the above, we see that the edge of stability for vector Bosons is instead defined by the nominally tachyonic value
\begin{equation}
	\boxed{m_{\mathrm{v}}^2\ellds^2\Big|_{\text{edge}} = -2\inp{D-1}}
\end{equation}
or
\begin{equation}
	\boxed{\mu^2_{\mathrm{v}}\ellds^2\Big|_{\text{edge}} = 0}
\end{equation}
with $\mu^2_{\mathrm{v}}$ defined as in \eqref{mu2} above. As $\mu^2_{\mathrm{v}} \to 0$ (at fixed $\ell_{\mathrm{dS}}$) $\Delta_- \to 0$ and the system develops an asymptotically (in fact, as we will see in the next section, exactly) static mode. As with the scalar, this edge of stability for the lowest angular momentum mode will control the edge of stability for the full (but compactified) field $\mathsf{A}_{\mu}$, a fact that we will carefully check in \S\ref{QNM} below. Recall that for scalars the system becomes underdamped and picks up an oscillatory factor for $m_{\mathrm{s}}^2\ellds^2 > \inp{\frac{D-1}{2}}^2$. For compactified vector Bosons, the corresponding transition point is at 
\begin{equation}
	\mu_{\mathrm{v}}^2\ellds^2 > \inp{\frac{D + 1}{2}}^2 \quad \Longleftrightarrow \quad m_{\mathrm{v}}^2\ellds^2 > \inp{\frac{D - 3}{2}}^2 \qquad \text{(underdamped)}
\end{equation}
``Light" (physical) masses, for which the system asymptotically exponentially decays, correspond to the physical mass range 
\begin{equation}
	0 < \mu_{\mathrm{v}}^2\ellds^2 < \inp{\frac{D + 1}{2}}^2
\end{equation}
or, equivalently, the Lagrangian mass range
\begin{equation}
	\boxed{-2\inp{D-1} < m_{\mathrm{v}}^2\ellds^2 < \inp{\frac{D-3}{2}}^2}
\end{equation} 

\subsubsection{Why Is This Allowed?} 
\quad \ 
We will see that in many respects $\mu^2_{\mathrm{v}}$ controls the physics of the (compactified) massive vector field in dS$_D$ in the same way that the scalar mass $m_{\mathrm{s}}^2$ governs the physics of a massive scalar field in dS$_D$. For (compactified) massive vector fields, the constant $m_{\mathrm{v}}^2$ is simply a coefficient in the Lagrangian; it is $\mu^2_{\mathrm{v}}$ which governs both stability and (as we will see) the emergence of static solutions, zero modes, global shift symmetries, and ``infrared" divergences, as well as the dynamical exponents of correlation functions. In this sense we might consider $\mu_{\mathrm{v}}$ to be the ``effective physical mass" of the compactified massive minimally-coupled vector Boson field $\mathsf{A}_{\mu}$. We emphasize the label \emph{compactified}, which reminds us that we have broken the de Sitter isometries \eqref{IsodS} in order to define our spherical decomposition \eqref{LTDecomp} and the subsequent decomposition of our matter fields into modes of fixed angular momentum. 

Since we are not working within an $\mathrm{SO}(D,1)$-covariant framework, we are not forced to furnish a unitary irreducible representation of $\mathrm{SO}(D,1)$ which is what would usually lead to the requirement that $m_{\mathrm{v}}^2\ellds^2 > 0$ and to the rejection of the range $m_{\mathrm{v}}^2\ellds^2 < 0$ as ``tachyonic". Here is another perspective: In the usual $\mathrm{SO}(D,1)$-covariant framework for dS field theory, we are not allowed to enter the parameter range $m_{\mathrm{v}}^2\ellds^2 < 0$ lest we give up unitarity or locality. But by the latter we mean \emph{$D$-dimensional locality}, which is precisely what we have given up in the process of compactification, since all fields have been replaced by towers of modes, with the members of each tower determined by weighted averages over the local $(D-2)$-spheres. What we are left with (we conjecture) is a theory which is unitary and perfectly local in the $(1 + 1)$-dimensional sense, but rather nonlocal in the $D$-dimensional sense. We see that for situations such as ours---in which the full de Sitter isometry group is explicitly broken down to a static patch subgroup---the ``edge of stability" requirement $m_{\mathrm{v}}^2\ellds^2 > - 2\inp{D-1}$ seems to replace the usual $\mathrm{SO}(D,1)$ ``Higuchi bound" $m_s^2\ellds^2 \geq \inp{s-1}\inp{D-4 + s}$ where $s$ is the ``spin" (tensor rank) (see e.g. \cite{Higuchi:1986py,Anninos:2020hfj,Lust:2019lmq}).

\subsection{Edge of Stability for the Electric Field in $D = 3$}
\label{EdgeE}
\quad \
\quad Recall that for the special case $D = 3$, the electric field $E_r$ of the $s$-wave vector resembled the $l = 1$ mode of a scalar field of mass 
\begin{equation}
	m_{\mathrm{s}}^2 = \mu_E^2 \equiv m_{\mathrm{v}}^2 + \ellds^{-2}
\end{equation}
One might initially be worried that taking $m_{\mathrm{v}}^2\ellds^2 < -1$ drives this scalar mass into what na\"ively looks like a tachyonic range $\mu_{E}^2\ellds^2 < 0$. To see that this is indeed fine, recall our comments in \eqref{1Decay}: Since $E_r$ behaves as the \emph{isolated} $l = 1$ component of a $D = 3$ scalar field, it decays via the subleading quasinormal mode, with decay coefficient
\begin{equation}
	\updelta_- = 1 + 1 - \sqrt{1-\mu_E^2\ellds^2}
\end{equation}
Its edge of stability is therefore given by 
\begin{equation}
	\boxed{\mu_E^2\ell_{\mathrm{dS}}^2\Big|_{\text{edge}}= - 1\big(1 + 2\big) = -3}
\end{equation}
Using that
\begin{equation}
	\mu_{\mathrm{v}}^2 = m_{\mathrm{v}}^2 + 4\ellds^{-2} = \mu_E^2 + 3\ellds^2
\end{equation}
we see that this agrees with the previously determined edge of stability $\mu_{\mathrm{v}}^2\ellds^2 = 0$. Indeed we have that
\begin{align}
	\updelta_{-} 
	&= 2 -\sqrt{1-\mu_E^2\ellds^2}\\
	&= 2 - \sqrt{4-\mu_{\mathrm{v}}^2\ellds^2}\\
	&= \Delta_-
\end{align}
and so we see that the electric field decays with precisely the same decay exponent as the $s$-wave vector. So we see that whether we work in terms of $A_a$ or $E_r$ (or $Q$ which, by Gauss's law, is simply proportional to $E_r$) we find the same late time decay exponent and therefore the same edge of stability
\begin{equation}
	\mu_{E}^2\ell_{\mathrm{dS}}^2\Big|_{\text{edge}} = - 3 \quad \Longleftrightarrow \quad \mu_{\mathrm{v}}^2\ell_{\mathrm{dS}}^2\Big|_{\text{edge}} = 0 \quad \Longleftrightarrow \quad m_{\mathrm{v}}^2\ell_{\mathrm{dS}}^2\Big|_{\text{edge}} = -4
\end{equation}

\section{Physics of the Edge of Stability}
\subsection{Static Solutions Exist At (And Only At) The Edge of Stability}
\label{Static}
\quad \ 
We saw that, at the edge of stability, there were $s$-wave solutions of the equations of motion which were asymptotically static. This begs the question: do there exist \emph{exactly} static solutions to the equations of motion? And, if so, what is the origin of the requirement that $m_{\mathrm{v}}^2\ellds^2 = -2\inp{D-1}$ (i.e. that $\mu^2_{\mathrm{v}}\ellds^2 = 0$)? For a static $s$-wave solution $\partial_tA_a = 0$, the equations of motion \eqref{AtStatic}, \eqref{ArStatic} reduce to\footnote{That we have $A_r = 0$ is consistent with the fact that, for a static solution, we must also have $\partial_tE_r \propto A_r = 0$.} 
\begin{align}
	A_t& = \frac{1}{m_{\mathrm{v}}^2}\frac{f(r)}{r^{D-2}}\,\tdrm{}{r}\inp{r^{D-2}\,\tdrm{A_t}{r}}
	\label{tstatic}\\
	A_r &= 0
	\label{Areq0}
\end{align}
Note that the Lorenz constraint is trivially satisfied, so all we have to do is solve the ODE \eqref{tstatic}. In the range $-2\inp{D-1} \leq m_{\mathrm{v}}^2\ellds^2 \leq \inp{\frac{D-3}{2}}^2$---i.e. $0 \leq \mu_{\mathrm{v}}^2\ellds^2 < \inp{\frac{D+1}{2}}^2$---a solution obeying the boundary conditions \eqref{DirichletE}, \eqref{DirichletH} only exists at the gauge theory point $m_{\mathrm{v}}^2\ellds^2 = 0$ and at the edge of stability $m_{\mathrm{v}}^2\ellds^2 = -2\inp{D-1}$. In both cases this solution is given by
\begin{equation}
	A_t^{(\mathrm{static})} = -\frac{\ellds}{2}\frac{Q_0}{\mathrm{Area}}\,f(r)
	\label{staticD}
\end{equation}
with $Q_0$ an arbitrary constant. The electric field of this static mode is given by 
\begin{equation}
	E_r^{(\mathrm{static})} = \partial_rA_t = \frac{Q_0}{\mathrm{Area}}\inp{\frac{r}{\ellds}}
\end{equation}
This mode therefore encodes a charge
\begin{equation}
	Q^{(\mathrm{static})} = Q_0
	\label{QStatic}
\end{equation}
in the static patch. We can also think of it as corresponding to an electric potential 
\begin{equation}
	V(r) \equiv - A_t(r) = \frac{\ellds}{2}\frac{Q_0}{\mathrm{Area}}\,f(r)
\end{equation}
Geometrically, this mode corresponds to the situation in which $\mathsf{A}^{\mu}$ is a constant multiple of the boost Killing field
\begin{equation}
	\mathsf{A}^{\mu}\inp{\pd{}{\mathsf{x}^{\mu}}}\ =  \ \frac{\ellds}{2}\frac{Q_0}{\mathrm{Area}}\inp{\pd{}{t}}
\end{equation}
We see that, at the edge of stability, there is an emergent one-parameter family of static $s$-wave solutions to the equations of motion.

\subsection{Emergent Zero Modes At the Edge of Stability}
\label{Zero}
\quad \ 
In fact, the static solution \eqref{staticD} encodes an emergent zero mode of the action; the solution \eqref{staticD} has vanishing action and energy for any value of the parameter $Q_0$. We have
\begin{align}
	+H[\text{static mode}]
	&= -I[\text{static mode}]\\
	\ &\propto \ \intf{0}{\ellds}\mathrm{d}r\,r^{D-2}\,\inp{\big(E_r^{(\mathrm{static})}\big)^2 + \frac{m_{\mathrm{v}}^2}{f(r)}\,\big(A_t^{(\mathrm{static})}\big)^2}\\
	\ &\propto \ \intf{0}{\ellds}\mathrm{d}r\,r^{D-2}\inp{\inp{-\frac{2r}{\ellds^2}}^2 + \frac{m_{\mathrm{v}}^2}{f(r)}\,f(r)^2}\\
	\ &\propto \ \intf{0}{\ellds}\mathrm{d}r\inp{4r^D + m_{\mathrm{v}}^2\ellds^4\,r^{D-2}f(r)}\\
	\ &\propto \ \intf{0}{\ellds}\mathrm{d}r\inp{\inp{4-m_{\mathrm{v}}^2\ellds^2}r^D + m_{\mathrm{v}}^2\ellds^4\,r^{D-2}}\\[0.25em]
	\ &\propto \ \frac{\inp{4-m_{\mathrm{v}}^2\ellds^2}}{D + 1} + \frac{m_{\mathrm{v}}^2\ellds^2}{D-1}\\[0.25em]
	\ &\propto \ \inp{D-1}\inp{4-m_{\mathrm{v}}^2\ellds^2} + \inp{D+1}m_{\mathrm{v}}^2\ellds^2\\
	&\propto 2\inp{D-1} + m_{\mathrm{v}}^2\ellds^2\\[0.25em]
	&= 0 
\end{align}
where in the last step we have used that, at the edge of stability, $m_{\mathrm{v}}^2\ellds^2 = - 2\inp{D-1}$. So we see that, at the edge of stability, there is an emergent one-parameter family of zero modes of the action (namely the static $s$-wave solutions parameterized by $Q_0 \in \R$).

\subsection{Emergent Shift Symmetry at the Edge of Stability} 
\label{Symmetry}
\quad We will now show that the emergent zero modes at edge of stability also govern corresponding emergent global symmetries of the action.

\subsubsection{A Warm Up: Scalar Field}
\quad \ 
At the edge of stability $m_{\mathrm{s}}^2 = 0$ for a scalar field, the action \eqref{IPhi} reduces to
\begin{equation}
	I[\upphi] = -\frac{1}{2}\int\mathrm{d}^{D}\mathsf{x}\sqrt{|\mathsf{g}|}\,\mathsf{g}^{\mu\nu}\nabla_{\mu}\upphi\nabla_{\nu}\upphi
\end{equation}
and consequently develops a zero mode corresponding to constant field configurations
\begin{equation}
	\upphi(\mathsf{x}) = \upphi_0 = \mathrm{constant}
\end{equation}
The action also develops an associated continuous global symmetry corresponding to shifts by the zero mode
\begin{equation}
	\upphi(\mathsf{x}) \to \upphi(\mathsf{x}) + \upphi_0, \qquad \upphi_0 = \mathrm{constant}
	\label{Shift}
\end{equation} 
We can think of the zero mode $\upphi_0$ as the Goldstone mode associated to the spontaneous breaking of the shift symmetry by any concrete field configuration $\upphi(\mathsf{x})$.

\subsubsection{The $s$-Wave Vector Boson} 
\quad \
Consider now the edge of stability $\mu^2_{\mathrm{v}} = 0$ for a vector field. The system again develops a zero mode
\begin{equation}
	\mathsf{A}_{\mu}^{(\mathrm{static})}(\mathsf{x})\,\mathrm{d}\mathsf{x}^{\mu} = - \frac{\ellds}{2}\frac{Q_0}{\mathrm{Area}}\,f(r)\,\mathrm{d}t
\end{equation}
There is again an associated continuous global symmetry corresponding to constant shifts
\begin{equation}
	\mathsf{A}_{\mu}(\mathsf{x})\,\mathrm{d}\mathsf{x}^{\mu} \to \mathsf{A}_{\mu}(\mathsf{x})\,\mathrm{d}\mathsf{x}^{\mu} + \mathsf{A}_{\mu}^{(\mathrm{static})}(\mathsf{x})\,\mathrm{d}\mathsf{x}^{\mu}
\end{equation}
or \cite{Bonifacio:2018zex}
\begin{align}
	\mathsf{A}^{\mu}(\mathsf{x})\inp{\pd{}{\mathsf{x}^{\mu}}} \to \mathsf{A}^{\mu}(\mathsf{x})\inp{\pd{}{\mathsf{x}^{\mu}}} + \frac{\ellds}{2}\frac{Q_0}{\mathrm{Area}}\inp{\pd{}{t}}
	\label{ShiftV}
\end{align}
That this is a symmetry simply follows from the fact that $\mathsf{A}^(\mathrm{static})$ is a solution of the equation of motion for a quadratic action:
\begin{align}
	I[\mathsf{A} + \mathsf{A}^{(\mathrm{static})}] 
	&= I[\mathsf{A}^{(\mathrm{static})} + \mathsf{A}]\\
	&= \underbrace{I[\mathsf{A}^{(\mathrm{static})}]}_{\text{$= 0$ (zero mode)}} + \underbrace{\int\mathrm{d}^D\mathsf{x}\,\mathsf{A}_{\mu}(\mathsf{x})\,\frac{\updelta I}{\updelta \mathsf{A}_{\mu}(\mathsf{x})}\bigg|_{\mathsf{A}^{(\mathrm{static})}}}_{\text{$= 0$ since $\mathsf{A}^{(\mathrm{static})}$ satisfies EOM}}\nonumber\\[0.5em]
	&\hspace{12em} + \underbrace{\frac{1}{2}\int\mathrm{d}^D\mathsf{x}\,\mathrm{d}^D\mathsf{y}\,\mathsf{A}_{\mu}(\mathsf{x})\frac{\updelta^2 I}{\updelta \mathsf{A}_{\nu}(\mathsf{y})\updelta \mathsf{A}_{\mu}(\mathsf{x})}\bigg|_{\mathsf{A}^{(\mathrm{static})}}\,\mathsf{A}_{\nu}(\mathsf{y})}_{\text{$= I[\mathsf{A}]$ since action is quadratic}}\\
	&= I[\mathsf{A}]
\end{align}
In going to the second line we have expanded the action about $\mathsf{A}^{(\mathrm{static})}$ and then used the various properties described inline. We see that, at the edge of stability, there is an emergent global shift symmetry \eqref{ShiftV} accompanied by emergent Goldstone modes $\mathsf{A}^{(\mathrm{static})}_{\mu}$ associated to the spontaneous breaking of this shift symmetry by any concrete field configuration $\mathsf{A}_{\mu}$. The fact that the equations of motion of the Proca field develop a shift symmetry at the (squared) Lagrangian mass value $m_{\mathrm{v}}^2 = - 2\inp{D-1}\ellds^{-2}$ was first noticed in \cite{Bonifacio:2018zex}. Here we have shown that this symmetry descends from a symmetry of the action, and have also explained the relation to static solutions, zero modes, and the ``edge of stability".

\subsection{Emergent Supersymmetry of the Equations of Motion in $D = 3$}
\quad \ 
In the special case of $D = 3$, there is another signature of the distinguished role played by the edge of stability $\mu^2_{\mathrm{v}} = 0$. Specifically, at the edge of stability the static equation of motion \eqref{tstatic} becomes the time-independent Schr\"odinger equation for the ground state of a supersymmetric\footnote{Note that we are \emph{not} asserting the emergence of any sort of ``real" spacetime supersymmetry; we are simply observing a neat feature of the structure of the static equation of motion at the edge of stability.} $(0+1)$-dimensional quantum mechanics.  To see this, let us return to the static equation of motion \eqref{tstatic} 
\begin{equation}
	-\frac{1}{r}\tdrm{}{r}\inp{r\,\tdrm{A_t}{r}} + \frac{m_{\mathrm{v}}^2}{f(r)}A_t = 0
	\label{tstaticpresusy}
\end{equation}
If we define
\begin{equation}
	u(r) \equiv -2\ln\Big(\frac{r}{\ellds}\Big) \quad \Longleftrightarrow \quad \frac{r(x)}{\ellds} =  e^{-u/2}
\end{equation}
and 
\begin{equation}
	A_t(r) \equiv \psi\big(u(r)\big)
\end{equation}
\eqref{tstaticpresusy} becomes
\begin{equation}
	\hat{H}(u)\,\psi(u) \equiv \inp{-\tdrmord{2}{}{u} + V(u)}\psi(u) = 0
\end{equation}
which we recognize as the time-independent Schr\"odinger equation for a zero energy eigenstate $\psi(u)$ on the half line $u \in [0,\infty)$ ($u$ runs from $0$ at the horizon to $+\infty$ at the pode) with potential
\begin{equation}
	V(u) = -\frac{1}{4}\frac{4 - \mu^2_{\mathrm{v}}\ellds^2}{e^u-1}
\end{equation}
For $\mu^2_{\mathrm{v}} = 0$ and only for $\mu^2_{\mathrm{v}} = 0$ the Hamiltonian becomes supersymmetric 
\begin{equation}
	\hat{H}(u) = \inp{\hat{p}_u + \mathrm{i}W(u)}\inp{\hat{p}_u-\mathrm{i}W(u)}
\end{equation}
with $\hat{p}_u = -\mathrm{i}\tdrm{}{u}$ and with superpotential
\begin{equation}
	W(u) = -\frac{1}{e^u - 1}
\end{equation}
The equation of motion therefore reduces to the \emph{linear} equation 
\begin{equation}
	\big(\hat{p}_u-\mathrm{i}W(u)\big)\psi(u) = -\mathrm{i}\inp{\tdrm{}{u} - \frac{1}{e^u-1}}\psi(u) = 0
\end{equation}
with solution 
\begin{equation}
	\psi(u) = -A^0\inp{1-e^{-u}} = -A^0f(r)
\end{equation}
with $A^0$ a constant, as expected. This ground state is normalizable with respect to the natural measure on the constant $t$ surface $\Sigma_t$:
\begin{align}
	2\pi\intf{0}{\ellds}\,\mathrm{d}r\,r\,\|\psi\big(u(r)\big)\|^2 
	&= \pi\ellds^2\inp{A^0}^2\,\intf{0}{\infty}\mathrm{d}u\,e^{-u}\inp{1-e^{-u}}^2\\
	&= \frac{\pi\ellds^2}{3}\inp{A^0}^2
\end{align}

\section{Beyond $s$-Wave Modes} 
\quad \ 
Let us now turn to the remaining two types of field configurations for the massive minimally-coupled vector Boson. In this section we will summarize results which will then be carefully derived in \S\ref{QNM} below. 

\subsection{Sphere-Transverse Modes} 
\quad 
The first remaining type of mode a pure sphere-transverse field $$\mathsf{A}_{\mu}(\mathsf{x})\,\mathrm{d}\mathsf{x}^{\mu} = \tilde{\mathsf{A}}_A(\mathsf{x})\,\mathrm{d}\theta^A$$ obeying 
\begin{equation}
	\Omega^{AB}D_A\tilde{\mathsf{A}}_B = 0
\end{equation}
There are two subcases to consider, namely $D = 3$ and $D > 3$. 

\subsubsection{$D = 3$: The Circularly Polarized Mode}
\label{CPM3}
\quad \
In $D = 3$ spacetime dimensions, the ``local $(D-2)$-spheres" are circles. Unlike in higher spacetime dimension, where the transverse spin-1 vector harmonics begin at angular momentum $j \geq 1$, in $D = 3$ spacetime dimensions there is a single transverse spin-1 vector harmonic with angular momentum $l = 0$:
\begin{equation}
	\tilde{\mathsf{A}}_{\theta}(\mathsf{x})\,\mathrm{d}\theta \ \underset{D =3 }{=} \ C(x)\,\mathrm{d}\theta
\end{equation}
We will call this exceptional $l = 0$ mode the ``circularly polarized" mode, since it gives rise to configurations of the electric field 
\begin{equation}
	\mathsf{E}_{\theta} = \partial_tC
\end{equation}
which propagate radially while being polarized along the supressed local circle. This mode does not couple to any others and is described by an effective action
\begin{equation}
	\frac{I[C]}{2\pi\ellds} = -\frac{1}{2}\int\mathrm{d}^{2}x\,\frac{1}{r}\,\bigg(g^{ab}\partial_aC\,\partial_bC + m_{\mathrm{v}}^2\,C^2\bigg)
\end{equation}
This is of course identical to the action \eqref{IQ} of the $s$-wave charge mode, meaning that $C(x)$ will be controlled---for identical reasons, and in an identical way---by the physical mass $\mu_{\mathrm{v}}$. In fact, we expect the circularly polarized mode $C(x)$ to posess all of the same features as the $s$-wave mode described above.

\subsubsection{$D > 3$: Nontrivial Sphere-Transverse Modes}
\quad \ 
Let us now consider the generic case of spacetime dimension $D \geq 4$. As we will show in  \S\ref{QNM} below, the sphere-transverse field $\tilde{\mathsf{A}}_A$ decays via quasinormal resonances of purely complex frequency
\begin{equation}
	\omega_{\pm,j,n} = -\mathrm{i}\inp{\Delta_{\pm} + \inp{j-1} + 2n}, \qquad (j-1), n \in \Z_{\geq 0}
\end{equation}
where $j \geq 1$ is the ``angular momentum" defined by the spin-1 transverse vector harmonics, and we have defined the vector weights $\Delta_{\pm}$ via
\begin{equation}
	\Delta_{\pm} \equiv \frac{D + 1}{2} \pm \sqrt{\inp{\frac{D+1}{2}}^2-\mu_{\mathrm{v}}^2\ellds^2}
\end{equation}
The lowest quasinormal frequency is given by $-\mathrm{i}\Delta_-$, identically with the $s$-wave mode. We therefore see that (at least classically) \emph{the edge of stability of the sphere-transverse mode $\tilde{\mathsf{A}}_A$ is controlled by the same effective physical mass $\mu_{\mathrm{v}}$.}

\subsection{The Remaining Modes}
\quad \ 
The remaining type of field configuration is the non-$s$-wave part of the nonspherical component coupled to the sphere-longitudinal component via the Lorenz constraint \eqref{LC}. As we will also show in  \S\ref{QNM} below, such a field decays via subleading quasinormal resonances of purely complex frequency 
\begin{equation}
	\omega_{\pm,l,n} = -\mathrm{i}\inp{\Delta_{\pm} + l + 2n}, \qquad l \in \Z_{\geq 1}, \quad n \in \Z_{\geq 0}
\end{equation}
Since the decay of these modes is subleading to the ones previously considered, they are not expected to interfere with the edge of stability (again, at least classically).

\section{Classical Solutions and Quasinormal Frequencies}
\label{QNM}
\quad \
In this section, we will fully solve the classical equation of motion \eqref{Proca} for the massive minimally-coupled spin-1 real vector Boson in the static patch of dS$_D$, generalizing previous work by Higuchi \cite{Higuchi:1986ww}---done for the special case of $D = 4$---to general spacetime dimension $D \geq 3$. We will then use the structure of these solutions to quantify the quasinormal frequency spectrum of the vector Boson as a function of mass and angular momentum in order to justify the statements made in the previous section. 

We begin by reviewing the well known case of the massive scalar field before turning to the case of interest, namely the massive minimally-coupled real vector Boson. While this work was being completed, we became aware of upcoming work by Grewal, Law, and Lochab \cite{Albert} which will contain some overlap with the following.

\subsection{A Warm Up: Massive Minimally-Coupled Real Scalar in the Static Patch}
\label{ScalarC}
\quad \ 
As a warm up, consider again the massive minimally-coupled real scalar field $\upphi$ of mass $m_{\mathrm{s}}$ in the static patch of dS$_D$, which is classically governed by the Klein-Gordon equation
\begin{equation}
	\inp{-\nabla^2_{(0)} + m_{\mathrm{s}}^2}\upphi = 0
	\label{KGApp}
\end{equation}
Here we have denoted by $\nabla^2_{(0)}$ the Laplace-Beltrami operator (covariant Laplacian) $\mathsf{g}^{\mu\nu}\nabla_{\mu}\nabla_{\nu}$ of dS$_D$ acting on (not necessarily $s$-wave) zero-forms/scalars:
\begin{align}
	\nabla^2_{(0)}\upphi
	&\equiv 
	\frac{1}{\sqrt{|\mathsf{g}|}}\,\partial_{\mu}\inp{\sqrt{|\mathsf{g}|}\,\mathsf{g}^{\mu\nu}\partial_{\nu}\upphi}\\[0.5em]
	&= \insb{-\frac{1}{f(r)}\,\partial_t^2 + f(r)\,\partial_r^2 + \inp{\frac{D-2}{r}-\frac{Dr}{\ellds^2}}\partial_r}\upphi
	\label{SL}
\end{align}
We would like to find the general solution to \eqref{KGApp} in the static patch
. Due to the time translation and spherical symmetries of the static patch as well as the linearity of the Klein-Gordon equation \eqref{KGApp}, we can expand a general solution of \eqref{KGApp} in ``normal modes"
\begin{equation}
	\upphi(\mathsf{x}) 
	= \sum_{lm}\intf{0}{\infty}\frac{\mathrm{d}\omega}{2\pi}\,\frac{1}{\sqrt{2\omega}}\inp{\mathsf{a}_{\omega lm}\,\phi_{\omega l}(r)\,Y_{lm}(\theta)\,e^{-\mathrm{i}\omega t/\ell_{\mathrm{dS}}} + \mathrm{c.c.}}
	\label{PhiDecompApp}
\end{equation}
of definite angular momentum $l \in \Z_{\geq 0}$ and 
dimensionless temporal frequency $\omega \in \R$ (here ``c.c." denotes the complex conjugate). 
The $\mathsf{a}_{\omega lm}$ are dimensionless constants expressing the relative contribution of each normal mode in the expansion \eqref{PhiDecompApp} (we absorb the dimensions of $\upphi$ into the ``radial functions"\footnote{Note that our usage of ``radial function" here differs slightly from what is usually called the ``radial function" in the literature. The latter notion of radial function, which we will denote by $\psi_{\omega l}$, is given by radiative part of $\phi_{\omega l}$:
	\begin{equation}
		\phi_{\omega l}(r) = \frac{\psi_{\omega l}(r)}{r^{\frac{D-2}{2}}}
\end{equation}} $\phi_{\omega l}$). The equation of motion \eqref{KGApp} implies that the radial functions $\phi_{\omega l}(r)$ satisfy the radial equation
\begin{equation}
	\insb{-\frac{1}{f(r)}\,\frac{\omega^2}{\ell_{\mathrm{dS}}^2} - f(r)\,\partial_r^2 - \inp{\frac{D-2}{r} - \frac{Dr}{\ell_{\mathrm{dS}}^2}}\partial_r + \frac{l\inp{l + D - 3}}{r^2} + m_{\mathrm{s}}^2}\phi_{\omega l}(r) = 0
	\label{radialScalar}
\end{equation} 
Requiring that $\upphi$ be regular at the pode $(r = 0)$ gives
\begin{equation}
	\phi_{\omega l}(r) 
	\ \propto \ \inp{\frac{r}{\ell_{\mathrm{dS}}}}^lf(r)^{+\mathrm{i}\omega/2}\, {}_2F_1\inp{\frac{\delta_+ + \mathrm{i}\omega + l}{2},\frac{\delta_- + \mathrm{i}\omega + l}{2};\frac{D-1}{2} + l; \frac{r^2}{\ell_{\mathrm{dS}}^2}}
	\label{phiModes}
\end{equation} 
where the scalar weights $\delta_{\pm}$ are given by
\begin{equation}
	\delta_{\pm} = \frac{D-1}{2} \pm \sqrt{\inp{\frac{D-1}{2}}^2 -m_{\mathrm{s}}^2\ell_{\mathrm{dS}}^2}
\end{equation}

\subsubsection{Quasinormal Frequencies}
\quad \
In order to analyze near-horizon behavior, it will be helpful to define the tortoise coordinate
\begin{align}
	\frac{r_*}{\ellds} &\equiv \mathrm{arctanh}\bigg(\frac{r}{\ellds}\bigg)
\end{align}
In terms of the tortoise coordinate, the cosmological horizon is pushed out to $r_* = \infty$ and the nonspherical part of the metric becomes conformally flat
\begin{equation}
	\mathsf{g}_{\mu\nu}(\mathsf{x})\,\mathrm{d}\mathsf{x}^{\mu}\mathrm{d}\mathsf{x}^{\nu} = f(r)\inp{-\mathrm{d}t^2 + \mathrm{d}r_*^2} + \ellds^2\tanh^2\Big(\frac{r_*}{\ellds}\Big)\,\mathrm{d}\Omega_{(D-2)}^2
\end{equation}
As we approach the horizon, 
the normal modes decompose into sums of ingoing and outgoing waves
\begin{equation}
\phi_{\omega l}(r)\,e^{-\mathrm{i}\omega t/\ellds} \ \underset{r \to \ellds}{\sim} \ \inp{T(\omega)\,e^{-\mathrm{i}\omega \inp{t-r_*}/\ellds} + R(\omega)\,e^{-\mathrm{i}\omega \inp{t + r_*}/\ellds}}
\end{equation} 
The static patch scattering phase is defined to be the ratio of the transmission and reflection coefficients
\begin{equation}
S(\omega) \equiv \frac{T(\omega)}{R(\omega)} = \frac{\Gamma\big(\frac{\delta_+ + l - \mathrm{i}\omega}{2}\big)\Gamma\big(\frac{\delta_- + l - \mathrm{i}\omega}{2}\big)}{\Gamma\big(\frac{\delta_+ + l + \mathrm{i}\omega}{2}\big)\Gamma\big(\frac{\delta_- + l + \mathrm{i}\omega}{2}\big)}\cdot\frac{\Gamma(+\mathrm{i}\omega)}{\Gamma(-\mathrm{i}\omega)}
\label{Sw}
\end{equation}
As explained in e.g. \cite{Grewal:2024emf}, the quasinormal frequencies (i.e. the poles of the retarded Green's function) can be read off as the poles of the first factor of \eqref{Sw}. For each fixed orbital angular momentum $l \geq 0$, there are two towers of such quasinormal frequencies 
\begin{equation}
\omega_{\pm,l,n} = -\mathrm{i}\inp{\delta_{\pm} + l + 2n}, \qquad n \in \Z_{\geq 0}
\end{equation}
In other words, the quasinormal frequency spectrum is given by 
\begin{equation}
	\boxed{\omega_{\pm,l,n} = -\mathrm{i}\inp{\delta_{\pm} + l + 2n}, \qquad l,n \in \Z_{\geq 0}}
\end{equation}
These quasinormal frequencies control the late time decay of the retarded Green's function $G^{(R)}(\mathsf{x};\mathsf{x}')$ via 
\begin{equation}
G^{(R)}(t,r,\theta;0,r',\theta') \ \underset{t\gg\ellds}{\sim} \  \sum_{\pm}\sum_{l,n}g_{\pm,l,n}(r,r',\theta-\theta')\,e^{-\inp{\delta_{\pm} + l + 2n}t/\ellds}
\end{equation}
Equivalently\footnote{As explained in \cite{Grewal:2024emf} this is a not exactly true (i.e. we have made a slight abuse of terminology). The late time decay of the ingoing part of the solution is described by the quasinormal frequencies \emph{along with} the Matsubara" frequencies $\omega^M_{k} = -\mathrm{i}k$ ($k> 0$) associated with the ``Rindler" part of the transmission phase (the second factor of \eqref{Sw}).}, they describe the late time decay of the purely ingoing part of the classical solution (here by ``ingoing" we mean the mode which is moving towards the horizon).

\subsection{Massive Minimally-Coupled Vector Boson in the Static Patch}
\label{General}
\quad \ 
Consider now the massive minimally-coupled real vector Boson $\mathsf{A}_{\mu}\,\mathrm{d}\mathsf{x}^{\mu}$ of squared ``Lagrangian mass" $m_{\mathrm{v}}^2$ in the static patch of dS$_{D}$. This field is classically governed by the Proca equation \eqref{Proca}: 
\begin{equation}
-\frac{1}{\sqrt{|\mathsf{g}|}}\,\mathsf{g}_{\mu\rho}\,\partial_{\nu}\inp{\sqrt{|\mathsf{g}|}\,\mathsf{g}^{\nu\sigma}\mathsf{g}^{\rho\lambda}\,\mathsf{F}_{\sigma\lambda}} + m_{\mathrm{v}}^2\mathsf{A}_{\mu} = 0
\label{ProcaApp}
\end{equation}
which contains the ``Lorenz constraint" \eqref{LC}: 
\begin{equation}
	-\frac{1}{\sqrt{|\mathsf{g}|}}\,\partial_{\mu}\inp{\sqrt{|\mathsf{g}|}\,\mathsf{g}^{\mu\nu}\,\mathsf{A}_{\nu}} = 0
	\label{LCApp}
\end{equation}
We would like to find the general solution to \eqref{ProcaApp} in the static patch. This was done for the special case of $D = 4$ spacetimes in \cite{Higuchi:1986ww}, and we will mainly follow and generalize the logic of that work. To begin, it will be helpful to divide $\mathsf{A}_{\mu}$ into nonspherical, sphere-longitudinal, and sphere-transverse parts
\begin{equation}
	\mathsf{A}_{\mu}(\mathsf{x})\,\mathrm{d}\mathsf{x}^{\mu} = \mathsf{A}_a(\mathsf{x})\,\mathrm{d}x^a + D_A\mathes{A}(\mathsf{x})\,\mathrm{d}\theta^A + \tilde{\mathsf{A}}_A(\mathsf{x})\,\mathrm{d}\theta^A
	\label{splitApp}
\end{equation}
We remind the reader that, by definition, the sphere-transverse part $\tilde{\mathsf{A}}_A$ obeys 
\begin{equation}
	\Omega^{AB}D_A\tilde{\mathsf{A}}_B
\end{equation}
with $D_A$ the covariant derivative operator associated to the Levi-Civita connection of $\Omega_{AB}$. In terms of the split \eqref{splitApp}, the Lorenz constraint \eqref{LCApp} becomes 
\begin{equation}
-\frac{1}{f(r)}\,\partial_t\mathsf{A}_t + \frac{1}{r^{D-2}}\,\partial_{r}\inp{r^{D-2}f(r)\,\mathsf{A}_{r}} + \frac{1}{r^2}\,\Omega^{AB}D_AD_B\mathes{A} = 0
\label{LCSplitApp}
\end{equation}
which completely fixes the longitudinal part of $D_A\mathes{A}$  in terms of the nonspherical (and non-$s$-wave) part $\mathsf{A}_a$ via 
\begin{equation}
	\mathes{A}_{lm} = \frac{r^2}{l\inp{l + D - 3}}\insb{-\frac{1}{f(r)}\,\partial_t\mathsf{A}^{lm}_t + \frac{1}{r^{D-2}}\,\partial_{r}\inp{r^{D-2}f(r)\,\mathsf{A}^{lm}_{r}}}, \qquad l \geq 1
\end{equation}
where we have expanded 
\begin{equation}
	D_A\mathes{A} = \sum_{l\geq 1, m}\mathes{A}_{lm}D_AY_{lm}, \qquad \mathsf{A}_a = \sum_{lm}A_a^{lm}\,Y_{lm}
\end{equation}
For $l = 0$, there is no sphere-longitudinal part of $\mathsf{A}_{\mu}$ and the Lorenz constraint serves to further constrain the $s$-wave mode $A_a \equiv A_a^{00}$, leading to the nontrivial constrained phase space structure explored in the previous sections. 

Let us denote by $\nabla^2_{(0)}$ the Laplace-Beltrami operator (covariant Laplacian) $\mathsf{g}^{\mu\nu}\nabla_{\mu}\nabla_{\nu}$ of dS$_D$ acting on the \emph{components} of the one-form $\mathsf{A}_{\mu}$ \emph{treated as zero-forms/scalars}, i.e.
\begin{equation}
\nabla^2_{(0)}
=
-\frac{1}{f(r)}\,\partial_t^2 + \frac{1}{r^{D-2}}\,\partial_r\inp{r^{D-2}\,f(r)\,\partial_r} + \frac{1}{r^2}\,\Omega^{AB}D_AD_B
\end{equation}
The $t$, $r$, and $A$ components of the Proca equation \eqref{ProcaApp} then read
\begin{equation}
\inp{-\nabla^2_{(0)} + m_{\mathrm{v}}^2}\mathsf{A}_t  - \frac{2r}{\ellds^2}\,\partial_r\mathsf{A}_{t} + \frac{2r}{\ellds^2}\,\partial_t\mathsf{A}_{r} + \partial_t\inp{\text{Lorenz constraint}}= 0
\label{Pt}
\end{equation}
\begin{multline}
\inp{-\nabla^2_{(0)} + m_{\mathrm{v}}^2}\mathsf{A}_r + \inp{\frac{D}{\ellds^2} + \frac{D-2}{r^2} + \frac{2r}{\ellds^2}\partial_r}\mathsf{A}_r  + \frac{1}{f(r)^2}\frac{2r}{\ellds^2}\partial_t\mathsf{A}_t + \frac{2}{r^3}\,\Omega^{AB}D_AD_B\mathes{A} \\+  \partial_r\inp{\text{Lorenz constraint}} = 0
\label{Pr}
\end{multline}
\begin{equation}
\inp{-\nabla^2_{(0)} + m_{\mathrm{v}}^2}\mathsf{A}_A + \frac{D-3}{r^2}\,\mathsf{A}_A + \frac{2f(r)}{r}\,\mathsf{F}_{rA}  + D_A\inp{\text{Lorenz constraint}} = 0
\label{PA}
\end{equation}
Solving the Lorenz constraint \eqref{LCSplitApp} for $\Omega^{AB}D_AD_B\mathes{A}$ and plugging into \eqref{Pr}, we find that\footnote{Note that \eqref{Pt} is---up to an overall minus sign---equation (3.6a) of \cite{Higuchi:1986ww}, which we see actually holds exactly independent of spacetime dimension $D \geq 3$. For $D = 4$ \eqref{Pr} and \eqref{315} reduce---again up to overall minus signs---to equations (3.6b) and (3.15) of \cite{Higuchi:1986ww} respectively.}  
\begin{equation}
\insb{-\nabla^2_{(0)}+\inp{m_{\mathrm{v}}^2 + \frac{3D}{\ell_{\mathrm{dS}}^2}} - \frac{D-2}{r^2} - \inp{\frac{2}{r}-\frac{4r}{\ell_{\mathrm{dS}}^2}}\partial_r}\mathsf{A}_r + \frac{2}{rf(r)^2}\,\partial_t\mathsf{A}_t = 0
\label{315}
\end{equation}

We now define the modes\footnote{Note that $A_+ = A_T$ with $T$ defined as in \eqref{Tflat}. We can recover the components $\mathsf{A}_t$ and $\mathsf{A}_r$ from the modes $\mathsf{A}_{\pm}$ via 
\begin{equation}
	\mathsf{A}_t = \frac{f(r)}{2}\inp{\mathsf{A}_+ + \mathsf{A}_-} \quad\text{and}\quad \mathsf{A}_r = \frac{\ell_{\mathrm{dS}}}{2r}\inp{\mathsf{A}_+ - \mathsf{A}_-}
	\label{trpm}
\end{equation}
} \cite{Higuchi:1986ww}
\begin{equation}
\mathsf{A}_{\pm} \equiv \frac{1}{f(r)}\,\mathsf{A}_{t} \pm \frac{r}{\ellds}\mathsf{A}_r
\label{Apm}
\end{equation}
in terms of 
which $\frac{2}{f(r)}$ times \eqref{Pt} becomes 
\begin{equation}
\inp{-\nabla^2_{(0)} + \mu_{\mathrm{v}}^2 + \frac{2r}{\ell_{\mathrm{dS}}^2}\,\partial_r}\inp{\mathsf{A}_+ + \mathsf{A}_-} + \frac{2}{\ell_{\mathrm{dS}} f(r)}\,\partial_t\inp{\mathsf{A}_+-\mathsf{A}_-} = 0
\label{p}
\end{equation}
while $2r$ times \eqref{315} becomes
\begin{equation}
\inp{-\nabla^2_{(0)}+\mu_{\mathrm{v}}^2 + \frac{2r}{\ell_{\mathrm{dS}}^2}\,\partial_r}\inp{\mathsf{A}_+-\mathsf{A}_-} + \frac{2}{\ell_{\mathrm{dS}} f(r)}\,\partial_t\inp{\mathsf{A}_+ + \mathsf{A}_-} = 0
\label{m}
\end{equation}
Here we have recalled the definition \eqref{mu2} of the (squared) effective physical mass:
\begin{equation}
	\mu_{\mathrm{v}}^2 \equiv m_{\mathrm{v}}^2 + \frac{2\inp{D-1}}{\ell_{\mathrm{dS}}^2}
\end{equation}
Taking the sum and difference of the equations \eqref{p} and \eqref{m} reveals the two equations\footnote{Note that \eqref{317} is equation (3.17) of \cite{Higuchi:1986ww}, which we see actually holds exactly independent of spacetime dimension $D \geq 3$.}
\begin{equation}
\inp{-\nabla^2_{(0)} + \mu^2 + \frac{2r}{\ell_{\mathrm{dS}}^2}\partial_r \pm \frac{2}{\ell_{\mathrm{dS}} f(r)}\partial_t}\mathsf{A}_{\pm} = 0
\label{317}
\end{equation} 
which are decoupled outside of the $s$-wave sector.

We would like to find the general solution to \eqref{317} in the static patch. Due to the time translation and spherical symmetries of the static patch as well as the linearity of \eqref{317}, we can again expand a general solution of \eqref{317} in normal modes
\begin{equation}
\mathsf{A}_{\pm}(\mathsf{x})
= \sum_{lm}\intf{0}{\infty}\frac{\mathrm{d}\omega}{2\pi}\,\frac{1}{\sqrt{2\omega}}\inp{\upalpha^{\pm}_{\omega lm}\,A_{\pm}^{\omega l}(r)\,Y_{lm}(\theta)\,e^{-\mathrm{i}\omega t/\ell_{\mathrm{dS}}} + \mathrm{c.c.}}
\label{ApmDecompApp}
\end{equation}
where $\upalpha^{\pm}_{\omega lm}$ are dimensionless constants expressing the relative contribution of each normal mode in the expansion (we again absorb the dimensions of $\mathsf{A}_{\pm}$ into the radial functions $A^{\omega l}_{\pm}$). \eqref{317} implies that the radial function $A_{\pm}^{\omega l}(r)$ satisfies the radial equation
\begin{equation}
\insb{-\frac{1}{f(r)}\frac{\omega^2}{\ellds^2} - f(r)\,\partial_r^2 - \inp{\frac{D-2}{r} - \frac{Dr}{\ell_{\mathrm{dS}}^2}}\partial_r 
+ \frac{l\inp{l + D - 3}}{r^2} + \mu^2_{\mathrm{v}} + \frac{2r}{\ellds^2}\,\partial_r \mp \frac{2\mathrm{i}}{\ellds^2 f(r)}\,\omega}A_{\pm(\omega)}(r) = 0
\end{equation}
Requiring that that $\mathsf{A}_{\pm}$ be regular at the pode ($r = 0$) gives
\begin{equation}
\boxed{A_{\pm}^{\omega l}(r) \ \propto \ \inp{\frac{r}{\ellds}}^lf(r)^{\mp\mathrm{i}\omega/2}\,{}_2F_1\inp{\frac{\Delta_+\mp\mathrm{i}\omega + l}{2},\frac{\Delta_-\mp\mathrm{i}\omega + l}{2};\frac{D-1}{2} + l;\frac{r^2}{\ell_{\mathrm{dS}}^2}}}
\label{Apmw}
\end{equation}
with the vector weights $\Delta_{\pm}$ given by
\begin{equation}
\Delta_{\pm} 
= \frac{D+1}{2} \pm \sqrt{\inp{\frac{D-3}{2}}^2 - m_{\mathrm{v}}^2\ellds^2}
\end{equation}
or 
\begin{equation}
\boxed{\Delta_{\pm} = \frac{D+1}{2} \pm \sqrt{\inp{\frac{D+1}{2}}^2-\mu_{\mathrm{v}}^2\ell_{\mathrm{dS}}^2}}
\end{equation}
Note these weights coincide with those of a scalar field of squared mass $\mu_{\mathrm{v}}^2$ in spacetime dimension $D + 2$. 
The sphere-longitudinal part of $\mathes{A}_{\mu}$ is then fixed by the Lorenz constaint \eqref{LCSplitApp} to be given by
\begin{equation}
	D_A\mathes{A}(\mathsf{x}) = \sum_{l\geq 1,m}\intf{0}{\infty}\frac{\mathrm{d}\omega}{2\pi}\,\frac{1}{\sqrt{2\omega}}\inp{\mathes{A}_{\omega l}(r)\,D_AY_{lm}\,e^{-\mathrm{i}\omega t/\ellds} + \mathrm{c.c.}}
	\label{DAA}
\end{equation}
with
\begin{equation}
	\boxed{\mathes{A}_{\omega l} = \frac{r^2}{2l\inp{l + D -3 }}\bigg[-\partial_t\inp{\upalpha^+_{\omega l m}A^{\omega l}_+ + \upalpha^-_{\omega l m}A^{\omega l}_-} + \frac{\ellds}{r^{D-2}}\,\partial_r\insb{r^{D-3}f(r)\inp{\upalpha^+_{\omega l m}A^{\omega l}_+ - \upalpha^-_{\omega l m}A^{\omega l}_-}}\bigg]}
	\label{LCpmApp}
\end{equation}
Note that the expansion \eqref{DAA} is completely determined by the expansion \eqref{ApmDecompApp} of the nonspherical part $\mathsf{A}_a$, which is why the normal modes in \eqref{DAA} are not premultiplied by independent expansion coefficients.

As we approach the horizon, these normal modes decompose into sums of ingoing and outgoing waves
\begin{equation}
	A_{\pm}^{\omega l}(r)\,e^{-\mathrm{i}\omega t/\ellds} \ \underset{r \to \ellds}{\sim} \ \inp{\mathsf{T}(\omega)\,e^{-\mathrm{i}\omega \inp{t-r_*}/\ellds} + \mathsf{R}(\omega)\,e^{-\mathrm{i}\omega \inp{t + r_*}/\ellds}}
\end{equation} 
leading to a scattering phase
\begin{equation}
	\mathsf{S}(\omega) \equiv \frac{\mathsf{T}(\omega)}{\mathsf{R}(\omega)} = \frac{\Gamma\big(\frac{\Delta_+ + l - \mathrm{i}\omega}{2}\big)\Gamma\big(\frac{\Delta_- + l - \mathrm{i}\omega}{2}\big)}{\Gamma\big(\frac{\Delta_+ + l + \mathrm{i}\omega}{2}\big)\Gamma\big(\frac{\Delta_- + l + \mathrm{i}\omega}{2}\big)}\cdot\frac{\Gamma(+\mathrm{i}\omega)}{\Gamma(-\mathrm{i}\omega)}
	\label{Sw}
\end{equation}
Note that we get the same asymptotics---and therefore the same scattering phase---regardless of whether we consider $\mathsf{A}_+$ or $\mathsf{A}_-$. There are again two towers of quasinormal frequencies\footnote{For the sphere-longitudinal constrained normal modes \eqref{LCpmApp}, we have (for $l \geq 1$)
	\begin{multline}
		\frac{1}{\ellds^2}\,\mathes{A}_{\omega l}(r)\,e^{-\mathrm{i}\omega t/\ellds}
		\ \underset{r \to \ellds}{\sim} \ \insb{\frac{\mathrm{i}\omega\inp{\upalpha^+_{\omega lm} + \upalpha^-_{\omega lm}}-\inp{\upalpha^+_{\omega lm} - \upalpha^-_{\omega lm}}}{l\inp{l + D -3 }}}\inp{\mathsf{T}(\omega)\,e^{-\mathrm{i}\omega \inp{t-r_*}/\ellds} + \mathsf{R}(\omega)\,e^{-\mathrm{i}\omega \inp{t + r_*}/\ellds}}
	\end{multline}
The overall multiplicative factor cancels out of the scattering phase, which is the same as that for $A_{\pm}^{\omega l}\,e^{-\mathrm{i}\omega t/\ellds}$, leading as expected to the same quasinormal frequency spectrum.
} 
\begin{equation}
	\boxed{\upomega_{\pm,l,n} = -\mathrm{i}\inp{\Delta_{\pm} + l + 2n}, \qquad l,n \in \Z_{\geq 0}}
	\label{QNMns}
\end{equation}
where the ``$\pm$" in \eqref{QNMns} refers to $\Delta_+$ and $\Delta_-$ (as opposed to $\mathsf{A}_+$ and $\mathsf{A}_-$, which correspond to the same quasinormal frequency spectrum). The lowest quasinormal mode is of course the $s$-wave mode which we have spent much of this paper considering, with quasinormal frequency $-\mathrm{i}\Delta_-$ controlled by the effective physical mass $\mu_{\mathrm{v}}$.

For the $l = 0$, the Lorenz constraint \eqref{LCApp} expresses the non-independence of the $s$-wave parts of $\mathsf{A}_+$ and $\mathsf{A}_-$. In terms of the expansions \eqref{ApmDecompApp} we can analyze the near-pode (i.e. $r \to 0$) behavior of the Lorenz constraint, which gives 
\begin{equation}
\begin{cases}
\dfrac{\inp{D-3}\ell_{\mathrm{dS}}^2}{r^2}\inp{\upalpha^{+}_{\omega 00}-\upalpha^{-}_{\omega 00}}  \ \underset{r \to 0}{\sim} \ 0 & D \neq 3 \vspace{1em}\\
\dfrac{\ell_{\mathrm{dS}}}{r}\inp{\upalpha^{+}_{\omega 00}-\upalpha^{-}_{\omega 00}} \ \underset{r \to 0}{\sim} \ 0 & D = 3 
\end{cases}
\end{equation}
In order to solve the constraint, we must have that
\begin{equation}
\upalpha_{\omega} \equiv \upalpha^{+}_{\omega 00} = \upalpha^{-}_{\omega 00}
\end{equation}
In other words, we find that in order to solve the full equation of motion, including the Lorenz constraint \eqref{LCApp}, \eqref{ApmDecompApp} should really read 
\begin{multline}
\mathsf{A}_{\pm}(\mathsf{x})
=
\intf{0}{\infty}\frac{\mathrm{d}\omega}{2\pi}\,\frac{1}{\sqrt{2\omega}}\inp{\upalpha_{\omega}\,A_{\pm}^{\omega}(r)\,e^{-\mathrm{i}\omega t/\ell_{\mathrm{dS}}} + \mathrm{c.c.}}\\
+ \sum_{l\geq 1, m}\intf{0}{\infty}\frac{\mathrm{d}\omega}{2\pi}\,\frac{1}{\sqrt{2\omega}}\inp{\upalpha^{\pm}_{\omega lm}\,A_{\pm}^{\omega l}(r)\,Y_{lm}(\theta)\,e^{-\mathrm{i}\omega t/\ell_{\mathrm{dS}}} + \mathrm{c.c.}}
\label{Amode}
\end{multline}

\subsubsection*{Brief Aside: Explicit Solution For The $s$-Wave Mode}
\quad \
\eqref{Amode} tells us that the general $s$-wave solution $A_a$ of the full equation of motion \eqref{ProcaApp} in the static patch can be expanded as 
\begin{equation}
A_{a}(x) = \intf{0}{+\infty}\frac{\mathrm{d}\omega}{2\pi}\,\frac{1}{\sqrt{2\omega}}\inp{\upalpha_{\omega}\,A_{a}^{\omega}(r)\,e^{-\mathrm{i}\omega t/\ell_{\mathrm{dS}}} + \mathrm{c.c.}}
\label{Amode}
\end{equation}
with
\begin{align}
A_{t}^{\omega} = \frac{f(r)}{2}\inp{A_{+}^{\omega} + A_{-}^{\omega}},\qquad
A_{r}^{\omega} = \frac{\ell_{\mathrm{dS}}}{2r}\inp{A_{+}^{\omega} - A_{-}^{\omega}}
\end{align}
Explicitly,
\begin{multline}
A_{t}^{\omega}(r) \ \propto \ \frac{f(r)}{2}\Bigg[f(r)^{+\mathrm{i}\omega/2}\,{}_2F_1\inp{\frac{\Delta_++\mathrm{i}\omega}{2};\frac{\Delta_-+\mathrm{i}\omega}{2};\frac{D-1}{2};\frac{r^2}{\ellds^2}}\\
+ f(r)^{-\mathrm{i}\omega/2}\,{}_2F_1\inp{\frac{\Delta_+-\mathrm{i}\omega}{2};\frac{\Delta_--\mathrm{i}\omega}{2};\frac{D-1}{2};\frac{r^2}{\ellds^2}}\Bigg]
\label{AtModesApp}
\end{multline}
and
\begin{multline}
A_{r}^{\omega}(r) \ \propto \ \frac{\ellds}{2r}\Bigg[f(r)^{+\mathrm{i}\omega/2}\,{}_2F_1\inp{\frac{\Delta_++\mathrm{i}\omega}{2};\frac{\Delta_-+\mathrm{i}\omega}{2};\frac{D-1}{2};\frac{r^2}{\ellds^2}}\\
- f(r)^{-\mathrm{i}\omega/2}\,{}_2F_1\inp{\frac{\Delta_+-\mathrm{i}\omega}{2};\frac{\Delta_--\mathrm{i}\omega}{2};\frac{D-1}{2};\frac{r^2}{\ellds^2}}\Bigg]
\label{ArModesApp}
\end{multline}
Note that setting $\mu_{\mathrm{v}}^2 = 0$ and taking the limit $\omega \to 0$ recovers the static solution \eqref{staticD} found in \S\ref{Static} above.
\subsubsection*{End Aside}

\subsubsection{Sphere-Transverse Modes}
\quad \
Finally let us solve for the sphere-transverse (i.e. sphere-divergence-free) part $\tilde{\mathsf{A}}_A$, which we can similarly expand as 
\begin{equation}
	\tilde{\mathsf{A}}_A = \sum_{j\geq 1, m}\intf{0}{\infty}\frac{\mathrm{d}\omega}{2\pi}\frac{1}{\sqrt{2\omega}}\inp{\tilde{\mathsf{a}}_{\omega j m}\,\tilde{A}_{\omega j}(r)\,\tilde{\mathsf{Y}}^{jm}_A\,e^{-\mathrm{i}\omega t/\ellds} + \mathrm{c.c.}}
\end{equation}
$\tilde{\mathsf{a}}_{\omega jm}$ are again dimensionless constants expressing the relative contribution of each normal mode in the above expansion and $\tilde{\mathsf{Y}}^{jm}_A$ are as before the ``spin-1" transverse vector harmonics on $\mathbb{S}^{(D-2)}$, which as a reminder satisfy 
\begin{equation}
	-\Omega^{BC}D_BD_C\tilde{\mathsf{Y}}^{jm}_A = \insb{j\inp{j + D - 3} - 1}\tilde{\mathsf{Y}}^{jm}_A
\end{equation}
(see Appendix \ref{SHApp} for details). Note that the sphere-transverse mode $\tilde{\mathsf{A}}_A$ is not involved in---and hence is not constrained by---the Lorenz constraint \eqref{LCSplitApp}, and does not couple to $A_a$ or $D_A\mathes{A}$. 
The equation of motion \eqref{ProcaApp} (specifically the component \eqref{PA}) implies that the radial functions $\tilde{A}_{\omega j}(r)$ satisfy the radial equation\footnote{Here we have used that
\begin{equation}
\insb{j\inp{j + D - 3} -1} + \inp{D-3} = (j + 1)\inp{j + 1 + D - 5}
\end{equation}} 
\begin{equation}
\insb{-\frac{1}{f(r)}\frac{\omega^2}{\ellds^2} -f(r)\,\partial_r^2 - \inp{\frac{D-4}{r} - \frac{\inp{D-2}r}{\ellds^2}}\partial_r +  \frac{(j + 1)\inp{j + 1 + D - 5}}{r^2} + m_{\mathrm{v}}^2}\tilde{A}_{\omega j}(r) = 0
\end{equation}
Comparing with \eqref{radialScalar}, we see that this is the radial equation for the angular momentum $l = j + 1$ mode of a scalar field of squared mass $m_{\mathrm{v}}^2$ in $(D -2)$ spacetime dimensions. We therefore find that 
\begin{equation}
\tilde{A}_{\omega j}(r) \ \propto \ \inp{\frac{r}{\ell_{\mathrm{dS}}}}^{j + 1}f(r)^{+\mathrm{i}\omega/2}\, {}_2F_1\inp{\frac{\tilde{\delta}_+ + \mathrm{i}\omega + \inp{j + 1}}{2},\frac{\tilde{\delta}_- + \mathrm{i}\omega + \inp{j+1}}{2};\frac{D-3}{2} + \inp{j + 1}; \frac{r^2}{\ell_{\mathrm{dS}}^2}}
\end{equation} 
with weights $\tilde{\delta}_{\pm}$ given by
\begin{equation}
\tilde{\delta}_{\pm} = \frac{D-3}{2} \pm \sqrt{\inp{\frac{D-3}{2}}^2 -m_{\mathrm{v}}^2\ell_{\mathrm{dS}}^2}
\end{equation}
Using our previous analysis, we see that the quasinormal spectrum will be given by two towers 
\begin{equation}
	\omega_{\pm,j,n} = -\mathrm{i}\inp{\tilde{\delta}_{\pm} + \inp{j+1} + 2n}, \qquad j \in \Z_{\geq 1}, \quad n \in \Z_{\geq 0}
\end{equation}
These two towers can be equicalently parameterized as 
\begin{equation}
	\omega_{\pm,j,n} = -\mathrm{i}\inp{\Delta_{\pm} + \inp{j-1} + 2n}, \qquad \inp{j-1}, n \in \Z_{\geq 0}
\end{equation}
where we have used that 
\begin{align}
	\tilde{\delta}_{\pm} + \inp{j + 1}
	&=\frac{D-3}{2} \pm \sqrt{\inp{\frac{D-3}{2}}^2 -m_{\mathrm{v}}^2\ell_{\mathrm{dS}}^2} + \inp{j + 1}\\[0.5em]
	&=\frac{D+1}{2} \pm \sqrt{\inp{\frac{D+1}{2}}^2 - \mu_{\mathrm{v}}^2\ell_{\mathrm{dS}}^2} + \inp{j - 1}[0.5em]\\
	&= \Delta_{\pm} + \inp{j-1}
\end{align}
The lowest quasinormal mode will therefore be given by $-\mathrm{i}\Delta_-$ which goes to zero as $\mu_{\mathrm{v}}^2\ellds^2 \to 0$.

\section{Canonical Quantization of the $s$-Wave Vector in dS$_3$}
\quad \
We would like to check our conjecture that the na\"ively tachyonic mass range $-2\inp{D-1} < m_{\mathrm{v}}^2\ellds^2 < 0$ remains quantum mechanically well-defined, at least for the $s$-wave mode in the computable case of $D = 3$. In this special case, our setup is simple enough that we may carry out a straightforward canonical quantization of the physical phase space (at least within the static patch). As explained in \S\ref{Prelim2}, on the physical phase space of the $s$-wave sector of the dS$_D$-Proca theory, we may take our physical degrees of freedom to be the modes of electric field $E$, along with the modes of its canonical conjugate 
\begin{equation}
	\boldsymbol{\pi}_E = -\frac{2\pi}{m_{\mathrm{v}}^2}\frac{r}{f(r)}\,\partial_tE
\end{equation}
which we see is well-defined for $m_{\mathrm{v}}^2 \neq 0$. In terms of the electric field $E$, the equation of motion \eqref{Proca} reduces to the single wave equation
\begin{equation}
	\insb{\frac{1}{f(r)}\,\partial_t^2 - f(r)\,\partial_r^2 - \inp{\frac{1}{r} - \frac{3r}{\ell_{\mathrm{dS}}^2}}\partial_r + \frac{1}{r^2} + \inp{m_{\mathrm{v}}^2 + \ellds^{-2}}}E = 0
	\label{EOME}
\end{equation} 
Which is of course the same as the equation of motion for the $l = 1$ mode of a minimally-coupled scalar field of squared mass $m_{\mathrm{v}}^2$. Using the results of the previous section, we see that the general solution to \eqref{EOME} satisfying the boundary condition \eqref{DirichletE} is given by
\begin{equation}
	E_{\omega}(r) = \frac{\mathcal{N}_{\omega}}{\ell_{\mathrm{dS}}^{3/2}}\,\inp{\frac{r}{\ellds}}f(r)^{+\mathrm{i}\omega/2} {}_2F_1\inp{\frac{\Delta_+ + \mathrm{i}\omega}{2},\frac{\Delta_- + \mathrm{i}\omega}{2}; 2; \frac{r^2}{\ell_{\mathrm{dS}}^2}}
	\label{EModes}
\end{equation} 
The normalization constant
\begin{equation}
	\mathcal{N}_{\omega} (\mu_{\mathrm{v}}^2)
	= \left\|\frac{\Gamma\big(\frac{\Delta_+ + \mathrm{i}\omega}{2})\,\Gamma(\frac{\Delta_- + \mathrm{i}\omega}{2})}{\Gamma(\mathrm{i}\omega)}\right\|
	\label{CScalar}
\end{equation}
is chosen so that the scalar-normalized normal modes\footnote{Recall that scalar fields and spacetime components of vector fields in $D = 3$ spacetime dimensions carry dimensions of $(\mathrm{length})^{-1/2}$. Meanwhile the electric field---which is given by a derivative of the vector field---has dimensions of $(\mathrm{length})^{-3/2}$.}
$$\frac{\ellds}{2\pi}\,\frac{1}{\sqrt{2\omega}}\,E_{\omega}(r)\,e^{-\mathrm{i}\omega t/\ell_{\mathrm{dS}}} $$
are orthonormal with respect to the Klein-Gordon inner product
\begin{equation}
	\frac{(\phi_1,\phi_2)_{\mathrm{KG}}}{2\pi} \equiv -\mathrm{i}\intf{0}{\ellds}\frac{\mathrm{d}r\,r}{f(r)}\inp{\phi_1^{}\partial_{t}\phi_2^* - \phi_2^*\partial_{t}\phi_1^{}}
	\label{KGN}
\end{equation}
In terms of the radial functions, the normalization condition reads
\begin{equation}
	\intf{0}{\ellds}\frac{r\,\mathrm{d}r}{f(r)}\,E^{}_{\omega}(r)\,E_{\omega'}^*(r) = \frac{2\pi\delta(\omega-\omega')}{\ellds}
	\label{KG}
\end{equation}
Note that our modes are real and symmetric under\footnote{This follows from the identities
	\begin{equation}
		\Gamma^*(z) = \Gamma(z^*)
	\end{equation}
	\begin{equation}
		{}_2F_1\inp{\theta,\beta;\gamma;z} = 	\inp{1-z}^{\gamma-\inp{\beta + \theta}}{}_2F_1\inp{\gamma-\theta,\gamma-\beta;\gamma;z} 
	\end{equation}
	and 
	\begin{equation}
		{}_2F_1\inp{a,b;c,z} = {}_2F_1\inp{b,a;c;z}
	\end{equation}
} $\omega \to -\omega$:
\begin{equation}
	E_{(-\omega)} = E_{\omega}^* = E_{\omega}
\end{equation}

Let us now quantize $E$ by promoting the expansion coefficients $\upalpha_{\omega}$ ($\omega > 0$) to operators on a Fock space $\mathes{F}$. The electric field then acts on this Fock space as 
\begin{equation}
	\boxed{\hat{E}(x) = \intf{0}{\infty}\frac{\mathrm{d}\omega}{2\pi}\frac{1}{\sqrt{2\omega}}\inp{\hat{\upalpha}_{\omega}\,E_{\omega}(r)\,e^{-\mathrm{i}\omega t/\ell_{\mathrm{dS}}} + \mathrm{h.c.}}}
	\label{Emode}
\end{equation}
where $\mathrm{h.c.}$ now denotes the Hermitian conjugate. Demanding that the canonical commutation relation 
\begin{equation}
	[\,\hat{E}(t,r)\,,\,\partial_t\hat{E}(t,r')\,] = -\frac{m_{\mathrm{v}}^2}{2\pi}\frac{f(r)}{r}\,\mathrm{i}\,\delta(r-r')
\end{equation}
hold on the Fock space requires that the annihilation and creation operators $\hat{\upalpha}_{\omega}$ and $\hat{\upalpha}^{\dagger}_{\omega}$ obey
\begin{equation}
	[\,\upalpha_{\omega}^{}\,,\,\upalpha_{\omega'}^{\dagger}\,] = -m_{\mathrm{v}}^2\ellds^2\,\delta(\omega-\omega')
\end{equation}
or 
\begin{equation}
	\boxed{[\,\upalpha_{\omega}^{}\,,\,\upalpha_{\omega'}^{\dagger}\,] =  (4-\mu_{\mathrm{v}}^2\ellds^2)\,\delta(\omega-\omega')}
\end{equation}
As usual, the Fock space is built about the ``Boulware" vacuum state $|\,0\,\rangle$ which is annihilated by all of the positive frequency annihilation operators 
\begin{equation}
	\hat{\upalpha}_{\omega}\,|\,0\,\rangle = 0, \qquad \omega > 0
\end{equation}
i.e. we have that
\begin{equation}
	\mathes{F} = \mathrm{Span}\left\{\hat{\upalpha}^{\dagger}_{\omega_1}\dots\hat{\upalpha}^{\dagger}_{\omega_n}|\,0\,\rangle\,\big|\,\omega_i > 0, n \geq 0\right\}
\end{equation}
The usual Bunch-Davies-Hartle-Hawking/``Euclidean vacuum" state is then represented as the thermal state on this Hilbert space at inverse temperature $2\pi\ellds$ (see e.g. \cite{Grewal:2024emf}). Note that our quantization is most sensible when $m_{\mathrm{v}}^2\ellds^2 < 0$ (in which case the Hamiltonian \eqref{HEr} is bounded from below and the annhilation and creation operators can be canonically normalized by a rescaling by a positive real factor). 
		\begin{center}
			\emph{We see that our quantization is good for $-2\inp{D-1} < m_{\mathrm{v}}^2\ellds^2 < 0$.}
		\end{center}
The upper bound comes from the fact that our construction explicitly breaks down at the gauge-symmetric point $m_{\mathrm{v}}^2 = 0$, whereas the lower bound comes from the fact that causal correlators asymptotically grow with time below the edge of stability (indicating an inconsistency with a unitary quantization).

\subsection{The Edge of Stability as an IR Scale} 
\quad \ 
We will now see one final role for the edge of stability. To this end, consider a timelike two-point function of the electric field mode in the Boulware vacuum
\begin{align}
	\langle\,0\,|\,\hat{E}(t,r)\,\hat{E}(0,r)\,|\,0\,\rangle
	&= \frac{(4-\mu_{\mathrm{v}}^2\ellds^2)}{4\pi}\intf{0}{\infty}\frac{\mathrm{d}\omega}{2\omega}\,\big(E_{\omega}(r)\big)^2
	\label{Q2Int}\\
	&= \frac{(4-\mu_{\mathrm{v}}^2\ellds^2)}{4\pi\ellds^3}\intf{0}{\infty}\frac{\mathrm{d}\omega}{2\omega}\,\left(\,\mathcal{N}_{\omega}\,f(r)^{+\mathrm{i}\omega/2}\,{}_2F_1\inp{\frac{\Delta_+ + \mathrm{i}\omega}{2},\frac{\Delta_- + \mathrm{i}\omega}{2}; 2; \frac{r^2}{\ell_{\mathrm{dS}}^2}}\right)^2
\end{align}
Let us study this correlator in the near-horizon limit. We find that
\begin{equation}
	\langle\,0\,|\,\hat{E}(t,r)\,\hat{E}(0,r)\,|\,0\,\rangle
	\ \underset{r \to \ellds}{\sim} \  \frac{(4-\mu_{\mathrm{v}}^2\ellds^2)}{4\pi\ellds^3}\,\intf{0}{\infty}\frac{\mathrm{d}\omega}{2\omega}\,\Big(\,T(\omega)\,f(r)^{-\mathrm{i}\omega/2} + \mathrm{c.c.}\Big)^2
	\label{Q2SC}
\end{equation}
where we have defined the horizon transmission phase
\begin{equation}
	T(\omega) \equiv \left\|\frac{\Gamma\big(\frac{\Delta_++ \mathrm{i}\omega}{2})\,\Gamma(\frac{\Delta_- + \mathrm{i}\omega}{2})}{\Gamma(\mathrm{i}\omega)}\right\|\,\frac{\Gamma(\mathrm{i}\omega)}{\Gamma\big(\frac{\Delta_++ \mathrm{i}\omega}{2}\big)\Gamma\big(\frac{\Delta_- + \mathrm{i}\omega}{2}\big)}
\end{equation}
Away from the edge of stability, 
we have that (ignoring for the moment overall numerical constants)
\begin{equation}
	\mathrm{integrand} \ \underset{\omega \to 0}{\sim} \ \frac{\mathrm{d}\omega}{\omega}\,\omega^2\times\big(1 + O(\omega)\big)
\end{equation}
which goes to zero sufficiently rapidly as $\omega \to 0$ for the integral to converge. Right at the edge of stability, this is no longer the case: If we first take $\mu_{\mathrm{v}}^2\ellds^2 \to 0$ \emph{before} evaluating the integral, we find an emergent ``infrared" divergence:
\begin{equation}
	\mathrm{integrand}\big|_{\mathrm{edge}} \ \underset{\omega \to 0}{\sim} \ \frac{16}{\ellds}\,\frac{\mathrm{d}\omega}{\omega} + \frac{4}{\ellds}\inp{1+\ln\big(f(r)\big)}^2\,\omega\,\mathrm{d}\omega
\end{equation}
In other words we see that \emph{the edge of stability and the IR limit do not commute}. We have
\begin{equation}
	\Big(\mathrm{integrand}\big|_{\mathrm{IR}}\Big)\Big|_{\mathrm{edge}} = \frac{16}{\ellds}\frac{1}{\Delta_-^2}\,\omega\,\mathrm{d}\omega -\frac{16}{\ellds}\frac{1}{\Delta_-}\Big(1 + \ln\big(f(r)\big)\Big)\,\omega\,\mathrm{d}\omega + \frac{4}{\ellds}\Big(1 + \ln\big(f(r)\big)\Big)^2\,\omega\,\mathrm{d}\omega
\end{equation}
and 
\begin{equation}
	\Big(\mathrm{integrand}\big|_{\mathrm{edge}}\Big)\Big|_{\mathrm{IR}} = \frac{16}{\ellds}\,\frac{\mathrm{d}\omega}{\omega} + \frac{4}{\ellds}\Big(1+\ln\big(f(r)\big)\Big)^2\,\omega\,\mathrm{d}\omega
\end{equation}
Demanding that these actually be equal requires that the integral away from the edge of stability have an effective IR cutoff $\omega_{\mathrm{IR}}$ given by
\begin{equation}
	\frac{1}{\omega_{\mathrm{IR}}^2} = \frac{1}{\Delta_-^2}\times\Big(1 + O\big(\Delta_-\big)\Big)
\end{equation}
i.e.
\begin{equation}
	\omega_{\mathrm{IR}} = \Delta_-\times\Big(1 + O\big(\Delta_-\big)\Big)
\end{equation}
or 
\begin{equation}
	\boxed{\omega_{\mathrm{IR}} = \frac{\mu_{\mathrm{v}}^2\ellds^2}{4}\times\Big(1 + O\big(\mu_{\mathrm{v}}^2\ellds^2\big)\Big)}
\end{equation}
So we find the natural condition that the physical mass $\mu_{\mathrm{v}}$ acts as an IR cutoff for the theory. This is analogous to finite mass regulation of the the well-known ``infrared" divergences of massless scalar correlation functions in de Sitter space (see e.g. \cite{Allen:1985ux,Allen:1987tz,Grewal:2024emf,Gorbenko:2018oov}). Just as in that case, the emergent IR divergence here is related to the emergent shift symmetry, which renders the value of the correlation functions ambiguous unless the zero mode is fixed by hand. 

\section{Conclusions and Future Directions}
\quad \ 
In this paper we have argued that the massive minimally-coupled ``spin-1" vector Boson in de Sitter space, when studied relative to a fixed static patch frame, is controlled by an effective physical mass $\mu_{\mathrm{v}}$ which differs from the na\"ive ``Lagrangian mass" $m_{\mathrm{v}}$ appearing in the usual form of the Proca Lagrangian/action. The relationship between the two is given by $$\mu_{\mathrm{v}}^2\ellds^2 = m_{\mathrm{v}}^2\ellds^2 + 2\inp{D-1}.$$ 
We identified a concept which we dubbed the ``edge of stability", which is essentially the concept which replaces the familiar $\mathrm{SO}(D,1)$ ``Higuchi bound" when the symmetries of the problem are reduced to those of a static patch $\mathrm{O}(1,1)\times\mathrm{O}(D-1)$. We argued that the edge of stability for our vector Boson theory was defined by the na\"ively tachyonic mass value $m_{\mathrm{v}}^2\ellds^2 = -2\inp{D-1}$, at which several interesting features emerge. Among these features are the emergence of static solutions, zero modes, global shift symmetries, and ``infrared" divergences. We also derived the classical solutions and quasinormal frequency spectrum of our theory in the static patch, generalizing previous work of Higuchi \cite{Higuchi:1986ww} and overlapping with upcoming work by Grewal, Law, and Lochab \cite{Albert}.

An obvious follow up to the present work would be to understand whether the quantum-mechanical well-definedness of our theory persists beyond narrow scope of the $s$-wave mode in $D = 3$ spacetime dimensions. This direction is currently being pursued by one of us (A.R.). Another follow up would be to return to the original context in which we began to discover these features, namely the conjectured duality between the high-temperature double-scaled SYK model (DSSYK$_{\infty}$) and dimensionally-reduced $(2 + 1)$-dimensional de Sitter space. In that context, we were studying the possibility of a concrete duality between the charge operator of the charged (i.e. $\mathrm{U}(1)$-symmetric) version of DSSYK$_{\infty}$ and the $s$-wave mode of a massive minimally-coupled vector at the edge of stability. The goal would be to find a match between the two operators which persists even when the U(1) symmetry of the quantum theory is slightly broken (rendering the charge operator dynamical) and the mass of the bulk field is correspondingly pushed slightly away from the edge of stability. This direction is currently being pursued by the two of us and Y. Sekino. 

\section*{Acknowledgements}

\quad \ We would like to thank Jonathan Maltz, Juan Maldacena, Yasuhiro Sekino, and especially Y.T. Albert Law for helpful discussions. A.R. and L.S. are supported in part by NSF
Grant PHY-1720397 and by the Stanford Institute of
Theoretical Physics. A.R. is additionally supported by a Mellam Family Foundation fellowship.

\appendix

\section{Scalar and Vector Spherical Harmonics} 
\label{SHApp}
\quad \ 
In this appendix, we recall the definition and properties of the scalar and vector spherical harmonics on the round unit $n$-sphere $\mathbb{S}^n$. 

\subsection{Scalar Spherical Harmonics}
\subsubsection{Scalar Spherical Harmonics for $n \geq 2$}
\quad \ 
Let $(\mathbb{S}^{n},\Omega_{AB})$ be the round unit $n$-sphere, which, by abuse of terminology, we will simply refer to as $\mathbb{S}^n$. Let us begin by fixing $n \geq 2$, deferring the exceptional case of $n = 1$ to the next subsubsection. The scalar spherical harmonics are easiest to introduce in the context of the Hilbert space $L^2_{(0)}(\mathbb{S}^n)$ of square-integrable zero-forms (functions) on $\mathbb{S}^n$: 
\begin{equation}
	L^2_{(0)}(\mathbb{S}^n) = \left\{\,f: \mathbb{S}^n \to \C \ \big|\ \langle\,f\,,\,f\,\rangle_{L^2}  < \infty\,\right\}
\end{equation}
Here we have defined the $L^2$ inner product 
\begin{equation}
	\langle\,g\,,\,f\,\rangle_{L^2} = \frac{1}{\Omega_n}\int_{\mathbb{S}^n}\mathrm{d}^n\theta\,\sqrt{\Omega}\ g^*f
	\label{L2sApp}
\end{equation}
where $\Omega_n \equiv \int_{\mathbb{S}^n}\mathrm{d}^n\theta\,\sqrt{\Omega}$ denotes the volume of $\mathbb{S}^n$. $L^2_{(0)}(\mathbb{S}^n)$ is the physically relevant function space for studying scalar field theory on a spacetime with a (possibly warped) $\mathbb{S}^n$ factor\footnote{If a scalar field $\upphi$ is not square integrable on the $\mathbb{S}^n$ factor, its action will be infinite and its contribution to the path integral vanishing.}.

Let $D_A$ denote the covariant derivative operator (Levi-Civita connection) of $\mathbb{S}^n$ and let $D^2_{(p)}$ denote the corresponding Laplace-Beltrami operator $\Omega^{AB}D_AD_B$ acting on $p$-forms.  The ``scalar Laplacian" $-D^2_{(0)}$ is essentially self-adjoint on $L^2_{(0)}(\mathbb{S}^n)$ and its eigenfunctions---the spherical harmonics---span a dense subspace\footnote{The spectrum of $-D^2_{(0)}$ is ``pure point", so that it has as associated orthonormal eigenbasis.}. The eigenvalues of $-D^2_{(0)}$ can be labelled by the ``orbital angular momentum" quantum number $l \geq 0$; at orbital angular momentum $l \geq 0$, $-D^2_{(0)}$ acts on the corresponding eigenspace as multiplication by the nonnegative eigenvalue 
\begin{equation}
	l\inp{l + n - 1} \geq 0
	\label{lambdalApp}
\end{equation}
with multiplicity/degeneracy (see e.g. \cite{Grewal:2024emf})
\begin{equation}
	d^{(0)}_l \equiv 
	\frac{\inp{2l + n - 1}}{\inp{n-1}}\binom{l + n - 2}{n-2}
\end{equation}
For any fixed angular momentum $l \geq 0$, the ``spherical harmonics" $Y_{lm}$ ($1 \leq m \leq d^{(0)}_l$) furnish an orthonormal basis of the corresponding eigenspace of $-D^2_{(0)}$: 
\begin{equation}
	\underbrace{-\frac{1}{\sqrt{\Omega}}\,\partial_A\inp{\sqrt{\Omega}\,\Omega^{AB}\partial_B}}_{-D^2_{(0)}}Y_{lm} = l\inp{l + n-1}Y_{lm}, \qquad 1 \leq m \leq d^{(0)}_l
	\label{YApp}
\end{equation}
We define these to be orthonormal with respect to \eqref{L2sApp}\footnote{The more common normalization convention omits the factor of $1/\Omega_{n}$ from the left hand side of \eqref{NCApp} so that the lowest spherical harmonic is given by the dimension-dependent expression $\Omega_{n}^{-1/2}$. In this paper we have normalized our spherical harmonics so that $Y_{00} = 1$ independent of dimension.} 
\begin{equation}
	\langle\,Y_{lm}\,,\,Y_{l'm'}\,\rangle_{L^2} = \frac{1}{\Omega_n}\int_{\mathbb{S}^{n}}\mathrm{d}^n\theta\,\sqrt{\Omega}\,Y^*_{lm}\,Y^{}_{l'm'} = \delta_{ll'}\delta_{mm'}
	\label{NCApp}
\end{equation}
As mentioned above, the direct sum of these eigenspaces---i.e. the span of the $Y_{ln}$'s for all $l \geq 0$---is dense in $L^2_{(0)}(\mathbb{S}^2)$. The upshot is that any square-integrable function on $\mathbb{S}^{n}$ can be approximated arbitrarily well by a linear combination of the spherical harmonics:
\begin{equation}
	f(\theta) = \sum_{lm}f_{lm}Y_{lm}(\theta)
	\label{fYApp}
\end{equation}
where, with our normalization conventions, 
\begin{equation}
	f_{ln} = \frac{1}{\Omega_{n}}\int_{\mathbb{S}^{n}}\mathrm{d}^n\theta\,\sqrt{\Omega}\,Y_{ln}^*(\theta)f(\theta)
\end{equation}
The completeness of the spherical harmonics is expressed via the ``resolution of the identity"\footnote{\eqref{RIApp} should be understood in the sense of being true up to a subset of $L^2_{(0)}(\mathbb{S}^n)$ of measure zero.}
\begin{align}
	\delta_{\Omega}(\theta;\theta') 
	= \frac{1}{\Omega_{n}}\sum_{ln}Y^*_{ln}(\theta')Y_{ln}(\theta)
	\label{RIApp}
\end{align}
where 
\begin{equation}
	\delta_{\Omega}(\theta;\theta') 
	\equiv \frac{1}{\sqrt{\Omega}}\,\delta(\theta^1-\delta'^1)\dots\delta(\theta^{(D-2)}-\delta'^{(D-2)})
\end{equation}
is the covariant $\delta$-function on $\mathbb{S}^n$. That the identity \eqref{RIApp} makes sense is a reflection of the fact that \eqref{fYApp} can itself be extended to the statement that in fact \emph{any distribution on $\mathbb{S}^n$ can be approximated arbitrarily well by a linear combination of the spherical harmonics}. To see this, note that any test function $f \in C^{\infty}(\mathbb{S}^n) \subset L^2_{(0)}(\mathbb{S}^n)$ can be approximated as in \eqref{fYApp}. By linearity, the action of $F$ on $f$ can therefore be approximated arbitrarily well as
\begin{equation}
	F[f] = \sum_{lm} f_{lm}\,F[Y_{lm}]
\end{equation}
This then means that $F$ can in turn be approximated arbitrarily well bt the distribution which acts via integration against the kernel
\begin{equation}
	F(\theta) \equiv \frac{1}{\Omega_n}\sum_{lm} F[Y_{lm}]\,Y_{lm}^*(\theta)
\end{equation}
since we have that
\begin{align}
	\int_{\mathbb{S}^n}\mathrm{d}^n\theta \sqrt{\Omega}\,F(\theta)f(\theta) 
	&= \sum_{lm}f_{lm}\int_{\mathbb{S}^n}\mathrm{d}^n\theta \sqrt{\Omega}\,F(\theta)\,Y_{lm}(\theta)\\
	&= \frac{1}{\Omega_n}\sum_{lm}\sum_{l'm'}f_{lm}\,F[Y_{l'm'}]\int_{\mathbb{S}^n}\mathrm{d}^n\theta \sqrt{\Omega}\,Y_{l'm'}^*(\theta)\,Y_{lm}(\theta)\\
	&= \sum_{lm} f_{lm}\,F[Y_{lm}]\\
	&= F[f]
\end{align}

\subsubsection{Scalar Spherical Harmonics for $n = 1$}
\quad \
Now let us take $n = 1$. Everything in the preceeding subsection remains true through \eqref{lambdalApp}. The eigenspace at angular momentum $l \geq 0$ now has dimension 
\begin{equation}
	d^{(0)}_0 = 1, \qquad d^{(0)}_{l\geq 1} = 2
\end{equation}
We can pick a real basis of spherical harmonics given by 
\begin{equation}
	Y_{00} = 1, \qquad Y_{(l\geq1)m}(\theta)
	=
	\begin{cases}
		\sqrt{2}\cos(l\theta) & m = 0\\[0.5em]
		\sqrt{2}\sin(l\theta) & m = 1
	\end{cases}
\end{equation}
which obey \eqref{YApp} and \eqref{NCApp} with $n = 1$. The harmonics $Y_{l0}$ are parity even while the harmonics $Y_{l1}$ are parity odd, i.e. 
\begin{equation}
	Y_{lm}(-\theta) = (-1)^mY_{lm}(\theta)
\end{equation}
We also have that 
\begin{equation}
	\partial_{\theta}Y_{lm} = l\,Y_{l(m+1)}
	\label{dtY1}
\end{equation}
where the addition should be interpreted as addition mod 2. 

We once again have the statement that any square-integrable function on $\mathbb{S}^{1}$ can be approximated arbitrarily well by a linear combination of the $Y_{ln}$, which is equivalent to the usual statement that any such function can also be approximated arbitrarily well by a Fourier series 
\begin{equation}
	f(\theta) = \sum_{lm}f_{lm}Y_{lm}(\theta) = \sum_{k \in \Z}f_k\,e^{\mathrm{i}k\theta}
\end{equation}
The relation between these expansions is given by
\begin{equation}
	f_{00} = f_0
\end{equation}
and, for $l\geq 1$,
\begin{equation}
	f_{l0} = \frac{f_{l}+ f_{-l}}{\sqrt{2}}, \qquad f_{l1} = \frac{\mathrm{i}\inp{f_{l}-f_{-l}}}{\sqrt{2}}
\end{equation}
Let us note before moving on that we can use the spherical harmonics to explicitly separate a field into its parity even and parity odd parts via 
\begin{equation}
	f(\theta) \ = \ \overbrace{\sum_{l\geq0}f_{l0}\,Y_{l0}(\theta)}^{\text{parity even}} \quad + \quad \overbrace{\sum_{l\geq1}f_{l1}\,Y_{l1}(\theta)}^{\text{parity odd}}
\end{equation}

\subsection{Vector Spherical Harmonics}
\quad \
Let us now consider the space $L^2_{(1)}(\mathbb{S}^{n})$ of square-integrable one-forms on $\mathbb{S}^{n}$: 
\begin{equation}
	L^2_{(0)}(\mathbb{S}^n) = \left\{\,\upomega \in  \Omega^{1}(\mathbb{S}^n) \ \big|\ \langle\,\upomega\,,\,\upomega\,\rangle_{L^2}  < \infty\,\right\}
\end{equation}
where we have defined the $L^2$ inner product 
\begin{equation}
	\langle\,\upnu\,,\,\upomega\,\rangle_{L^2} = \frac{1}{\Omega_n}\int_{\mathbb{S}^n}\mathrm{d}^n\theta\,\sqrt{\Omega}\ \Omega^{AB}\,\bar{\upnu}_A\,\upomega^{}_B
	\label{L2vApp}
\end{equation}
This is the physically relevant function space for spin-1 vector Boson field theory on a spacetime with a (possibly warped) $\mathbb{S}^n$ factor. This space is the direct sum of the Hilbert space $\tilde{L}^2_{(1)}(\mathbb{S}^n)$ of square-integrable ``transverse" (i.e. divergence-free\footnote{For the mathematician: ``co-closed".}) one forms and the Hilbert space of square-integrable exact one-forms. In other words, any square-integrable one-form on $\mathbb{S}^{n}$ can be written as the sum of a divergence-free one-form on $\mathbb{S}^n$ and the derivative of a scalar field on $\mathbb{S}^{n}$:
\begin{equation}
	\upomega_A = \tilde{\upomega}_A + \partial_Af , \qquad D^A\tilde{\upomega}_A = 0
\end{equation}
In differential forms notation, 
\begin{equation}
	\upomega = \tilde{\upomega} + \mathrm{d}f, \qquad \mathrm{d}\star_{\Omega}\tilde{\upomega} = 0
\end{equation}
where $\star_{\Omega}$ is the Hodge star operator of $\mathbb{S}^n$ taking $p$ forms to $(n-p)$-forms.

Any square-integrable exact one-form on $\mathbb{S}^n$ is the derivative of a square-integrable function\footnote{
	Let $f$ be a function on $\mathbb{S}^n$ such that $\mathrm{d}f$ is square integrable. This means that 
	\begin{equation}
		\|\mathrm{d}f\|^2_{L^2} = \frac{1}{\Omega_n}\int_{\mathbb{S}^{n}}\mathrm{d}^n\theta\,\sqrt{\Omega}\,\Omega^{AB}\partial_Af\,\partial_Bf = \frac{1}{\Omega_n}\int_{\mathbb{S}^{n}}\mathrm{d}^n\theta\,\sqrt{\Omega}\,f\,(-D^2_{(0)})f = \sum_{lm}l\inp{l + n-1}\|f_{lm}\|^2 < \infty
	\end{equation}
	Recalling that 
	\begin{equation}
		\|f\|^2_{L^2} = \sum_{lm}\|f_{lm}\|^2 
	\end{equation}
	we see that we must have 
	\begin{equation}
		\|f\|^2_{L^2} < (f_{00})^2 +  \|\mathrm{d}f\|^2_{L^2} < \infty
	\end{equation}
	i.e. that $f \in L^2_{(0)}(\mathbb{S}^n)$.} on $\mathbb{S}^n$, and hence can be approximated arbitrarily well by a linear combination of the derivatives of the spherical harmonics with $l \geq 1$: 
\begin{equation}
	\partial_Af(\theta) = \sum_{l\geq 1,m} f_{lm}\,\partial_AY_{lm}(\theta)
\end{equation}
We are therefore motivated to define the spin-1 \emph{longitudinal} vector harmonics of angular momentum $j \geq 1$ (which is bounded below by the spin) via 
\begin{equation}
	\mathsf{Y}_{jm} \equiv \frac{1}{\sqrt{j\inp{j + n - 1}}}\,\mathrm{d}Y_{jm}, \qquad j \geq 1
	\label{LVH}
\end{equation}
which are orthonormal with respect to \eqref{L2vApp}. Any square-integrable exact one-form on $\mathbb{S}^n$ can be approximated arbitrarily well by a linear combination of the $\mathsf{Y}_{jm}$. Note that these are \emph{not} transverse, as 
\begin{equation}
	D^A(\mathsf{Y}_{jm})_A = -D^2_{(0)}Y_{jm} = j\inp{j + n}Y_{jm} \neq 0 \quad \text{(since $j \geq 1$)}
\end{equation}
The longitudinal vector harmonics are eigenvectors of the ``vector Laplacian" $-D^2_{(1)}$ with respective eigenvalues 
\begin{equation}
	j\inp{j + n -1 }-1
\end{equation}
i.e.
\begin{equation}
	\underbrace{-\Omega^{BC}D_BD_C}_{-D^2_{(1)}}\mathsf{Y}_{jm} = \insb{j\inp{j + n - 1} - 1}\mathsf{Y}_{jm}
\end{equation}

To analyze the space $\tilde{L}^2_{(1)}(\mathbb{S}^n)$, we will treat the cases $n = 1$, $n = 2$, and $n \geq 3$ separately. 

\subsubsection{Vector Spherical Harmonics for $n = 1$}
\quad \
Let us begin by considering the ``trivial" case $n = 1$. In this case the statement that $\tilde{\omega}$ be a transverse one-form is equivalent to the statement that $\star_{\Omega}\tilde{\omega}$ be a closed $0$-form, i.e. a constant function. This is in turn equivalent to the statement that $\tilde{\omega}$ be a constant multiple of the volume form $\mathrm{d}\theta$. Using \eqref{dtY1}, we have that 
\begin{equation}
	\mathsf{Y}_{jm} = Y_{j(m + 1)}\,\mathrm{d}\theta
\end{equation}
where the addition should be interpreted as addition mod 2. In other words, we have 
\begin{equation}
	\mathsf{Y}_{(j\geq1)m}(\theta)
	=
	\begin{cases}
		\sqrt{2}\sin(l\theta)\,\mathrm{d}\theta & m = 0\\[0.5em]
		\sqrt{2}\cos(l\theta)\,\mathrm{d}\theta & m = 1
	\end{cases}
\end{equation}
Note that just as in the scalar case, the longitudinal vector harmonics $\mathsf{Y}_{j0}$ are parity even while the harmonics $\mathsf{Y}_{j1}$ are parity odd; the transverse vector harmonic $\tilde{\mathsf{Y}}_{00}$ is, however, parity odd. 

We see that any square-integrable one-form on $\mathbb{S}^1$ can be approximated arbitrarily well by a linear combination of the spherical harmonics since the expansion
\begin{equation}
	\upomega(\theta) = \omega_{00}\,\mathrm{d}\theta + \sum_{j\geq 1,m}\omega_{jm}\,\mathsf{Y}_{jm}(\theta)
\end{equation}
can also be written as 
\begin{equation}
	\upomega(\theta) = \sum_{lm}\omega_{l(m + 1)}Y_{lm}(\theta)\,\mathrm{d}\theta
\end{equation}
In other words we have $L^2_{(0)} \simeq L^2_{(1)}$ under the map $f \mapsto f\,\mathrm{d}\theta$. This could have been guessed from the fact that the space of functions on $\mathbb{S}^1$ is isomorphic to the space of one-forms on $\mathbb{S}^1$ (in one dimension there is no difference between a function and a vector field apart from their opposite behavior under parity). We can again split $\omega$ into its parity even and parity odd parts via

\begin{equation}
	\upomega(\theta) \ = \ \overbrace{\sum_{j\geq1}\omega_{l0}\,\mathsf{Y}_{l0}(\theta)}^{\text{parity even}} \quad + \quad \overbrace{\omega_{00}\,\mathrm{d}\theta + \sum_{j\geq1}\omega_{l1}\,\mathsf{Y}_{l1}(\theta)}^{\text{parity odd}}
\end{equation}

\subsubsection{Transverse Vector Spherical Harmonics for $n = 2$}
\quad \ 
Let us now consider the case $n = 2$. In this case the statement that $\tilde{\upomega}$ be a transverse one-form is equivalent to the statement that $\star_{\Omega}\tilde{\upomega}$ be a closed $1$-form. Since $H^1(\mathbb{S}^2) = 0$, this means that $\star_{\Omega}\tilde{\upomega}$ is an exact one-form, i.e. that\footnote{In other words, $\tilde{\upomega}^A$ is a Hamiltonian vector field on $\mathbb{S}^2$ (viewed as a symplectic manifold).} 
\begin{equation}
	\tilde{\upomega}_A = -\epsilon\indices{_A^B}\partial_B\tilde{f}
\end{equation}
for some square-integrable function\footnote{This follows from the fact that $\|\tilde{\upomega}\|^2_{L^2} = \|\mathrm{d}\tilde{f}\|^2_{L^2} < \infty$ implies that $\|\tilde{f}\|^2_{L^2} < (\tilde{f}_{00})^2 + \|\mathrm{d}\tilde{f}\|^2_{L^2} < \infty$.} $\tilde{f}$, where $\epsilon_{AB}$ is the volume form of $\mathbb{S}^2$ and all indices have been raised and lowered using $\Omega_{AB}$. We see that any square-integrable transverse one-form on $\mathbb{S}^2$ can be approximated arbitrarily well by a linear combination of the co-derivatives of the spherical harmonics with $l \geq 1$: 
\begin{equation}
	\tilde{\upomega}_A(\theta) = -\sum_{l\geq 1,m} \tilde{f}_{lm}\,\epsilon\indices{_A^B}\partial_BY_{lm}(\theta)
\end{equation}
We are therefore motivated to define the spin-1 \emph{transverse} vector harmonics of angular momentum $j \geq 1$ (which is again bounded below by the spin) via 
\begin{equation}
	\mathsf{Y}^{(-)}_{jm} \equiv \star_{\Omega}\mathsf{Y}_{jm},  \qquad j \geq 1
\end{equation}
which are orthonormal with respect to \eqref{L2vApp}. Any that any square-integrable transverse one-form on the $\mathbb{S}^2$ can be approximated arbitrarily well by a linear combination of the $\mathsf{Y}^{(-)}_{jm}$. Note that these are manifestly transverse and are again eigenvectors of the vector Laplacian, again with respective eigenvalues $j\inp{j + 1}-1$:
\begin{equation}
	\underbrace{-\Omega^{BC}D_BD_C}_{-D^2_{(1)}}\mathsf{Y}^{(-)}_{jm} = \insb{j\inp{j+1}-1}\mathsf{Y}^{(-)}_{jm}
\end{equation}
We see that the eigenvalues of the vector Laplacian $-D^2_{(1)}$ on $\mathbb{S}^2$ are labeled by the angular momentum $j \geq 1$ with nonnegative eigenvalues
\begin{equation}
	j\inp{j + 1} \geq 2 \quad\text{with multiplicity/degeneracy}\quad 2d_j^{(0)}
\end{equation}
We see that \emph{any} square-integrable one-form on $\mathbb{S}^2$ can be approximated arbitrarily well by a linear combination of the form 
\begin{align}
	\upomega = \sum_{j\geq 1,m}\inp{\omega^{(+)}_{jm} + \omega^{(-)}_{jm}\star_{\Omega}}\mathrm{d}Y_{jm}
\end{align}

\subsubsection{Vector Spherical Harmonics for $n \geq 3$}
\quad \
Let us finally consider the case $n \geq 3$. In this case, appealing to Hodge duality will force us to deal with $(n-1) \geq 2$-forms, and it is more convenient to instead analyze the spectrum of the ``vector Laplacian" $-D^2_{(1)}$ (as we did for the scalar Laplacian $-D^2_{(0)}$ above). Acting on $\tilde{L}^2_{(1)}(\mathbb{S}^n)$, this operator is again essentially self-adjoint and moreover positive; its eigenfunctions---the spin-1 transverse vector harmonics\footnote{These are a special case of the spin-$s$ symmetric transverse traceless tensor harmonics.}---span a dense subspace of $\tilde{L}^2_{(1)}(\mathbb{S}^n)$. The spectrum of $-D^2_{(1)}$ on this space (which is again ``pure point") can be labelled by the angular momentum $j \geq 1$ which is bounded below by the spin; at angular momentum $j$, $-D^2_{(1)}$ acts on the corresponding eigenspace as multiplication by the nonnegative eigenvalue 
\begin{equation}
	j\inp{j + n - 1} - 1 \geq n
\end{equation}
with multiplicity/degeneracy\footnote{Including the longitudinal vector harmonics \eqref{LVH}, the total multiplicity of the eigenvalue $j\inp{j + n - 1}$ on $L^2_{(1)}(\mathbb{S}^n)$ is given by
\begin{equation}
	d^{(1)}_j = d^{(0)}_j + \tilde{d}^{(1)}_j \equiv C^{(1)}_jd^{(0)}_j
\end{equation}
\begin{equation}
	C^{(1)}_j = 1 + \tilde{C}^{(1)}_j = \frac{nj\inp{j + 1} + \inp{n-2}\inp{nj + 1}}{\inp{j+1}\inp{j + n - 2}}
	\end{equation}}
\begin{equation}
	\tilde{d}^{(1)}_j\equiv 
	\tilde{C}^{(1)}_j\,d_j^{(0)}
\end{equation}
where 
\begin{equation}
	\tilde{C}^{(1)}_j = \inp{n-1}\frac{j}{\inp{j + n-2}}\frac{\inp{j + n - 1}}{\inp{j + 1}}
\end{equation}
See e.g. \cite{Higuchi:1986wu} for details. For any fixed angular momentum $j \geq 1$, the ``spin-1 transverse vector harmonics" $\tilde{\mathsf{Y}}^{jm}$ ($1 \leq m \leq \tilde{d}^{(1)}_j$) furnish an orthonormal basis of the corresponding eigenspace (within $\tilde{L}^1_{(1)}(\mathbb{S}^n)$) of $-D^2_{(1)}$: 
\begin{equation}
	\underbrace{-\Omega^{BC}D_BD_C}_{-D^2_{(1)}}\tilde{\mathsf{Y}}^{jm}_A = \insb{j\inp{j + n-1}-1}\tilde{\mathsf{Y}}^{jm}_A, \qquad 1 \leq m \leq d^{(1)}_j
	\label{YApp}
\end{equation}
We define these to be orthonormal with respect to \eqref{L2vApp}
\begin{equation}
	\langle\,\tilde{\mathsf{Y}}^{jm}\,,\,\tilde{\mathsf{Y}}^{j'm'}\,\rangle_{L^2} = \frac{1}{\Omega_n}\int_{\mathbb{S}^{n}}\mathrm{d}^n\theta\,\sqrt{\Omega}\,\Omega^{AB}\tilde{\mathsf{Y}}^{*jm}_A\,\tilde{\mathsf{Y}}^{j'm'}_B = \delta^{jj'}\delta^{mm'}
	\label{NCApp}
\end{equation}

The upshot is that any square-integrable one-form on $\mathbb{S}^{n\geq3}$ can be approximated arbitrarily well by a linear combination of the form 
\begin{equation}
	\upomega = \sum_{j\geq 1,m}\inp{\omega_{jm}\mathsf{Y}_{jm} + \tilde{\omega}^{jm}\tilde{\mathsf{Y}}^{jm}}
\end{equation}

\section{The $s$-Wave Charge Mode from the Constraints}
\label{Qdensity}
\quad \
In this Appendix will give another derivation of the differential relations \eqref{ArFromEr}, which may be familiar to those who have studied 2D gauge theory. 

\subsection{The Current as a Density}
\label{DensityRed}
\quad \
We begin by recalling an alternative but equivalent formalism to the one presented in \S\ref{charge}, which is in some sense more convenient for a static coordinate system. We begin by recalling the equation of motion \eqref{Proca}
\begin{equation}
	\partial_{\nu}\inp{\sqrt{|\mathsf{g}|}\,\mathsf{F}^{\nu\mu}} = \sqrt{|\mathsf{g}|}\,m_{\mathrm{v}}^2\,\mathsf{A}^{\mu}
	\label{ProcaCoord}
\end{equation}
containing the Lorenz constraint \eqref{LC}
\begin{equation}
	\partial_{\mu}\inp{\sqrt{|\mathsf{g}|}\,m_{\mathrm{v}}^2\,\mathsf{A}^{\mu}} = 0
	\label{LCCoord}
\end{equation}
The Lorenz constraint \eqref{LCCoord} allows us to interpret the Proca equation \eqref{ProcaCoord} as Maxwell's equations with a $D$-dimensional \emph{vector density} source $\boldsymbol{J}^{\mu}$
\begin{equation}
	\boldsymbol{J}^{\mu} \equiv \sqrt{|\mathsf{g}|}\,m_{\mathrm{v}}^2\,A^{\mu}
\end{equation}
which is ``ordinarily" conserved (i.e. conserved in local coordinates) 
\begin{equation}
	\partial_{\mu}\boldsymbol{J}^{\mu} = 0
	\label{Cons}
\end{equation}
Provided we choose coordinates $\mathsf{x}^{\mu}$ such that $\Sigma$ is a slice of constant $\mathsf{x}^0$, the charge \eqref{Q(t)} can equivalently be written as
\begin{equation}
	Q(t) = \int_{\mathsf{x}^0 = t}\,\boldsymbol{J}^{0}
	\label{ChargeDef}
\end{equation}
We will denote the $s$-wave reduction of $\boldsymbol{J}^{\mu}(\mathsf{x})$ by $\boldsymbol{j}^a(x)$, which is defined by
\begin{equation}
	\boldsymbol{j}^a(x) = \frac{1}{\Omega_{(D-2)}}\int_{\mathbb{S}^{(D-2)}}\mathrm{d}^{D-2}\theta\,\boldsymbol{J}^a(x,\theta)
	\label{jfromJ}
\end{equation}
and is explicitly given in terms of the $s$-wave mode $A_a$ by
\begin{equation}
	\boldsymbol{j}^a(x) = r(x)^{D-2}\,\sqrt{|g(x)|}\,m_{\mathrm{v}}^2\,A^a(x)
\end{equation}
The conservation law \eqref{Cons} implies the corresponding conservation law 
\begin{equation}
	\partial_a\boldsymbol{j}^a = 0
	\label{Conss}
\end{equation}

\subsection{A New Definition of the Bulk Charge Mode}
\quad \ 
The $(1 + 1)$-dimensional ``ordinary" (i.e. coordinate) conservation law \eqref{Conss} means that we must have
\begin{equation}
	\boldsymbol{j}^{a}(x) = \frac{1}{\Omega_{(D-2)}}\,\ve^{ab}\partial_{b}Q(x)
	\label{Efromj}
\end{equation}
with $Q$ the $s$-wave reduction of a $D$-dimensional scalar density (i.e. a $(1 + 1)$-dimensional \emph{Dilaton-weighted} scalar density). Here $\ve^{ab}$ denotes the non-covariant $(1 + 1)$-dimensional \emph{coordinate} Levi-Civita symbol, which obeys 
\begin{equation}
	\ve^{10} = -\ve^{01} = 1
\end{equation}
This shows rather explicitly how there is only one field's worth of degrees of freedom---namely $Q$---contained within the $s$-wave mode $A_a$, since the above means that we must have
\begin{align}
	A_a 
	&= \frac{1}{m_{\mathrm{v}}^2}\frac{1}{\Omega_{(D-2)}}\frac{1}{r(x)^{D-2}}\frac{1}{\sqrt{|g|}}\,g_{ab}\,\ve^{bc}\partial_cQ
	\label{QdefApp}
\end{align}
In static patch coordinates, we find
\begin{align}
	A_t(t,r) &= +\frac{1}{m_{\mathrm{v}}^2}\frac{1}{\V}\frac{1}{r^{D-2}}\,f(r)\,\partial_rQ(t,r) \nonumber\\
	A_r(t,r) &= +\frac{1}{m_{\mathrm{v}}^2}\frac{1}{\V}\frac{1}{r^{D-2}}\,\frac{1}{f(r)}\,\partial_tQ(t,r)
	\label{AfromE}
\end{align}
Comparing with \eqref{AtQ}, \eqref{ArQ} of the main text, we see that---with an appropriate choice of integration constant---$Q(t,r)$ is indeed the bulk charge mode defined in the main text. 

To see how to pick the right integration constant, we note that we may calculate the charge $Q_r$ contained within a spatial $2$-ball of radius $r$ centered at the pode via
\begin{align}
	Q_r(t) 
	&\equiv -\intf{0}{r}\mathrm{d}^{D-2}\theta\,\mathrm{d}r'\,\boldsymbol{J}^{t}(t,r')\\
	&= -\Omega_{(D-2)}\intf{0}{r}\mathrm{d}r'\,\boldsymbol{j}^{t}(t,r')\\
	&= +\intf{0}{r}\mathrm{d}r'\,\partial_rQ(t,r')\\
	&= Q(t,r) - Q(t,0)
	\label{QE0}
\end{align}
We can fix the ambiguous constant in the definition \eqref{QdefApp} of $Q(t,r)$ by setting $$Q(0,0) = 0$$ Due to the boundary condition $\partial_tQ|_{\mathrm{pode}} \propto \boldsymbol{j}^r|_{\mathrm{pode}} = 0$, this actually sets $$Q|_{\mathrm{pode}} = 0$$ for all $t$, and allows us to reduce \eqref{QE0} to
\begin{equation}
	Q_r(t) = Q(t,r)
\end{equation}
So we see that $Q$ (with this choice of integration constant) is just the same old charge operator considered in the main text. \eqref{QdefApp} can be regarded as an alternative derivation of \eqref{AtQ}, \eqref{ArQ} of the main text.

\bibliographystyle{JHEP}
\bibliography{ScalarVector_bib} 

\providecommand{\href}[2]{#2}\begingroup\raggedright\begin{thebibliography}{10}

\bibitem{Higuchi:1986ww}
A.~Higuchi, \emph{{Quantization of Scalar and Vector Fields Inside the
  Cosmological Event Horizon and Its Application to Hawking Effect}},
  \href{https://doi.org/10.1088/0264-9381/4/3/029}{\emph{Class. Quant. Grav.}
  {\bfseries 4} (1987) 721}.

\bibitem{Albert}
M.~Grewal, Y.~T.~A. Law and V.~Lochab, \emph{{Spinning bulk and edge
  correlators in de Sitter static patch}}, {\emph{In Preparation} }.

\bibitem{Goheer:2002vf}
N.~Goheer, M.~Kleban and L.~Susskind, \emph{{The Trouble with de Sitter
  space}}, \href{https://doi.org/10.1088/1126-6708/2003/07/056}{\emph{JHEP}
  {\bfseries 07} (2003) 056}
  [\href{https://arxiv.org/abs/hep-th/0212209}{{\ttfamily hep-th/0212209}}].

\bibitem{Marolf:2008hg}
D.~Marolf and I.~A. Morrison, \emph{{Group Averaging for de Sitter free
  fields}}, \href{https://doi.org/10.1088/0264-9381/26/23/235003}{\emph{Class.
  Quant. Grav.} {\bfseries 26} (2009) 235003}
  [\href{https://arxiv.org/abs/0810.5163}{{\ttfamily 0810.5163}}].

\bibitem{Susskind:2021omt}
L.~Susskind, \emph{{De Sitter Holography: Fluctuations, Anomalous Symmetry, and
  Wormholes}}, \href{https://doi.org/10.3390/universe7120464}{\emph{Universe}
  {\bfseries 7} (2021) 464} [\href{https://arxiv.org/abs/2106.03964}{{\ttfamily
  2106.03964}}].

\bibitem{Chandrasekaran:2022cip}
V.~Chandrasekaran, R.~Longo, G.~Penington and E.~Witten, \emph{{An Algebra of
  Observables for de Sitter Space}},
  \href{https://arxiv.org/abs/2206.10780}{{\ttfamily 2206.10780}}.

\bibitem{HV}
H.~Verlinde, ``{A duality between SYK and (2+1)D de Sitter Gravity}.'' 12,
  2019.

\bibitem{Susskind:2021esx}
L.~Susskind, \emph{{Entanglement and Chaos in De Sitter Space Holography: An
  SYK Example}}, \href{https://doi.org/10.22128/jhap.2021.455.1005}{\emph{JHAP}
  {\bfseries 1} (2021) 1} [\href{https://arxiv.org/abs/2109.14104}{{\ttfamily
  2109.14104}}].

\bibitem{Susskind:2022dfz}
L.~Susskind, \emph{{Scrambling in Double-Scaled SYK and De Sitter Space}},
  \href{https://arxiv.org/abs/2205.00315}{{\ttfamily 2205.00315}}.

\bibitem{Susskind:2022bia}
L.~Susskind, \emph{{De Sitter Space, Double-Scaled SYK, and the Separation of
  Scales in the Semiclassical Limit}},
  \href{https://arxiv.org/abs/2209.09999}{{\ttfamily 2209.09999}}.

\bibitem{Susskind:2023hnj}
L.~Susskind, \emph{{De Sitter Space has no Chords. Almost Everything is
  Confined.}}, \href{https://doi.org/10.22128/jhap.2023.661.1043}{\emph{JHAP}
  {\bfseries 3} (2023) 1} [\href{https://arxiv.org/abs/2303.00792}{{\ttfamily
  2303.00792}}].

\bibitem{Narovlansky:2023lfz}
V.~Narovlansky and H.~Verlinde, \emph{{Double-scaled SYK and de Sitter
  Holography}},  \href{https://arxiv.org/abs/2310.16994}{{\ttfamily
  2310.16994}}.

\bibitem{Rahman:2023pgt}
A.~A. Rahman and L.~Susskind, \emph{{Comments on a Paper by Narovlansky and
  Verlinde}},  \href{https://arxiv.org/abs/2312.04097}{{\ttfamily 2312.04097}}.

\bibitem{Rahman:2024vyg}
A.~A. Rahman and L.~Susskind, \emph{{Infinite Temperature is Not So Infinite:
  The Many Temperatures of de Sitter Space}},
  \href{https://arxiv.org/abs/2401.08555}{{\ttfamily 2401.08555}}.

\bibitem{Verlinde:2024znh}
H.~Verlinde, \emph{{Double-scaled SYK, Chords and de Sitter Gravity}},
  \href{https://arxiv.org/abs/2402.00635}{{\ttfamily 2402.00635}}.

\bibitem{Rahman:2024iiu}
A.~A. Rahman and L.~Susskind, \emph{{$p$-Chords, Wee-Chords, and de Sitter
  Space}},  \href{https://arxiv.org/abs/2407.12988}{{\ttfamily 2407.12988}}.

\bibitem{Rahman:2022jsf}
A.~A. Rahman, \emph{{dS JT Gravity and Double-Scaled SYK}},
  \href{https://arxiv.org/abs/2209.09997}{{\ttfamily 2209.09997}}.

\bibitem{Banks:2003cg}
T.~Banks, \emph{{Some thoughts on the quantum theory of de sitter space}},  in
  \emph{{The Davis Meeting on Cosmic Inflation}}, 5, 2003,
  \href{https://arxiv.org/abs/astro-ph/0305037}{{\ttfamily astro-ph/0305037}}.

\bibitem{Banks:2006rx}
T.~Banks, B.~Fiol and A.~Morisse, \emph{{Towards a quantum theory of de Sitter
  space}}, \href{https://doi.org/10.1088/1126-6708/2006/12/004}{\emph{JHEP}
  {\bfseries 12} (2006) 004}
  [\href{https://arxiv.org/abs/hep-th/0609062}{{\ttfamily hep-th/0609062}}].

\bibitem{Anninos:2011af}
D.~Anninos, S.~A. Hartnoll and D.~M. Hofman, \emph{{Static Patch Solipsism:
  Conformal Symmetry of the de Sitter Worldline}},
  \href{https://doi.org/10.1088/0264-9381/29/7/075002}{\emph{Class. Quant.
  Grav.} {\bfseries 29} (2012) 075002}
  [\href{https://arxiv.org/abs/1109.4942}{{\ttfamily 1109.4942}}].

\bibitem{Gorbenko:2018oov}
V.~Gorbenko, E.~Silverstein and G.~Torroba, \emph{{dS/dS and $ T\overline{T}
  $}}, \href{https://doi.org/10.1007/JHEP03(2019)085}{\emph{JHEP} {\bfseries
  03} (2019) 085} [\href{https://arxiv.org/abs/1811.07965}{{\ttfamily
  1811.07965}}].

\bibitem{Coleman:2021nor}
E.~Coleman, E.~A. Mazenc, V.~Shyam, E.~Silverstein, R.~M. Soni, G.~Torroba
  et~al., \emph{{De Sitter microstates from T$ \overline{T} $ +
  \ensuremath{\Lambda}$_{2}$ and the Hawking-Page transition}},
  \href{https://doi.org/10.1007/JHEP07(2022)140}{\emph{JHEP} {\bfseries 07}
  (2022) 140} [\href{https://arxiv.org/abs/2110.14670}{{\ttfamily
  2110.14670}}].

\bibitem{Batra:2024kjl}
G.~Batra, G.~B. De~Luca, E.~Silverstein, G.~Torroba and S.~Yang,
  \emph{{Bulk-local dS$_{3}$ holography: the matter with $ T\overline{T} $ +
  \ensuremath{\Lambda}$_{2}$}},
  \href{https://doi.org/10.1007/JHEP10(2024)072}{\emph{JHEP} {\bfseries 10}
  (2024) 072} [\href{https://arxiv.org/abs/2403.01040}{{\ttfamily
  2403.01040}}].

\bibitem{Batra:2024qju}
G.~Batra, \emph{{Timelike boundaries in de Sitter JT gravity and the Gao-Wald
  theorem}},  \href{https://arxiv.org/abs/2407.08913}{{\ttfamily 2407.08913}}.

\bibitem{Ball:2024hqe}
A.~Ball, Y.~T.~A. Law and G.~Wong, \emph{{Dynamical edge modes and entanglement
  in Maxwell theory}},
  \href{https://doi.org/10.1007/JHEP09(2024)032}{\emph{JHEP} {\bfseries 09}
  (2024) 032} [\href{https://arxiv.org/abs/2403.14542}{{\ttfamily
  2403.14542}}].

\bibitem{Ball:2024xhf}
A.~Ball and Y.~T.~A. Law, \emph{{Dynamical Edge Modes in $p$-form Gauge
  Theories}},  \href{https://arxiv.org/abs/2411.02555}{{\ttfamily 2411.02555}}.

\bibitem{Grewal:2024emf}
M.~Grewal and Y.~T.~A. Law, \emph{{Real-time observables in de Sitter
  thermodynamics}},  \href{https://arxiv.org/abs/2403.06006}{{\ttfamily
  2403.06006}}.

\bibitem{Banks:2022irh}
T.~Banks and P.~Draper, \emph{{Comments on the entanglement spectrum of de
  Sitter space}}, \href{https://doi.org/10.1007/JHEP01(2023)135}{\emph{JHEP}
  {\bfseries 01} (2023) 135}
  [\href{https://arxiv.org/abs/2209.08991}{{\ttfamily 2209.08991}}].

\bibitem{A:2023psv}
S.~A, T.~Banks and W.~Fischler, \emph{{Quantum theory of three-dimensional de
  Sitter space}},
  \href{https://doi.org/10.1103/PhysRevD.109.025011}{\emph{Phys. Rev. D}
  {\bfseries 109} (2024) 025011}
  [\href{https://arxiv.org/abs/2306.05264}{{\ttfamily 2306.05264}}].

\bibitem{Banks:2024cqo}
T.~Banks and P.~Draper, \emph{{Generalized entanglement capacity of de Sitter
  space}}, \href{https://doi.org/10.1103/PhysRevD.110.045025}{\emph{Phys. Rev.
  D} {\bfseries 110} (2024) 045025}
  [\href{https://arxiv.org/abs/2404.13684}{{\ttfamily 2404.13684}}].

\bibitem{Banks:2024lvl}
T.~Banks, \emph{{''Observables'' in de Sitter Quantum Gravity: in Perturbation
  Theory and Beyond}},  \href{https://arxiv.org/abs/2405.01773}{{\ttfamily
  2405.01773}}.

\bibitem{Bonifacio:2018zex}
J.~Bonifacio, K.~Hinterbichler, A.~Joyce and R.~A. Rosen, \emph{{Shift
  Symmetries in (Anti) de Sitter Space}},
  \href{https://doi.org/10.1007/JHEP02(2019)178}{\emph{JHEP} {\bfseries 02}
  (2019) 178} [\href{https://arxiv.org/abs/1812.08167}{{\ttfamily
  1812.08167}}].

\bibitem{Allen:1985ux}
B.~Allen, \emph{{Vacuum States in de Sitter Space}},
  \href{https://doi.org/10.1103/PhysRevD.32.3136}{\emph{Phys. Rev. D}
  {\bfseries 32} (1985) 3136}.

\bibitem{Allen:1987tz}
B.~Allen and A.~Folacci, \emph{{The Massless Minimally Coupled Scalar Field in
  De Sitter Space}},
  \href{https://doi.org/10.1103/PhysRevD.35.3771}{\emph{Phys. Rev. D}
  {\bfseries 35} (1987) 3771}.

\bibitem{Chernikov:1968zm}
N.~A. Chernikov and E.~A. Tagirov, \emph{{Quantum theory of scalar fields in de
  Sitter space-time}}, {\emph{Ann. Inst. H. Poincare Phys. Theor. A} {\bfseries
  9} (1968) 109}.

\bibitem{Bunch:1978yq}
T.~S. Bunch and P.~C.~W. Davies, \emph{{Quantum Field Theory in de Sitter
  Space: Renormalization by Point Splitting}},
  \href{https://doi.org/10.1098/rspa.1978.0060}{\emph{Proc. Roy. Soc. Lond. A}
  {\bfseries 360} (1978) 117}.

\bibitem{Hartle:1983ai}
J.~B. Hartle and S.~W. Hawking, \emph{{Wave Function of the Universe}},
  \href{https://doi.org/10.1103/PhysRevD.28.2960}{\emph{Phys. Rev. D}
  {\bfseries 28} (1983) 2960}.

\bibitem{Add}
A.~A. Rahman, \emph{{More Remarks on Vector Bosons in the Static Patch}},
  {\emph{In Preparation} }.

\bibitem{Higuchi:1986py}
A.~Higuchi, \emph{{Forbidden Mass Range for Spin-2 Field Theory in De Sitter
  Space-time}}, \href{https://doi.org/10.1016/0550-3213(87)90691-2}{\emph{Nucl.
  Phys. B} {\bfseries 282} (1987) 397}.

\bibitem{Anninos:2020hfj}
D.~Anninos, F.~Denef, Y.~T.~A. Law and Z.~Sun, \emph{{Quantum de Sitter horizon
  entropy from quasicanonical bulk, edge, sphere and topological string
  partition functions}},
  \href{https://doi.org/10.1007/JHEP01(2022)088}{\emph{JHEP} {\bfseries 01}
  (2022) 088} [\href{https://arxiv.org/abs/2009.12464}{{\ttfamily
  2009.12464}}].

\bibitem{Lust:2019lmq}
D.~L\"ust and E.~Palti, \emph{{A Note on String Excitations and the Higuchi
  Bound}}, \href{https://doi.org/10.1016/j.physletb.2019.135067}{\emph{Phys.
  Lett. B} {\bfseries 799} (2019) 135067}
  [\href{https://arxiv.org/abs/1907.04161}{{\ttfamily 1907.04161}}].

\bibitem{Higuchi:1986wu}
A.~Higuchi, \emph{{Symmetric Tensor Spherical Harmonics on the $N$ Sphere and
  Their Application to the De Sitter Group SO($N$,1)}},
  \href{https://doi.org/10.1063/1.527513}{\emph{J. Math. Phys.} {\bfseries 28}
  (1987) 1553}.

\end{thebibliography}\endgroup

\end{document}